\documentclass[preprint,pra,superscriptaddress]{revtex4}
\usepackage{amsfonts}
\usepackage{amsmath}
\usepackage{amsthm}
\usepackage{amscd}
\usepackage{amssymb}
\usepackage{subfigure}
\usepackage{amsxtra}
\usepackage{gensymb}
\usepackage{bm}           
\usepackage{bbm}
\usepackage{graphicx}
\usepackage{epstopdf}
\usepackage{color}
\usepackage{times}
\usepackage{mdframed}
\usepackage{array}
\usepackage[colorlinks=true,citecolor=blue,urlcolor=black]{hyperref}

\def\bi#1\ei {\begin{itemize}#1\end{itemize}}
\def\bn#1\en {\begin{enumerate}#1\end{enumerate}}
\def\bea#1\eea {\begin{align}#1\end{align}}
\def\bean#1\eean {\begin{align*}#1\end{align*}}
\def\ben#1\een {\begin{equation*}#1\end{equation*}}
\def\be#1\ee {\begin{equation}#1\end{equation}}
\def\bes#1\ees {\begin{equation}\begin{split}#1\end{split}\end{equation}}
\def\bear#1\eear {\begin{eqnarray}#1\end{eqnarray}}
\def\bear#1\eear {\begin{eqnarray*}#1\end{eqnarray*}}

\emergencystretch=\maxdimen
\newcommand{\beq}{\begin{equation}}
\newcommand{\eeq}{\end{equation}}

\begin{document}

\title{Measurement-Device-Independent Quantum Key Distribution with Leaky Sources}

\author{Weilong Wang}
\email{wwang@com.uvigo.es}
\affiliation{EI~Telecomunicaci\'on,~Department~of~Signal~Theory~and~Communications, University~of~Vigo,~Vigo~E-36310,~Spain }
\affiliation{State Key Laboratory of Mathematical Engineering and Advanced Computing, Zhengzhou, Henan, 450001, China}
\affiliation{Henan Key Laboratory of Network Cryptography Technology, Zhengzhou, Henan, 450001, China}

\author{Kiyoshi Tamaki}
\affiliation{Faculty of engineering, University of Toyama, Gofuku 3190, Toyama 930-8555, Japan}

\author{Marcos Curty}
\affiliation{EI~Telecomunicaci\'on,~Department~of~Signal~Theory~and~Communications, University~of~Vigo,~Vigo~E-36310,~Spain }

\begin{abstract}
Measurement-device-independent quantum key distribution (MDI-QKD) can remove all detection side-channels from quantum communication systems. The security proofs  require, however, that certain assumptions on the sources are satisfied. This includes, for instance, the requirement that there is no information leakage from the transmitters of the senders, which unfortunately is very difficult to guarantee in practice. In this paper we relax this unrealistic assumption by presenting a general formalism to prove the security of MDI-QKD with leaky sources. With this formalism, we analyze the finite-key security of two prominent MDI-QKD schemes \--- a symmetric three-intensity decoy-state MDI-QKD protocol and a four-intensity decoy-state MDI-QKD protocol \--- and determine their robustness against information leakage from both the intensity modulator and the phase modulator of the transmitters. Our work shows that MDI-QKD is feasible within a reasonable time frame of signal transmission given that the sources are sufficiently isolated. Thus, it provides an essential reference for experimentalists to ensure the security of experimental implementations of MDI-QKD in the presence of information leakage.
\end{abstract}
\maketitle


\section{Introduction}\label{intr}
In theory, quantum key distribution (QKD)~\cite{bennett1984quantum, ekert1991quantum,scarani2009security, lo2014secure} provides an information-theoretically secure way to distribute secret keys between two distant parties (commonly known as Alice and Bob). In practice, however, this is not the case. This is so because real devices do not typically conform to the requirements imposed by the security proofs. Indeed, various types of quantum hacking attacks have been proposed and experimentally demonstrated recently, which exploit devices' imperfections in practical QKD systems~\cite{lo2014secure}. To tackle these implementation security loopholes, many efforts have been made, among which device-independent (DI) QKD~\cite{barrett2005no, acin2007device, vazirani2014fully} and measurement-device-independent (MDI) QKD~\cite{lo2012measurement} are two prominent approaches. The security of DI-QKD relies on the violation of a Bell inequality~\cite{bell1964einstein,brunner2014bell} and no knowledge about the inner working of the quantum apparatuses is needed given that the apparatuses are `honest'~\cite{curty2019foiling}, {\it i.e.}, given that they follow the prescriptions of the protocol and not those of Eve. DI-QKD is, however, difficult to implement experimentally with current technology, especially for long distances~\cite{PhysRevLett.105.070501,PhysRevA.84.010304,zapatero2019long}. On the other hand, thanks to its feasibility, MDI-QKD has attracted great attention and has been widely experimentally demonstrated in recent years~\cite{rubenok2013real, da2013proof, liu2013experimental,tang2014experimental,tang2014measurement,yin2016measurement,tang2016measurement,comandar2016quantum}. In terms of security, MDI-QKD closes all side-channels in the detection unit, which significantly simplifies the path towards achieving implementation security, as now one only needs to secure the source. In particular, MDI-QKD typically requires that certain assumptions on the sources are satisfied. Particularly, it requires that Alice's and Bob's transmitters do not leak any unwanted information out of their security zones.

Inspired by the results introduced in~\cite{lucamarini2015practical,tamaki2016decoy,Weilong2018Finite}, which study the information leakage problem in standard decoy-state QKD systems, here we relax such an unrealistic requirement and perform a finite-key security analysis of MDI-QKD with leaky sources. In particular, we focus on information leakage from two main apparatuses within the transmitters, the intensity modulator (IM), which is used to generate decoy states, and the phase modulator (PM), which is used to encode the basis and bit information. For instance, such information leakage might be due to a Trojan-horse attack (THA)~\cite{gisin2006trojan} performed by Eve. In this framework, we evaluate the security of two prominent MDI-QKD protocols: the symmetric three-intensity decoy-state MDI-QKD scheme~\cite{curty2014finite}, and the efficient four-intensity decoy-state MDI-QKD protocol introduced in~\cite{zhou2016making}, which has recently been implemented over a distance of 404 km~\cite{yin2016measurement}. As expected, our results show that MDI-QKD is more sensitive to information leakage than standard decoy-state QKD. Still, we show that MDI-QKD is feasible within a reasonable time frame of signal transmission given that Alice's and Bob's sources are sufficiently isolated. Moreover, we find that when the amount of information leakage is small enough, its effect has a slightly bigger impact on the four-intensity decoy-state MDI-QKD protocol than on the symmetric three-intensity decoy-state MDI-QKD protocol. However, when the amount of information leakage increases, the four-intensity protocol becomes more robust against information leakage than the symmetric three-intensity protocol.

The paper is organized as follows. In Sec.~\ref{ass} we summarize the assumptions that we make in the security analysis. In this section, we also define the symmetric three-intensity decoy-state MDI-QKD protocol. Then, in Sec.~\ref{para} we present the parameter estimation method to determine the secret key rate of the protocol in the presence of information leakage from the IM and the PM. The simulation results for different practical cases are shown in Sec.~\ref{sim}. In Sec.~\ref{XBP}, we evaluate the security of the four-intensity decoy-state MDI-QKD protocol and simulate its secret key rate. Finally, Sec.~\ref{CON} summarizes the main contributions of this paper. The detailed calculations to estimate the relevant parameters that are needed to evaluate the lower bound on the secret key rate are shown in Appendixes.

\section{Assumptions and Protocol Description}\label{ass}

We begin by providing a brief summary of the assumptions that we make on the users' devices in the absence of information leakage:
\begin{enumerate}
 \item Each of Alice and Bob generates perfect phase-randomized weak coherent pulses (WCPs) of the form
 \begin{equation}
{\rho ^{{\gamma ^j}}} = \sum\limits_{n = 0}^\infty {p_n^j|n\rangle\langle n|,}
\end{equation}
where $p_{n}^{j} = {\left( {{\gamma ^j}} \right)^n}{e^{ - {{\gamma ^j} } }}/ {n!} $ is the probability that Alice (Bob) sends an $n$-photon pulse given that she (he) selects the intensity setting $\gamma^j$, and $|n\rangle$ denotes a Fock state with $n$ photons.
 \item The state of a pulse generated by Alice (Bob) is in a single mode and the joint state of all the pulses generated by Alice (Bob) is in a tensor product. That is, we assume that the pulses are not correlated with each other when there is no information leakage. We note that the scenario where the users generate multi-mode signals could be evaluated with the techniques recently introduced in~\cite{Marga2019}, while the case of correlated pulses could be analyzed with the techniques developed in~\cite{Margaridacorrelated}.
 \item There are no intensity fluctuations, {\it i.e.}, the intensity of the pulses generated by Alice and Bob is precisely $\gamma^j$.
 \item Alice and Bob can perfectly encode the bit and basis information, {\it i.e.}, there are no state preparation flaws. We remark that the case of state preparation flaws could be studied using the methods presented in~\cite{tamaki2014loss,Marga2019}
 \item Alice's and Bob's phase modulators modulate only the phase of the pulses, and their intensity modulators modulate only the intensity of the pulses. That is, the information is only encoded in the desired degrees of freedom of the signals.
\end{enumerate}

With these assumptions in mind, next we describe the specific steps of the symmetric three-intensity decoy-state MDI-QKD protocol in detail. Here, we consider a sifting strategy which protects the protocol against the sifting attack~\cite{pfister2016sifting}. This is so because the total number of pulses sent by Alice and Bob is fixed {\it a priori} and, moreover, the termination condition is basis independent~\cite{tamaki2018security}. More specifically, the steps of the protocol are as follows:
\begin{enumerate}
  \item \emph{State preparation:} The first two steps of the protocol are repeated $N$ times, where $N$ is a prefixed number. In each round, Alice and Bob select a basis $\chi \in\{\rm Z,~\rm X\}$ with probabilities $p_{\rm Z}$ and $p_{\rm X}=1-p_{\rm Z}$, and select an intensity setting $\gamma^{j_{\rm A}}$ and $\gamma^{j_{\rm B}}$ with $j_{\rm A},~j_{\rm B} \in \{\rm s,v,w\}$, with probability $p_{j_{\rm A}}$ and $p_{j_{\rm B}}$, respectively. Afterwards, each of them encodes a random bit in a phase-randomized WCP of the chosen intensity in the chosen basis and sends it to the untrusted relay via the quantum channel.
  \item \emph{Measurement:} The untrusted relay is supposed to perform a Bell state measurement on the states received from Alice and Bob and then record the measurement outcomes. However, the relay can behave as Eve decides.
  \item \emph{Announcement of the measurement outcome and random data post-selection:} Once the $N$ rounds of steps 1 and 2 have finished, the relay announces in which rounds he obtained successful measurements together with the corresponding measurement outcomes. For each successful measurement event, Alice selects a fictitious basis ${\rm Z}_{\rm A_c}$ or ${\rm X}_{\rm A_c}$ with probability $p_{\rm Z_{A_c}}$ and $p_{\rm X_{A_c}}=1-p_{\rm Z_{A_c}}$, respectively, and then she announces her fictitious basis choices.
  \item \emph{Sifting:} If Alice's choice is the ${\rm X}_{\rm A_c}$ basis, Bob announces his state preparation basis choice but Alice does not announce hers and then they discard the corresponding data. If Alice's choice is the ${\rm Z}_{\rm A_c}$ basis, both Alice and Bob announce their state preparation basis choices as well as their intensity settings. We denote by $Z^{j_{\rm A}j_{\rm B}}$ ($X^{j_{\rm A}j_{\rm B}}$) the set of indexes that identify the successful measurement events when Alice and Bob select the intensity settings $\gamma^{j_{\rm A}}$ and $\gamma^{j_{\rm B}}$, respectively, Alice chooses the fictitious basis ${\rm Z}_{\rm A_c}$, and both of them select the Z (X) basis. If the sifting conditions $| Z^{j_{\rm A}j_{\rm B}} | \ge N^{j_{\rm A}j_{\rm B}}_{\rm Z}$ and $| X^{j_{\rm A}j_{\rm B}} |\ge N^{j_{\rm A}j_{\rm B}}_{\rm X}$ are satisfied for all ${j_{\rm A},~j_{\rm B}}\in \{\rm s,v,w\}$, where $N^{j_{\rm A}j_{\rm B}}_{\rm Z}$ and $N^{j_{\rm A}j_{\rm B}}_{\rm X}$ are prefixed threshold values, Alice and Bob proceed to execute the next steps of the protocol. If the sifting conditions are not satisfied, the protocol aborts.
  \item \emph{Parameter estimation:} Alice and Bob estimate a lower bound, which we denote by $N^{L}_{{\rm click,00,\rm {ss}}|\rm Z}$ ($N^{L}_{{\rm click,11,\rm {ss}}|\rm Z}$), on the number of successful measurement events in the sifted key data set ${Z}^{\rm ss}$, in which both of them sent vacuum (single-photon) pulses. Also they use all the data in the sets ${Z}^{k_{\rm A}k_{\rm B}}$ and ${X}^{j_{\rm A}j_{\rm B}}$, except that in the set $ Z^{\rm ss} $, to estimate an upper bound on the single-photon phase error rate in the sifted key date set ${Z}^{\rm ss}$, which we denote by $e^{\rm U}_{\rm ph}$.
  \item {\it Information reconciliation and privacy amplification:} Alice and Bob perform an error correction step for a predetermined quantum bit error rate (QBER), which we denote by $E_{\rm Z}^{\rm ss}$. Then Alice computes a hash of the sifted key data in $Z^{\rm ss}$ by using a random universal$_{\rm 2}$ hash function~\cite{carter1979universal} and sends Bob the hash value together with the hash function. Bob uses the hash function to compute a hash of his corrected sifted key data and checks if the hash value coincides with that of Alice. If both hash values coincide, this error verification step guarantees that they share identical keys after error correction except for an exponentially small probability. If this step succeeds, then they perform a privacy amplification step by applying a random universal$_{\rm 2}$ hash function to distill the final secret key.
\end{enumerate}

Note that the sifting condition in Step 4 of the above protocol is only for data processing, and it is not related to the termination of the quantum communication steps, {\it i.e.}, Steps 1 and 2, which is basis independent. Therefore, as indicated above, the protocol is secure against the sifting attack~\cite{tamaki2018security}.
\section{Parameter estimation method for leaky sources}\label{para}

In this section we present a method to estimate the relevant parameters that are required to evaluate the secret key rate formula in the presence of information leakage. For concreteness, we consider the security analysis introduced in~\cite{mizutani2015finite}, which provides a lower bound on the secret key length, $\ell$, given by
\begin{equation}
\ell \geq  N^{L}_{{\rm click,00,\rm {ss}}|\rm Z} + N^{L}_{{\rm click,11,\rm {ss}}|\rm Z} \left[1 - H\left({e^{\rm U}_{\rm ph}}\right)\right] - leak{_{\rm EC}} - {\log _2}\frac{2}{{{\varepsilon^2 _{{\rm{sec}}}} - \varepsilon }} - {\log _2}\frac{2}{{{\varepsilon _{{\rm{cor}}}}}} ,\label{MDIkey}
\end{equation}
where $H( x ) =  - x{\log _2}( x ) - ( {1 - x} ){\log _2}( {1 - x} )$ is the binary Shannon entropy function. The parameter $leak_{\rm EC}$ is the amount of syndrome information declared by Alice in the error correction step of the protocol, given by $leak_{\rm EC}=|Z^{\rm ss}|f_{\rm EC}H(E_{\rm Z}^{\rm ss})$ for simplicity, where the parameter $f_{\rm EC}$ is the efficiency of the error correction code. The quantities $\varepsilon _{{\rm{sec}}}$ and $\varepsilon _{{\rm{cor}}}$ are the secrecy and correctness parameters of the protocol, respectively, and $\varepsilon  \leq 1 - {\varepsilon _{\rm Z,00}}{\varepsilon _{\rm Z,11}}{\varepsilon _{\rm ph,11}}$ with $\varepsilon _{\rm Z,00}$, $\varepsilon _{\rm Z,11}$ and $\varepsilon _{\rm ph,11}$ being defined as the success probabilities when estimating the quantities $N^{L}_{{\rm click,00,\rm {ss}}|\rm Z}$, $N^{L}_{{\rm click,11,\rm {ss}}|\rm Z}$ and $e^{\rm U}_{\rm ph}$, respectively. In other words, $\varepsilon$ denotes the failure probability that at least one of the estimations of $N^{L}_{{\rm click,00,\rm {ss}}|\rm Z}$, $N^{L}_{{\rm click,11,\rm {ss}}|\rm Z}$ and $e^{\rm U}_{\rm ph}$ is incorrect.

\begin{figure}[!t]
\centerline{\includegraphics*[scale=0.55]{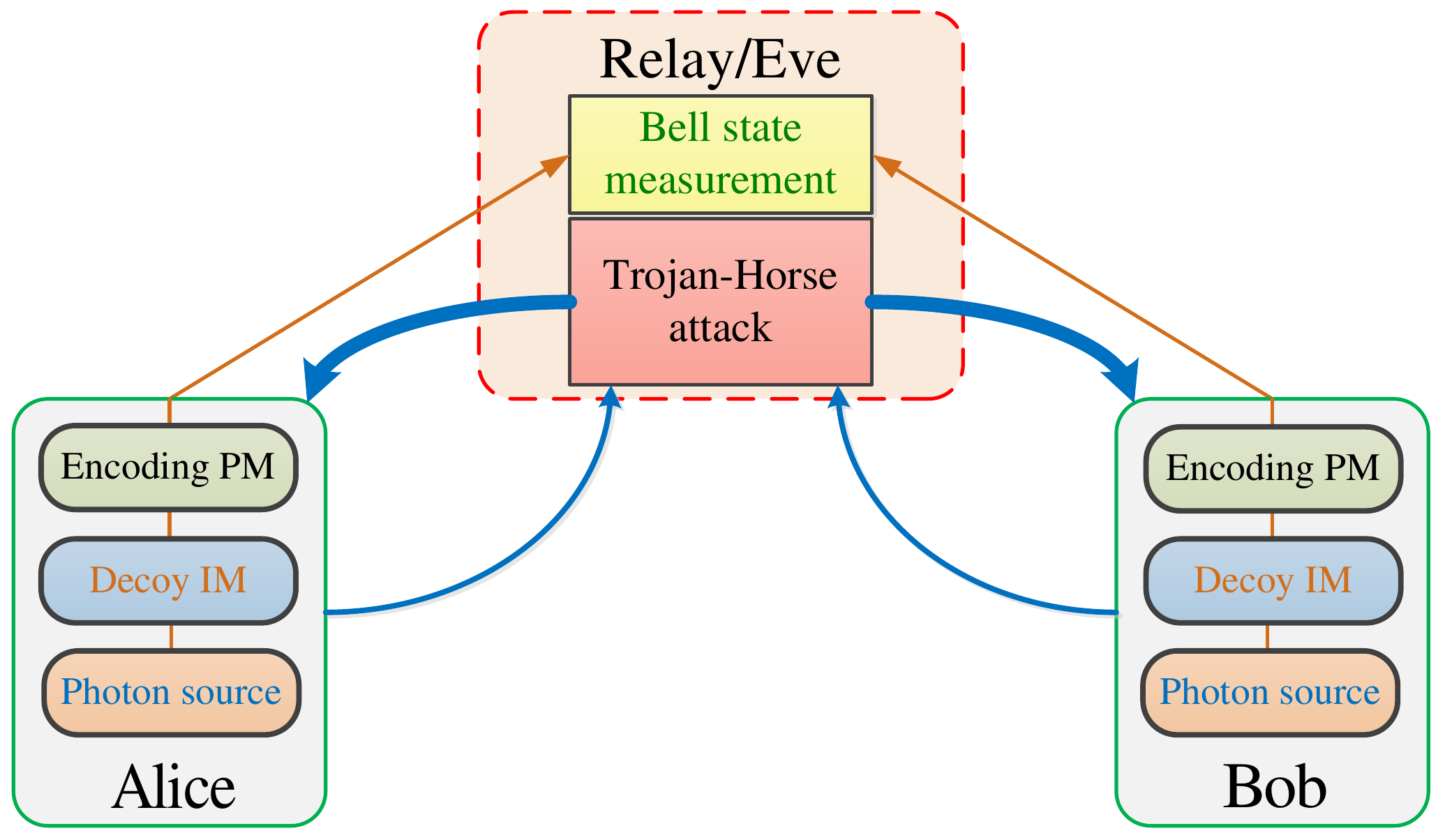}}
\caption{\footnotesize Each of Alice and Bob uses a photon source to prepare phase-randomised WCPs. Decoy states are generated by means of an intensity modulator (Decoy IM). The bit and basis information of the pulses are encoded with a state encoding setup (Encoding PM). The relay is supposed to perform a Bell state measurement on the incoming pulse pairs. In a THA, Eve actively sends bright light pulses (thick blue arrows) into Alice's and Bob's devices to trigger the emission of side-channel signals. Then, Eve measures the back-reflected light (thin blue arrows) to extract information about Alice's and Bob's internal settings. Note that since the relay is untrusted ({\it i.e.}, it can be even Eve), in this figure we consider that it is the relay who performs the THA.}
\label{MDITha}
\end{figure}

In the following we show how to estimate the quantities $N^{L}_{{\rm click,00,\rm {ss}}|\rm Z}$, $N^{L}_{{\rm click,11,\rm {ss}}|\rm Z}$ and $e^{\rm U}_{\rm ph}$ in the presence of information leakage. For concreteness, we shall assume that the information leakage is due to a THA performed by an active Eve. In this THA against the MDI-QKD system, Eve separately sends bright light into Alice's and Bob's devices and then measures the back-reflected light. In so doing, she can obtain partial information about Alice's and Bob's internal settings for each experimental trial. See Fig. \ref{MDITha} for an illustration of Eve's THA. We remark, however, that our method is general and can be applied to analyze passive information leakage scenarios as well.

Importantly, and similar to the analysis performed in~\cite{Weilong2018Finite}, a THA against the IM affects the estimation of the parameters $N^{L}_{{\rm click,00,ss}|\rm Z}$, $N^{L}_{{\rm click,11,ss}|\rm Z}$ and $e^{\rm U}_{\rm ph}$, while a THA against the PM only influences the value of $e^{\rm U}_{\rm ph}$. Below, we follow the structure introduced in~\cite{Weilong2018Finite} to present the estimation method in different steps. First, we mathematically relate the expected numbers of events associated with different intensity settings in the presence of information leakage. Second, we use Azuma's inequality~\cite{azuma1967weighted} to relate these expected numbers of events to the corresponding actual numbers of events taking into account some bounded deviation terms. These relations impose some constraints on the relevant parameters needed to evaluate Eq.~(\ref{MDIkey}). Note that, to use Azuma's inequality, the number of trials must be fixed {\it a priori, i.e.}, before the trials start~\cite{tamaki2018security}. Finally, we estimate these relevant parameters given the constraints provided by the mathematical relations obtained in the previous step. This last step can be done by using, for instance, linear programming techniques~\cite{vanderbei2015linear}.

\subsection{THA against the intensity modulator}\label{THAIM}
Here, we analyze a THA targeted against the intensity modulator (IM), which is used to generate decoy states. To simplify the analysis, we first consider an asymptotic scenario where Alice and Bob send an infinite number of pulses.
\subsubsection{The Asymptotic Limit}\label{THAIMa}
Let us denote the intensity settings of Alice and Bob in the $i$th trial of the protocol by $\gamma^{j_{\rm A},i}$ and $\gamma^{j_{\rm B},i}$ with $j_{\rm A},j_{\rm B}\in \left\{ {{{\rm s}},~ {{\rm v}},~ {{\rm w}}} \right\}$. Also, suppose that Eve prepares and sends Alice (Bob) a probe system $E_{\rm p}^{\rm A}$ ($E_{\rm p}^{\rm B}$) which could be entangled with an ancilla system $E_{\rm a}^{\rm A}$ ($E_{\rm a}^{\rm B}$) stored in her quantum memory. Here, for simplicity, we shall assume that the state of systems $E_{\rm p}^{\rm A}$ and $E_{\rm a}^{\rm A}$ is not correlated with that of systems $E_{\rm p}^{\rm B}$ and $E_{\rm a}^{\rm B}$. However, we remark that our formalism can be adapted to the correlated case as well. Afterwards, Eve performs a joint measurement on the pulses emitted by Alice (Bob) together with the system $E_{\rm a}^{\rm A}$ ($E_{\rm a}^{\rm B}$) and the back-reflected light from $E_{\rm p}^{\rm A}$ ($E_{\rm p}^{\rm B}$), which is denoted by $E_{\rm p}^{\rm A'}$ ($E_{\rm p}^{\rm B'}$). Let ${\rho _{\chi,nm}^{\gamma ^{j_{\rm A}}\gamma ^{j_{\rm B}},i}}$ represent the normalized joint state of Alice's $n$-photon pulse and Bob's $m$-photon pulse when they select the intensity settings $\gamma ^{j_{\rm A}}$ and $\gamma ^{j_{\rm B}}$, respectively, with the same basis choice $\chi$, together with the systems $E_{\rm a}^{\rm A}$, $E_{\rm a}^{\rm B}$, $E_{\rm p}^{\rm A'}$ and $E_{\rm p}^{\rm B'}$ in the $i$th trial. Now, it is important to determine how well Eve can distinguish the states ${\rho _{\chi,nm}^{\gamma ^{j_{\rm A}}\gamma ^{j_{\rm B}},i}}$ for different intensity settings. In particular, we consider how well she can distinguish the intensity settings $\gamma^{j_{\rm A}}$ and $\gamma^{j_{\rm B}}$ from, say, $\gamma^{k_{\rm A}}$ and ${\gamma^{k_{\rm B}}}$ or $\gamma^{l_{\rm A}}$ and ${\gamma^{l_{\rm B}}}$ in each trial, where $j_{\rm A},j_{\rm B},k_{\rm A},k_{\rm B},l_{\rm A},l_{\rm B} \in \left\{\rm s, v,w \right\}$ and $j_{\rm A} \ne k_{\rm A},l_{\rm A}$ and $j_{\rm B} \ne k_{\rm B},l_{\rm B}$. All these quantum systems are listed in Table.~\ref{sys1}.

\begin{table}[tbp]
\caption{\footnotesize {Quantum systems defined in a THA against the IM}}
\centering \label{sys1}
\begin{tabular}{|m{2cm}<{\centering}|m{12cm}<{\centering}|}
\hline
  $E_{\rm p}^{\rm A}$ ($E_{\rm p}^{\rm B}$) & Eve's probe system sent to Alice (Bob)\\
  \hline
  $E_{\rm a}^{\rm A}$ ($E_{\rm a}^{\rm B}$) & Eve's ancillary system which could be entangled with $E_{\rm p}^{\rm A}$ ($E_{\rm p}^{\rm B}$)\\
  \hline
  $E_{\rm p}^{\rm A'}$ ($E_{\rm p}^{\rm B'}$) & Back-reflected light from $E_{\rm p}^{\rm A}$ ($E_{\rm p}^{\rm B}$)\\
  \hline
  ${\rho _{\chi,nm}^{\gamma ^{j_{\rm A}}\gamma ^{j_{\rm B}},i}}$  & Normalized joint state of Alice's $n$-photon pulse and Bob's $m$-photon pulse when they select the intensity settings $\gamma ^{j_{\rm A}}$ and $\gamma ^{j_{\rm B}}$, respectively, with the same basis choice $\chi$, and the systems $E_{\rm a}^{\rm A}$, $E_{\rm a}^{\rm B}$, $E_{\rm p}^{\rm A'}$, $E_{\rm p}^{\rm B'}$ in the $i$th trial\\
  \hline
\end{tabular}
\end{table}

According to the trace distance argument~\cite{nielsen2000quantum,tamaki2016decoy,Weilong2018Finite}, we have that
\begin{equation}\label{Trace}
\sum\limits_{\omega  \in \Omega } {\left| {P( {\omega | {{\rho _{\chi,nm}^{\gamma^{j_{\rm A}}\gamma^{j _{\rm B}},i}}}} ) - P( {\omega | {{\sigma _{\chi,nm}^{\gamma ^{k_{\rm A}l_{\rm A}}\gamma ^{k_{\rm B}l_{\rm B}},i}}} } )} \right| \le D^{j_{\rm A}j_{\rm B},k_{\rm A}k_{\rm B}l_{\rm A}l_{\rm B},i}_{\chi,nm}},
\end{equation}
where $\Omega$ is a set of physical events that satisfies $\sum\limits_{\omega  \in \Omega } {P \left( \omega  \right)}  = 1$, $P\left( {\omega \left| \rho  \right.} \right)$ is the conditional probability that the event $\omega$ occurs given a state $\rho$, and $\sigma _{\chi,nm}^{\gamma ^{k_{\rm A}l_{\rm A}}\gamma ^{k_{\rm B}l_{\rm B}},i}: = {q_{nmkl}}{\rho _{\chi,nm}^{\gamma^{k_{\rm A}}\gamma^{k_{\rm B}},i}} + \left( {1 - {q_{nmkl}}} \right){\rho _{\chi,nm}^{\gamma^{l_{\rm A}}\gamma^{l_{\rm B}},i}}$, where ${q_{nmkl}} = {{{p_{{k_{\rm{A}}}}}{p_{k_{\rm{B}}}}p_n^{{k_{\rm{A}}}}p_m^{{k_{\rm{B}}}}}}/({{{p_{k_{\rm{A}}}}{p_k}_{_{\rm{B}}}
p_n^{{k_{\rm{A}}}}p_m^{{k_{\rm{B}}}} +{p_{{l_{\rm{A}}}}}{p_{{l_{\rm{B}}}}}p_n^{{l_{\rm{A}}}}p_m^{{l_{\rm{B}}}}}})$ is a normalization factor. That is, $\sigma _{\chi,nm}^{\gamma ^{k_{\rm A}l_{\rm A}}\gamma ^{k_{\rm B}l_{\rm B}},i}$ corresponds to the normalized joint state of Alice's $n$-photon pulse and Bob's $m$-photon pulse when they select the intensity settings $\gamma^{k_{\rm A}}$ and $\gamma^{k_{\rm B}}$ or $\gamma^{l_{\rm A}}$ and $\gamma^{l_{\rm B}}$, respectively, with the same basis choice $\chi$, together with the systems $E_{\rm a}^{\rm A}$, $E_{\rm a}^{\rm B}$, $E_{\rm p}^{\rm A'}$ and $E_{\rm p}^{\rm B'}$ in the $i$th trial. We remark that more general cases with up to eight different combinations of intensity settings ({\it i.e.}, all combinations of intensity settings except $\gamma^{j_{\rm A}}$ and $\gamma^{j_{\rm B}}$) could be considered here. The parameter $D^{j_{\rm A}j_{\rm B},k_{\rm A}k_{\rm B}l_{\rm A}l_{\rm B},i}_{\chi,nm}$ denotes the trace distance between the states $\rho _{\chi,nm}^{\gamma^{j_{\rm A}}\gamma^{j _{\rm B}},i}$ and $\sigma _{\chi,nm}^{\gamma ^{k_{\rm A}l_{\rm A}}\gamma ^{k_{\rm B}l_{\rm B}},i}$ and it is given by
\begin{equation}
D^{j_{\rm A}j_{\rm B},k_{\rm A}k_{\rm B}l_{\rm A}l_{\rm B},i}_{\chi,nm}=\frac{1}{2}{\rm Tr}\left[\sqrt{(\rho _{\chi,nm}^{\gamma^{j_{\rm A}}\gamma^{j _{\rm B}},i}-\sigma _{\chi,nm}^{\gamma ^{k_{\rm A}l_{\rm A}}\gamma ^{k_{\rm B}l_{\rm B}},i})^2}\right].
\end{equation}Basically, Eq.~(\ref{Trace}) quantifies how well the states ${{\rho _{\chi,nm}^{\gamma^{j_{\rm A}}\gamma^{j _{\rm B}},i}}}$ and ${{\sigma _{\chi,nm}^{\gamma ^{k_{\rm A}l_{\rm A}}\gamma ^{k_{\rm B}l_{\rm B}},i}}}$ can be distinguished from each other.

Specially, let $\Omega  = \left\{ {\rm{click},~no~ click} \right\}$, where ``click'' (``no click'') represents a successful (unsuccessful) measurement event at the relay and let ${\rm{P}}{{\rm{r}}^i}\left( {{\rm{click}}\left| {nm,{j_{\rm{A}}}{j_{\rm{B}}},\chi } \right.} \right)$ denote the conditional probability that the relay obtains a ``click'' given the state ${\rho _{\chi,nm}^{\gamma ^{j_{\rm A}}\gamma ^{j_{\rm B}},i}}$. Then according to Eq.~(\ref{Trace}) we have that
\begin{equation}\label{MDIPD}
\begin{array}{*{20}{l}}
{\left| {{\rm{P}}{{\rm{r}}^i}\left( {{\rm{click}}\left| {nm,{j_{\rm{A}}}{j_{\rm{B}}},\chi } \right.} \right) - \left[ {{q_{nmkl}}{\rm{P}}{{\rm{r}}^i}\left( {{\rm{click}}\left| {nm,{k_{\rm{A}}}{k_{\rm{B}}},\chi } \right.} \right) + \left( {1 - {q_{nmkl}}} \right){\rm{P}}{{\rm{r}}^i}\left( {{\rm{click}}\left| {nm,{l_{\rm{A}}}{l_{\rm{B}}},\chi } \right.} \right)} \right]} \right|}\\
{ \le D_{\chi,nm }^{{j_{\rm{A}}}{j_{\rm{B}}},{k_{\rm{A}}}{k_{\rm{B}}}{l_{\rm{A}}}{l_{\rm{B}}},i}.}
\end{array}
\end{equation}

By multiplying both sides of Eq.~(\ref{MDIPD}) by $p_{j_{\rm A}}p_{j_{\rm B}}p_n^{j_{\rm A}}p_m^{j_{\rm B}}$ and taking the sum over $i = \{ 1,2,...,{N_\chi }\}$, where $N_{\chi}$ denotes the number of events when both Alice and Bob choose the $\chi$ basis, we obtain
\begin{equation}
\begin{array}{*{20}{l}}\label{MDITHA}
\Bigg| \sum\limits_{i = 1}^{{N_\chi }} {{\rm{P}}{{\rm{r}}^i}\left( {{\rm{click}},nm,{j_{\rm{A}}}{j_{\rm{B}}}\left| \chi  \right.} \right)}  - {p_{{j_{\rm{A}}}}}{p_{{j_{\rm{B}}}}}p_n^{{j_{{\rm{A}}}}}p_m^{{j_{{\rm{B}}}}}\times\\
\sum\limits_{i = 1}^{{N_\chi }} {\left[ {{q_{nmkl}}\frac{{{\rm{P}}{{\rm{r}}^i}\left( {{\rm{click}},nm,{k_{\rm{A}}}{k_{\rm{B}}}\left| \chi  \right.} \right)}}{{{p_{{k_{\rm{A}}}}}{p_{{k_{\rm{B}}}}}p_n^{{k_{_{\rm{A}}}}}p_n^{{k_{_{\rm{B}}}}}}} + \left( {1 - {q_{nmkl}}} \right)\frac{{{\rm{P}}{{\rm{r}}^i}\left( {{\rm{click}},nm,{l_{\rm{A}}}{l_{\rm{B}}}\left| \chi  \right.} \right)}}{{{p_{{l_{\rm{A}}}}}{p_{{l_{\rm{B}}}}}p_n^{{l_{_{\rm{A}}}}}p_m^{{l_{_{\rm{B}}}}}}}} \right]} \Bigg|\\
{ \le {p_{{j_{\rm{A}}}}}{p_{{j_{\rm{B}}}}}p_n^{{j_{_{\rm{A}}}}}p_m^{{j_{_{\rm{B}}}}}{N_\chi }D^{{j_{\rm{A}}}{j_{\rm{B}}},{k_{\rm{A}}}{k_{\rm{B}}}{l_{\rm{A}}}{l_{\rm{B}}}}_{\chi,nm },}
\end{array}
\end{equation}
where ${{\rm{P}}{{\rm{r}}^i}\left( {{\rm{click}},nm,{j_{\rm{A}}}{j_{\rm{B}}}\left| \chi  \right.} \right)}$ is the conditional probability that Alice selects the intensity setting $\gamma^{j_{\rm A}}$ and sends an $n$-photon pulse, Bob selects the intensity setting $\gamma^{j_{\rm B}}$ and sends an $m$-photon pulse, and the relay obtains a successful measurement result given that they both select the $\chi$ basis. In Eq.~(\ref{MDITHA}) we have used the definition $D^{{j_{\rm{A}}}{j_{\rm{B}}},{k_{\rm{A}}}{k_{\rm{B}}}{l_{\rm{A}}}{l_{\rm{B}}}}_{\chi,nm }=\frac{1}{N_{}\chi}\sum\limits_{i = 1}^{{N_\chi }} {D_{\chi,nm }^{{j_{\rm{A}}}{j_{\rm{B}}},{k_{\rm{A}}}{k_{\rm{B}}}{l_{\rm{A}}}{l_{\rm{B}}},i}} $.

The quantity $\sum\limits_{i = 1}^{{N_\chi }} {{\rm{P}}{{\rm{r}}^i}\left( {{\rm{click}},nm,{j_{\rm{A}}}{j_{\rm{B}}}\left| \chi  \right.} \right)}$ corresponds to the conditional expected number of events, which we shall denote by $\mathcal{E}_{{\rm{click}},nm,{j_{\rm{A}}}{j_{\rm{B}}}|\chi}$. Then, we have that Eq.~(\ref{MDITHA}) can be rewritten as
\begin{equation}\label{MDIND}
\begin{array}{*{20}{l}}
\left| \mathcal{E}_{{\rm{click}},nm,{j_{\rm{A}}}{j_{\rm{B}}}|\chi} - \left[ {q_{nmkl}
\frac{p_{{j_{\rm{A}}}}{p_{{j_{\rm{B}}}}}p_n^{{j_{{\rm{A}}}}}p_m^{{j_{{\rm{B}}}}}}{{{p_{{k_{\rm{A}}}}}
{p_{{k_{\rm{B}}}}}p_n^{{k_{{\rm{A}}}}}p_m^{{k_{{\rm{B}}}}}}}\mathcal{E}_{{\rm{click}},nm,{k_{\rm{A}}}
{k_{\rm{B}}}|\chi} + \left( {1 - {q_{nmkl}}} \right)\frac{p_{{j_{\rm{A}}}}{p_{{j_{\rm{B}}}}}p_n^{{j_{{\rm{A}}}}}p_m^{{j_{{\rm{B}}}}}}{{{p_{{l_{\rm{A}}}}}
{p_{{l_{\rm{B}}}}}p_n^{{l_{{\rm{A}}}}}p_m^{{l_{{\rm{B}}}}}}}\mathcal{E}_{{\rm{click}},nm,{l_{\rm{A}}}{l_{\rm{B}}}|\chi}} \right] \right| \\
\le {p_{{j_{\rm{A}}}}}{p_{{j_{\rm{B}}}}}p_n^{{j_{_{\rm{A}}}}}p_m^{{j_{_{\rm{B}}}}}{N_\chi }D^{{j_{\rm{A}}}{j_{\rm{B}}},{k_{\rm{A}}}{k_{\rm{B}}}{l_{\rm{A}}}{l_{\rm{B}}}}_{\chi,nm }.
\end{array}
\end{equation}

If we take, for instance, the particular case where $l_{\rm{A}}=k_{\rm{A}}$ and $l_{\rm{B}}=k_{\rm{B}}$, then Eq. (\ref{MDIND}) can be written as:
\begin{equation}\label{MDIYD}
\left| \mathcal{E}_{{\rm{click}},nm,{j_{\rm{A}}}{j_{\rm{B}}}|\chi} - \frac{p_{{j_{\rm{A}}}}{p_{{j_{\rm{B}}}}}p_n^{{j_{{\rm{A}}}}}p_m^{{j_{{\rm{B}}}}}}{{{p_{{k_{\rm{A}}}}}
{p_{{k_{\rm{B}}}}}p_n^{{k_{{\rm{A}}}}}p_m^{{k_{{\rm{B}}}}}}}\mathcal{E}_{{\rm{click}},nm,{k_{\rm{A}}}
{k_{\rm{B}}}|\chi} \right| \le {p_{{j_{\rm{A}}}}}{p_{{j_{\rm{B}}}}}p_n^{{j_{_{\rm{A}}}}}p_m^{{j_{_{\rm{B}}}}}{N_\chi }D^{{j_{\rm{A}}}{j_{\rm{B}}},{k_{\rm{A}}}{k_{\rm{B}}}}_{\chi,nm },
\end{equation}
where
\begin{equation}
D^{{j_{\rm{A}}}{j_{\rm{B}}},{k_{\rm{A}}}{k_{\rm{B}}}}_{\chi,nm } = \frac{1}{{{N_\chi }}}\sum\limits_{i = 1}^{{N_\chi }} {D_{\chi,nm }^{{j_{\rm{A}}}{j_{\rm{B}}} ,{k_{\rm{A}}}{k_{\rm{B}}} ,i}} : = \frac{1}{{{N_\chi }}}\sum\limits_{i = 1}^{{N_\chi }} {{\rm{Tr}}} \left[ \sqrt{\left({\rho _{\chi,nm}^{\gamma^{j_{\rm A}}\gamma^{j _{\rm B}},i}}-\rho _{\chi,nm}^{\gamma^{k_{\rm A}}\gamma^{k _{\rm B}},i}\right)^2} \right].
\end{equation}
Equivalently, Eq~(\ref{MDIYD}) can be written as:
\begin{equation}\label{MDIYDel}
\mathcal{E}_{{\rm{click}},nm,{j_{\rm{A}}}{j_{\rm{B}}}|\chi} = \frac{p_{{j_{\rm{A}}}}{p_{{j_{\rm{B}}}}}p_n^{{j_{{\rm{A}}}}}p_m^{{j_{{\rm{B}}}}}}{{{p_{{k_{\rm{A}}}}}
{p_{{k_{\rm{B}}}}}p_n^{{k_{{\rm{A}}}}}p_m^{{k_{{\rm{B}}}}}}}\mathcal{E}_{{\rm{click}},nm,{k_{\rm{A}}}
{k_{\rm{B}}}|\chi} + {\Delta_{\chi,nm} ^{{j_{\rm{A}}}{j_{\rm{B}}}{k_{\rm{A}}}{k_{\rm{B}}}}},
\end{equation}
where the parameter ${\Delta_{\chi,nm} ^{{j_{\rm{A}}}{j_{\rm{B}}}{k_{\rm{A}}}{k_{\rm{B}}}}} \in \left[ { - p_{{j_{\rm{A}}}}{p_{{j_{\rm{B}}}}}p_n^{{j_{{\rm{A}}}}}p_m^{{j_{{\rm{B}}}}}N_{\chi}{D^{{j_{\rm{A}}}
{j_{\rm{B}}} ,{k_{\rm{A}}}{k_{\rm{B}}} }_{\chi,nm }},~p_{{j_{\rm{A}}}}{p_{{j_{\rm{B}}}}}p_n^{{j_{{\rm{A}}}}}p_m^{{j_{{\rm{B}}}}}N_{\chi}{D^{{j_{\rm{A}}}
{j_{\rm{B}}} ,{k_{\rm{A}}}{k_{\rm{B}}}}_{\chi,nm }}} \right]$.

Note that, by considering different values for the parameters $\{l_{\rm{A}},k_{\rm{A}},l_{\rm{B}},k_{\rm{B}}\}$, one can obtain
similar equations to Eq.~(\ref{MDIYDel}) that relate the expected numbers of events corresponding to different intensity settings. These equations can be used as linear constraints to the estimation procedure.

Note that in the asymptotic limit where $N_{\chi}\rightarrow \infty$, we have that the actual number of events converge to the expected number of events and therefore one can directly use the constraints given by Eq.~(\ref{MDIYDel}).

\subsubsection{The Finite-Key Regime}
In the previous section, we have derived mathematical relations between the expected numbers of events associated with different intensity settings in the asymptotic limit. By applying Azuma's inequality~\cite{azuma1967weighted}, this analysis can be extended to the realistic regime where Alice and Bob send a finite number ($N$) of pulses. For this, one can consider a virtual scenario in which Alice and Bob first decide the basis $\chi$ for each of the total $N$ rounds. And thus the value of the quantity $N_{\chi}$ is now fixed. To be precise, in such a fictitious scenario we perform a delayed choice of the intensity setting ``\emph{after}" finishing all the basis choices in the actual protocol. Therefore, $N_{\chi}$ is fixed, and then we can use Azuma's inequality for the decoy-state method. Note that the Kraus operator acting on the $i$th pulse depends on all the previous intensity choices, all the basis choices and Eve's arbitrary operation that is dependent on all the announcements she made. Importantly, the trace distance argument is still valid thanks to the generality of the trace distance as well as the fact that Eve does not know Alice and Bob's intensity information for the $i$th pulse in advance. According to Azuma's inequality, we have that
\begin{equation}\label{MDID}
{{\mathcal E}_{{\rm{click}},nm,{j_{\rm{A}}}{j_{\rm{B}}}|\chi }} \equiv \sum\limits_{i = 1}^{{N_\chi }} {{\rm{P}}{{\rm{r}}^i}({\rm{click}},nm,{j_{\rm{A}}}{j_{\rm{B}}}\left| {\chi } \right.)}  = {N_{{\rm{click}},nm,{j_{\rm{A}}}{j_{\rm{B}}}|\chi }} + \delta _{\chi ,nm}^{{j_{\rm{A}}}{j_{\rm{B}}}}, \end{equation}
where $N_{\chi}$ is the $actual$ number of trials, and $ {N_{{\rm{click}},nm,{j_{\rm{A}}}{j_{\rm{B}}}|\chi }} $ is the actual number of such events. The deviation term $\delta _{\chi ,nm}^{{j_{\rm{A}}}{j_{\rm{B}}}}$ lies in the interval $[ { - \Delta _{\chi,nm}^{{j_{\rm{A}}}{j_{\rm{B}}}},\widehat{\Delta} _{\chi,nm}^{{j_{\rm{A}}}{j_{\rm{B}}}}} ]$ except for a small error probability $\varepsilon_{\chi,nm}^{{j_{\rm{A}}}{j_{\rm{B}}}}+\widehat{\varepsilon}_{\chi,nm}^{{j_{\rm{A}}}{j_{\rm{B}}}}$
where the bounds $ \Delta _{\chi,nm}^{{j_{\rm{A}}}{j_{\rm{B}}}}$ and $\widehat{\Delta} _{\chi,nm}^{{j_{\rm{A}}}{j_{\rm{B}}}}$ are given by $\Delta_{\chi,nm}^{{j_{\rm{A}}}{j_{\rm{B}}}} = f( {{N_{\chi} },\varepsilon_{\chi,nm}
^{{j_{\rm{A}}}{j_{\rm{B}}}}} )$ and $\widehat{\Delta}_{\chi,nm}^{{j_{\rm{A}}}{j_{\rm{B}}}} = f( {{N_{\chi}},\widehat{\varepsilon}_{\chi,nm}^{{j_{\rm{A}}}{j_{\rm{B}}}}} )$, respectively. The function $f( {x,y} )$ is given by
\begin{equation}
 f( {x,y} ) = \sqrt {2x\ln (1/y)}.
\end{equation}
 That is, $\varepsilon_{\chi,nm}^{{j_{\rm{A}}}{j_{\rm{B}}}}$ quantifies the error probability that
$\delta _{\chi ,nm}^{{j_{\rm{A}}}{j_{\rm{B}}}}$ is not lower bounded by ${ - \Delta _{\chi,nm}^{{j_{\rm{A}}}{j_{\rm{B}}}}} $ and $\widehat{\varepsilon}_{\chi,nm}^{{j_{\rm{A}}}{j_{\rm{B}}}}$ quantifies the error
probability that the parameter $\delta _{\chi ,nm}^{{j_{\rm{A}}}{j_{\rm{B}}}}$ is not upper bounded by ${ \widehat{\Delta} _{\chi,nm}^{{j_{\rm{A}}}{j_{\rm{B}}}}}$. Note that Azuma's inequality takes any correlation between the probabilities ${\rm Pr}^i({\rm{click}},nm,{j_{\rm{A}}}{j_{\rm{B}}}\left| {\chi } \right.)$ associated with different trials into account. To simplify our notation, here we omit, however, the explicit dependence of the probabilities ${\rm Pr}^i({\rm{click}},nm,{j_{\rm{A}}}{j_{\rm{B}}}\left| {\chi } \right.)$ with all the previous events, all the basis choices, and Eve's arbitrary operation that is dependent on all the announcements she made.

On the other hand, we also have that
\begin{equation}\label{MDINn}
{{\mathcal E}_{{\rm{click}},{j_{\rm{A}}}{j_{\rm{B}}}|\chi }} \equiv \sum\limits_{i = 1}^{{N_\chi }} {{\rm{P}}{{\rm{r}}^i}({\rm{click}},{j_{\rm{A}}}{j_{\rm{B}}}\left| {\chi } \right.)}  = {N_{{\rm{click}},{j_{\rm{A}}}{j_{\rm{B}}}|\chi }} + \delta _\chi ^{{j_{\rm{A}}}{j_{\rm{B}}}},
\end{equation}
where $ {{\mathcal E}_{{\rm{click}},{j_{\rm{A}}}{j_{\rm{B}}}|\chi }}$ denotes the expected number of events when Alice and Bob select the intensity settings $\gamma^{j_{\rm A}}$ and $\gamma^{j_{\rm B}}$, respectively, and the relay obtains a successful measurement result given that both Alice and Bob select the $\chi$ basis in $N_{\chi}$ trials. ${N_{{\rm{click}},{j_{\rm{A}}}{j_{\rm{B}}}|\chi }}$ is the
corresponding actual number. The deviation term $\delta _{\chi }^{{j_{\rm{A}}}{j_{\rm{B}}}}$ lies in the interval $[ { - \Delta _{\chi}^{{j_{\rm{A}}}{j_{\rm{B}}}},\widehat{\Delta} _{\chi}^{{j_{\rm{A}}}{j_{\rm{B}}}}} ]$ except for a small error probability $\varepsilon_{\chi}^{{j_{\rm{A}}}{j_{\rm{B}}}}+\widehat{\varepsilon}_{\chi}^{{j_{\rm{A}}}{j_{\rm{B}}}}$
where the bounds $ \Delta _{\chi}^{{j_{\rm{A}}}{j_{\rm{B}}}}$ and $\widehat{\Delta} _{\chi}^{{j_{\rm{A}}}{j_{\rm{B}}}}$ are given by $\Delta_{\chi}^{{j_{\rm{A}}}{j_{\rm{B}}}} = f( {{N_{\chi} },\varepsilon_{\chi}
^{{j_{\rm{A}}}{j_{\rm{B}}}}} )$ and $\widehat{\Delta}_{\chi}^{{j_{\rm{A}}}{j_{\rm{B}}}} = f( {{N_{\chi}},\widehat{\varepsilon}_{\chi}^{{j_{\rm{A}}}{j_{\rm{B}}}}} )$, respectively. That is, $\varepsilon_{\chi}^{{j_{\rm{A}}}{j_{\rm{B}}}}$ quantifies the error probability that
$\delta _{\chi }^{{j_{\rm{A}}}{j_{\rm{B}}}}$ is not lower bounded by ${ - \Delta _{\chi}^{{j_{\rm{A}}}{j_{\rm{B}}}}} $ and $\widehat{\varepsilon}_{\chi}^{{j_{\rm{A}}}{j_{\rm{B}}}}$ quantifies the error
probability that the parameter $\delta _{\chi }^{{j_{\rm{A}}}{j_{\rm{B}}}}$ is not upper bounded by ${ \widehat{\Delta} _{\chi}^{{j_{\rm{A}}}{j_{\rm{B}}}}}$.

By combining Eqs. (\ref{MDID}) and (\ref{MDINn}), we obtain the following equation:
\begin{equation}\label{MDIDn}
\begin{array}{*{20}{l}}
{{\mathcal E}_{{\rm{click}},{j_{\rm{A}}}{j_{\rm{B}}}|\chi }}& =& \sum\limits_{n,m = 0}^\infty {{\mathcal E}_{{\rm{click}},nm,{j_{\rm{A}}}{j_{\rm{B}}}|\chi }}\\& =& \sum\limits_{n,m = 0}^\infty { {N_{{\rm{click}},nm,{j_{\rm{A}}}{j_{\rm{B}}}|\chi }}} + \sum\limits_{n,m = 0}^\infty \delta _{\chi ,nm}^{{j_{\rm{A}}}{j_{\rm{B}}}}\\
 &=& {N_{{\rm{click}},{j_{\rm{A}}}{j_{\rm{B}}}|\chi }} + \delta _\chi ^{{j_{\rm{A}}}{j_{\rm{B}}}}.
\end{array}
\end{equation}
That is, we have that ${N_{{\rm{click}},{j_{\rm{A}}}{j_{\rm{B}}}|\chi }}=\sum\limits_{n,m = 0}^\infty { {N_{{\rm{click}},nm,{j_{\rm{A}}}{j_{\rm{B}}}|\chi }} }$ and $\delta _{\chi }^{{j_{\rm{A}}}{j_{\rm{B}}}}=\sum\limits_{n,m = 0}^\infty \delta _{\chi ,nm}^{{j_{\rm{A}}}{j_{\rm{B}}}}$.

The equations above relate the expected number of events to the corresponding actual numbers. These equations provide linear constraints on the quantities that we want to estimate. More precisely, by combining Eqs~(\ref{MDIYDel}) and (\ref{MDIDn}) and by taking $k_{\rm A}=k_{\rm B}=\rm s$, we obtain
\begin{equation}\label{MDINsvw}
\begin{array}{*{20}{l}}
{{{\mathcal E}_{{\rm{click}},{j_{\rm{A}}}{j_{\rm{B}}}{\rm{|}}\chi }}}& = &\sum\limits_{n,m = 0}^\infty  {{{\mathcal E}_{{\rm{click}},nm,{j_{\rm{A}}}{j_{\rm{B}}}|\chi }}} \\& =& \sum\limits_{n,m = 0}^\infty  {\left( {\frac{{{p_{{j_{\rm{A}}}}}{p_{{j_{\rm{B}}}}}p_n^{{j_{\rm{A}}}}p_m^{{j_{\rm{B}}}}}}{{{p_{\rm{s}}}{p_{\rm{s}}}
p_n^{\rm{s}}p_m^{\rm{s}}}}{{\mathcal E}_{{\rm{click}},nm,{\rm{ss}}|\chi }} + \Delta _{\chi ,nm}^{{j_{\rm{A}}}{j_{\rm{B}}}{\rm{ss}}}} \right)}\\
{}& = &{\sum\limits_{n,m = 0}^\infty  {\left( {\frac{{{p_{{j_{\rm{A}}}}}{p_{{j_{\rm{B}}}}}p_n^{{j_{\rm{A}}}}p_m^{{j_{\rm{B}}}}}}{{{p_{\rm{s}}}{p_{\rm{s}}}
p_n^{\rm{s}}p_m^{\rm{s}}}}{{\mathcal E}_{{\rm{click}},nm,{\rm{ss}}|\chi }}} \right)}  + \Delta _\chi ^{{j_{\rm{A}}}{j_{\rm{B}}}{\rm{ss}}}} \\&=& {\sum\limits_{n,m = 0}^\infty  {\frac{{{p_{{j_{\rm{A}}}}}{p_{{j_{\rm{B}}}}}p_n^{{j_{\rm{A}}}}p_m^{{j_{\rm{B}}}}}}{{{p_{\rm{s}}}{p_{\rm{s}}}
p_n^{\rm{s}}p_m^{\rm{s}}}}} \left( {{N_{{\rm{click}},nm,{\rm{ss}}|\chi }} + \delta _{\chi ,nm}^{{\rm{ss}}}} \right) + \Delta _\chi ^{{j_{\rm{A}}}{j_{\rm{B}}}{\rm{ss}}},}
\end{array}
\end{equation}
where $\Delta _{\chi}^{{j_{\rm{A}}}{j_{\rm{B}}}{\rm{ss}}} = \sum\limits_{n,m = 0}^\infty {\Delta _{{\chi},nm}^{{j_{\rm{A}}}{j_{\rm{B}}}{\rm{ss}}}}$.

Then, by combining Eq.~(\ref{MDINsvw}) with Eq~(\ref{MDINn}), we obtain the following linear constraints:
\begin{equation}\label{MDILProg}
{N_{{\rm{click}},{j_{\rm{A}}}{j_{\rm{B}}}|\chi }} = \sum\limits_{n,m = 0}^\infty  {\frac{{{p_{{j_{\rm{A}}}}}{p_{{j_{\rm{B}}}}}p_n^{{j_{\rm{A}}}}p_m^{{j_{\rm{B}}}}}}{{{p_{\rm{s}}}{p_{\rm{s}}}
p_n^{\rm{s}}p_m^{\rm{s}}}}} \left( {{N_{{\rm{click}},nm,{\rm{ss}}|\chi }} + \delta _{\chi ,nm}^{{\rm{ss}}}} \right) + \Delta _\chi ^{{j_{\rm{A}}}{j_{\rm{B}}}{\rm{ss}}} - \delta _\chi ^{{j_{\rm{A}}}{j_{\rm{B}}}}.
\end{equation}
Thus, Eq.~(\ref{MDILProg}) relates the actual observed quantities ${N_{{\rm{click}},{j_{\rm{A}}}{j_{\rm{B}}}|\chi }}$ to the quantities to be estimated, ${N_{{\rm{click}},nm,{\rm{ss}}|\chi }}$.

The equations above contain an infinite number of unknown variables. To numerically estimate the quantities $N^{L}_{{\rm click,00,\rm {ss}}|\rm Z}$, $N^{L}_{{\rm click,11,\rm {ss}}|\rm Z}$ and $e^{\rm U}_{\rm ph}$ by using linear programming techniques, we need to reduce the number of unknowns to a finite set. Due to the fact that $0\leq \mathcal{E}_{{\rm click},nm,{j_{\rm{A}}}{j_{\rm{B}}}|\chi} \leq N_{\chi}{{p_{{j_{\rm{A}}}}}{p_{{j_{\rm{B}}}}}p_n^{{j_{\rm{A}}}}p_m^{{j_{\rm{B}}}}}$ for all $n,~m$ and $j_{\rm{A}},j_{\rm{B}} \in \{\rm s,v,w\}$, we have that
\begin{equation}
\begin{array}{*{20}{l}}\label{MDIcut}
{\sum\limits_{n,m = 0}^\infty  {\left( {\frac{{{p_{{j_{\rm{A}}}}}{p_{{j_{\rm{B}}}}}p_n^{{j_{\rm{A}}}}p_m^{{j_{\rm{B}}}}}}{{{p_{\rm{s}}}{p_{\rm{s}}}
p_n^{\rm{s}}p_m^{\rm{s}}}}{{\mathcal E}_{{\rm{click}},nm,{\rm{ss}}|\chi }}} \right)} }&{ \ge \sum\limits_{n,m = 0}^{S{\rm{cut}}} {\left( {\frac{{{p_{{j_{\rm{A}}}}}{p_{{j_{\rm{B}}}}}p_n^{{j_{\rm{A}}}}p_m^{{j_{\rm{B}}}}}}{{{p_{\rm{s}}}{p_{\rm{s}}}
p_n^{\rm{s}}p_m^{\rm{s}}}}{{\mathcal E}_{{\rm{click}},nm,{\rm{ss}}|\chi }}} \right)} ,}\\
{\sum\limits_{n,m = 0}^\infty  {\left( {\frac{{{p_{{j_{\rm{A}}}}}{p_{{j_{\rm{B}}}}}p_n^{{j_{\rm{A}}}}p_m^{{j_{\rm{B}}}}}}{{{p_{\rm{s}}}{p_{\rm{s}}}
p_n^{\rm{s}}p_m^{\rm{s}}}}{{\mathcal E}_{{\rm{click}},nm,{\rm{ss}}|\chi }}} \right)} }& \le \sum\limits_{n,m = 0}^{S{\rm{cut}}} {\left( {\frac{{{p_{{j_{\rm{A}}}}}{p_{{j_{\rm{B}}}}}p_n^{{j_{\rm{A}}}}p_m^{{j_{\rm{B}}}}}}{{{p_{\rm{s}}}{p_{\rm{s}}}
p_n^{\rm{s}}p_m^{\rm{s}}}}{{\mathcal E}_{{\rm{click}},nm,{\rm{ss}}|\chi }}} \right)} \\
 &{\kern 11pt}+ \sum\limits_{n,m = S{\rm{cut}} + 1}^\infty  {\left( {\frac{{{p_{{j_{\rm{A}}}}}{p_{{j_{\rm{B}}}}}p_n^{{j_{\rm{A}}}}p_m^{{j_{\rm{B}}}}}}{{{p_{\rm{s}}}{p_{\rm{s}}}
 p_n^{\rm{s}}p_m^{\rm{s}}}}{N_\chi }{p_{\rm{s}}}{p_{\rm{s}}}p_n^{\rm{s}}p_m^{\rm{s}}} \right)} \\
{}&{ = \sum\limits_{n,m = 0}^{S{\rm{cut}}} {\left( {\frac{{{p_{{j_{\rm{A}}}}}{p_{{j_{\rm{B}}}}}p_n^{{j_{\rm{A}}}}p_m^{{j_{\rm{B}}}}}}{{{p_{\rm{s}}}{p_{\rm{s}}}
p_n^{\rm{s}}p_m^{\rm{s}}}}{{\mathcal E}_{{\rm{click}},nm,{\rm{ss}}|\chi }}} \right)}  + {N_\chi }{p_{{j_{\rm{A}}}}}{p_{{j_{\rm{B}}}}}T_{{{\rm{S}}_{{\rm{cut}}}}}^{{j_{\rm{A}}}{j_{\rm{B}}}},}
\end{array}
\end{equation}
where $T_{{{\rm{S}}_{{\rm{cut}}}}}^{{j_{\rm{A}}}{j_{\rm{B}}}} = \sum\limits_{n,m = S_{\rm cut} + 1}^\infty  {p_{n}^{j_{\rm{A}}}p_{m}^{j_{\rm{B}}} = } 1 - \sum\limits_{n,m = 0}^{S_{\rm cut}} p_{n}^{j_{\rm{A}}}p_{m}^{j_{\rm{B}}}$ for any natural number $S_{\rm cut} \geq0 $.

Therefore, we obtain the following constraints:
\begin{equation}\label{MDINcut}
\begin{array}{l}
{{\mathcal E}_{{\rm{click}},{\rm{ss|}}\chi }} \ge \sum\limits_{n,m = 0}^{{S_{{\rm{cut}}}}} {{{\mathcal E}_{{\rm{click}},nm,{\rm{ss}}|\chi }}} ,\\
{{\mathcal E}_{{\rm{click}},{\rm{ss|}}\chi }} \le \sum\limits_{n,m = 0}^{{S_{{\rm{cut}}}}} {{{\mathcal E}_{{\rm{click}},nm,{\rm{ss}}|\chi }}}  + {N_\chi }{p_{\rm{s}}}{p_{\rm{s}}}T_{{{\rm{S}}_{{\rm{cut}}}}}^{{\rm{ss}}},\\
{{\mathcal E}_{{\rm{click}},{j_{\rm{A}}}{j_{\rm{B}}}{\rm{|}}\chi }} \ge \sum\limits_{n,m = 0}^{{S_{{\rm{cut}}}}} {\frac{{{p_{{j_{\rm{A}}}}}{p_{{j_{\rm{B}}}}}p_n^{{j_{\rm{A}}}}p_m^{{j_{\rm{B}}}}}}{{{p_{\rm{s}}}{p_{\rm{s}}}
p_n^{\rm{s}}p_m^{\rm{s}}}}} {{\mathcal E}_{{\rm{click}},nm,{\rm{ss}}|\chi }} + \Delta _\chi ^{{j_{\rm{A}}}{j_{\rm{B}}}{\rm{ss}}},\\
{{\mathcal E}_{{\rm{click}},{j_{\rm{A}}}{j_{\rm{B}}}{\rm{|}}\chi }} \le \sum\limits_{n,m = 0}^{{S_{{\rm{cut}}}}} {\frac{{{p_{{j_{\rm{A}}}}}{p_{{j_{\rm{B}}}}}p_n^{{j_{\rm{A}}}}p_m^{{j_{\rm{B}}}}}}{{{p_{\rm{s}}}{p_{\rm{s}}}
p_n^{\rm{s}}p_m^{\rm{s}}}}} {{\mathcal E}_{{\rm{click}},nm,{\rm{ss}}|\chi }} + {N_\chi }{p_{{j_{\rm{A}}}}}{p_{{j_{\rm{B}}}}}T_{{{\rm{S}}_{{\rm{cut}}}}}^{{j_{\rm{A}}}{j_{\rm{B}}}} + \Delta _\chi ^{{j_{\rm{A}}}{j_{\rm{B}}}{\rm{ss}}},
\end{array}
\end{equation}
for any ${j_{\rm{A}}},{j_{\rm{B}}} \in \{\rm s,v,w\}$ and ${j_{\rm{A}}}{j_{\rm{B}}}\neq {\rm ss}$. Importantly, these equations now have a finite number of unknown variables.

Finally, according to Eqs.~(\ref{MDID}) and (\ref{MDINn}) we can replace the expected values in Eq.~(\ref{MDINcut}) with the corresponding actual numbers plus their deviation terms. In so doing, for example, we find that the parameter $N^{L}_{{\rm click,00,ss}|\rm Z}$ can be estimated by using the following linear program:
\begin{equation}\label{MDILP}
\begin{array}{*{20}{l}}
{\min }&{{N_{{\rm{click}},00,{\rm{ss}}|Z}}}\\
{s.t.}&{{N_{{\rm{click}},{\rm{ss|Z}}}} \ge \sum\limits_{n,m = 0}^{{S_{{\rm{cut}}}}} {\left( {{N_{{\rm{click}},nm,{\rm{ss}}|Z}} + \delta _{{\rm{Z}},nm}^{{\rm{ss}}}} \right)}  - \delta _{\rm{Z}}^{{\rm{ss}}},}\\
{}&{{N_{{\rm{click}},{\rm{ss|Z}}}} \le \sum\limits_{n,m = 0}^{{S_{{\rm{cut}}}}} {\left( {{N_{{\rm{click}},nm,{\rm{ss}}|Z}} + \delta _{{\rm{Z}},nm}^{{\rm{ss}}}} \right)}  - \delta _{\rm{Z}}^{{\rm{ss}}} + {N_{\rm{Z}}}p_{\rm{s}}p_{\rm{s}}{\rm{T}}_{{S_{{\rm{cut}}}}}^{{\rm{ss}}},}\\
{}&{{N_{{\rm{click}},{j_{\rm{A}}}{j_{\rm{B}}}{\rm{|Z}}}} \ge \sum\limits_{n,m = 0}^{{S_{{\rm{cut}}}}} {\frac{{{p_{{j_{\rm{A}}}}}{p_{{j_{\rm{B}}}}}p_n^{{j_{\rm{A}}}}p_m^{{j_{\rm{B}}}}}}{{{p_{\rm{s}}}{p_{\rm{s}}}
p_n^{\rm{s}}p_m^{\rm{s}}}}\left( {{N_{{\rm{click}},nm,{\rm{ss}}|Z}} + \delta _{{\rm{Z}},nm}^{{\rm{ss}}}} \right) - \delta _{\rm{Z}}^{{j_{\rm{A}}}{j_{\rm{B}}}} + \Delta _{\rm{Z}}^{{j_{\rm{A}}}{j_{\rm{B}}}{\rm{ss}}},} }\\
{}&{N_{{\rm{click}},{j_{\rm{A}}}{j_{\rm{B}}}{\rm{|Z}}}} \le \sum\limits_{n,m = 0}^{{S_{{\rm{cut}}}}} { {\frac{{{p_{{j_{\rm{A}}}}}{p_{{j_{\rm{B}}}}}p_n^{{j_{\rm{A}}}}p_m^{{j_{\rm{B}}}}}}{{{p_{\rm{s}}}{p_{\rm{s}}}
p_n^{\rm{s}}p_m^{\rm{s}}}}\left( {{N_{{\rm{click}},nm,{\rm{ss}}|Z}} + \delta _{{\rm{Z}},nm}^{{\rm{ss}}}} \right) - \delta _{\rm{Z}}^{{j_{\rm{A}}}{j_{\rm{B}}}} + \Delta _{\rm{Z}}^{{j_{\rm{A}}}{j_{\rm{B}}}{\rm{ss}}}}}  \\& {\kern 70pt}+ {{N_{\rm{Z}}}{p_{{j_{\rm{A}}}}}{p_{{j_{\rm{B}}}}}{\rm{T}}_{{S_{{\rm{cut}}}}}^{{j_{\rm{A}}}{j_{\rm{B}}}}} ,\\
{}& - {p_{{j_{\rm{A}}}}}{p_{{j_{\rm{B}}}}}{N_{\rm{Z}}}\sum\limits_{n,m = 0}^\infty  {p_n^{{j_{\rm{A}}}}p_m^{{j_{\rm{B}}}}{D^{{j_{\rm{A}}}
{j_{\rm{B}}},{\rm{ss}}}_{{\rm Z},nm}}}  \le \Delta _{\rm{Z}}^{{j_{\rm{A}}}{j_{\rm{B}}}{\rm{ss}}} \le {p_{{j_{\rm{A}}}}}{p_{{j_{\rm{B}}}}}{N_{\rm{Z}}}\sum\limits_{n,m = 0}^\infty  {p_n^{{j_{\rm{A}}}}p_m^{{j_{\rm{B}}}}{D^{{j_{\rm{A}}}
{j_{\rm{B}}},{\rm{ss}}}_{{\rm Z},nm}}} ,\\
 &- \Delta _{\rm{Z}}^{{j_{\rm{A}}}{j_{\rm{B}}}} \le \delta _{\rm{Z}}^{{j_{\rm{A}}}{j_{\rm{B}}}} \le \hat \Delta _{\rm{Z}}^{{j_{\rm{A}}}{j_{\rm{B}}}},\:- \Delta _{{\rm{Z}}}^{{\rm{ss}}} \le \delta _{{\rm{Z}}}^{{\rm{ss}}} \le \hat \Delta _{{\rm{Z}}}^{{\rm{ss}}},\; - \Delta _{{\rm{Z}},nm}^{{\rm{ss}}} \le \delta _{{\rm{Z}},nm}^{{\rm{ss}}} \le \hat \Delta _{{\rm{Z}},nm}^{{\rm{ss}}},
\end{array}
\end{equation}
where $j_{\rm{A}},j_{\rm{B}}\in\{\rm s,v,w\}$ and $j_{\rm{A}}j_{\rm{B}}\neq{\rm ss}$. The unknown variables in the linear program above are: $N_{{\rm{click}},nm,{{\rm ss}}|\rm Z} $, $\delta _{{\rm Z} ,nm}^{\rm{ss}}$, $\delta _{\rm Z} ^{\rm ss}$, $\delta _{\rm Z} ^{{j_{\rm{A}}}{j_{\rm{B}}}}$, and $\Delta_{\rm Z}^{{j_{\rm{A}}}{j_{\rm{B}}}{\rm ss}}$. The calculation of the parameters $D^{{j_{\rm{A}}}
{j_{\rm{B}}},{\rm{ss}}}_{{\rm Z},nm}$ is presented in Appendix A.

The solution to the linear program above is exactly $N^{L}_{{\rm click,00,ss}|\rm Z}$, with a total error probability
\begin{equation}
{\varepsilon _{{\rm{Z}},0{\rm{0}}}} = \sum\limits_{{j_{\rm{A}}},{j_{\rm{B}}} = {\rm{s}},{\rm{v}},{\rm{w}}} {\left( {\varepsilon _{\rm{Z}}^{{j_{\rm{A}}}{j_{\rm{B}}}} + \hat \varepsilon _{\rm{Z}}^{{j_{\rm{A}}}{j_{\rm{B}}}}} \right)}  + \sum\limits_{n,m = 0}^{{S_{{\rm{cut}}}}} {\left( {\varepsilon _{{\rm{Z}},nm}^{{\rm{ss}}} + \hat \varepsilon _{{\rm{Z}},nm}^{{\rm{ss}}}} \right)} ,
\end{equation}
where ${\varepsilon _{\rm{Z}}^{{j_{\rm{A}}}{j_{\rm{B}}}}}$ and $\hat \varepsilon _{\rm{Z}}^{{j_{\rm{A}}}{j_{\rm{B}}}}$ are the error probabilities associated with the estimation of the bounds on ${\delta_{\rm Z}^{{j_{\rm{A}}}{j_{\rm{B}}}}} $ with $j_{\rm{A}},j_{\rm{B}}\in\{\rm s,v,w\}$. The terms $\varepsilon _{{\rm{Z}},nm}^{\rm{ss}}$ and $\hat \varepsilon _{{\rm{Z}},nm}^{\rm{ss}}$ are the error probabilities associated with the estimation of the bounds on ${\delta_{{\rm Z},nm}^{\rm ss}}$.

To estimate the parameter $N^{L}_{{\rm click,11,ss}|\rm Z}$, one can reuse the same linear program given by Eq.~ (\ref{MDILP}) after replacing the objective with ``min $ N_{{\rm{click}},11,{{\rm ss}}|\rm Z} $''.

The steps to estimate the phase error rate $e^{\rm U}_{\rm ph}$ are as follows. First, one can redo the analysis above for those ``click'' events in the relay which are associated with an error (see~\cite{tamaki2016decoy,Weilong2018Finite}). That is, now we focus on the quantities $\mathcal{E}_{{\rm error},nm,j_{\rm A}j_{\rm B}|\chi}$ which denote the expected number of events when Alice and Bob select the intensity settings $\gamma^{j_{\rm A}}$ and $\gamma^{j_{\rm B}}$ to send an $n$-photon pulse and an $m$-photon pulse, respectively, and  the relay's detectors provide a click corresponding to an error given that both Alice and Bob select the $\chi$ basis. In addition, we can use the fact that $0\leq \mathcal{E}_{{\rm error},nm,j_{\rm A}j_{\rm B}|\chi} \leq N_{\chi}p_{j_{\rm A}}p_{j_{\rm B}}p_n^{j_{\rm A}}p_m^{j_{\rm B}}$ for all $n,~m$ and $j_{\rm A},j_{\rm B} \in \{\rm s,v,w\}$. In so doing, one can obtain constraints on error events which are similar to the ones given by Eq.~(\ref{MDINcut}). Second, with these equations as well as the constraints given by Eq.~(\ref{MDILP}) but now applied to the X basis events, one can estimate a lower bound on the number of single-photon click events when both Alice and Bob select the intensity setting $\gamma^{\rm s}$ given that they both select the X basis, which we denote by $N^{L}_{{\rm click,11,ss}|\rm X}$, as well as an upper bound on the corresponding number of errors, $N^{U}_{{\rm error,11,ss}|\rm X}$.

Note that the derivation of $N^{L}_{{\rm click,11,ss}|\rm X}$ is similar to the one of $N^{L}_{{\rm click,11,ss}|\rm Z}$. For this, one can use the linear program used to estimate $N^{L}_{{\rm click,11,ss}|\rm Z}$ after replacing all the parameters and variables in the Z basis with the corresponding ones in the X basis. Similarly, one can further modify the program for $N^{L}_{{\rm click,11,ss}|\rm X}$ to calculate $N^{U}_{{\rm error,11,ss}|\rm X}$. Specifically, one can simply replace all the numbers of click events with those of error events. In addition, one replaces ``min $N_{{\rm click,11,ss}|\rm X}$'' with ``max $N_{{\rm error,11,ss}|\rm X}$'' to obtain an upper bound on $N_{{\rm error,11,ss}|\rm X}$. Finally, given the values of $N^{L}_{{\rm click,11,ss}|\rm Z}$, $N^{L}_{{\rm click,11,ss}|\rm X}$ and $N^{U}_{{\rm error,11,ss}|\rm X}$, one can use a random sampling argument to relate the number of errors in the single-photon events in the X basis to the number of phase errors associated with the single-photon events in the Z basis and thus estimate $e^{\rm U}_{\rm ph}$~\cite{tomamichel2012tight,curty2014finite}. More precisely, by using Serfling's inequality~\cite{serfling1974probability}, we obtain that
\begin{equation}
\begin{array}{*{20}{l}}
{e_{{\rm{ph}}}^{\rm{U}} = \frac{1}{{N_{{\rm{click}},{\rm{11}},{\rm{ss}}|{\rm{Z}}}^L}}\min \left\{ {\left[ {N_{{\rm{click}},{\rm{11}},{\rm{ss}}|{\rm{Z}}}^L\frac{{N_{{\rm{error}},1{\rm{1}},{\rm{ss}}|{\rm{X}}}^U}}
{{N_{{\rm{click}},{\rm{11}},{\rm{ss}}|{\rm{X}}}^L}} + \left( {N_{{\rm{click}},{\rm{11}},{\rm{ss}}|{\rm{Z}}}^L + N_{{\rm{click}},{\rm{11}},{\rm{ss}}|{\rm{X}}}^L} \right)} \right.} \right.}\\
{\kern 106pt}{\left. {\left. { \times \Upsilon \left( {N_{{\rm{click}},{\rm{11}},{\rm{ss}}|{\rm{Z}}}^L,N_{{\rm{click}},{\rm{11}},{\rm{ss}}|{\rm{X}}}^L,\varepsilon '} \right)} \right],N_{{\rm{click}},{\rm{11}},{\rm{ss}}|{\rm{Z}}}^L} \right\},}
\end{array}
\end{equation}
except for a failure probability
\begin{equation}
{\varepsilon _{{\rm{ph}},{\rm{11}}}} \le {\varepsilon'} + {\varepsilon _{{\rm{X}}{\rm{,11}}}} + {\varepsilon _{{{\rm{E}}_{\rm{X}}}{\rm{,11}}}},
\end{equation}
where the function $\Upsilon \left( x,y,z \right)$ is defined as $\Upsilon \left( {x,y,z} \right) = \sqrt {\left( {x + 1} \right)\ln \left( {{z^{ - 1}}} \right)/\left[ {2y\left( {x + y} \right)} \right]}$, and ${\varepsilon _{{\rm{X}}{\rm{,11}}}}$ and ${\varepsilon _{{{\rm{E}}_{\rm{X}}}{\rm{,11}}}}$ are the failure probabilities corresponding to the estimation of $N^{L}_{{\rm click,11,ss}|\rm X}$ and $N^{U}_{{\rm error,11,ss}|\rm X}$, respectively.

\subsection{THA against the phase modulator}\label{PM3}
A THA against the PM might render Alice's and Bob's output states (which now also contain Eve's systems) \emph{basis dependent}. As a result, Eve might be able to learn partial information about Alice's and Bob's basis and bit value choices each given time. The security of the standard BB84 protocol with a basis-dependent flaw has been analyzed in a previous work~\cite{lo2007security} by using the idea of a quantum coin~\cite{gottesman2004security,koashi2009simple}. This idea was then generalized to phase encoding schemes for MDI-QKD where both Alice and Bob have basis-dependent flaws~\cite{tamaki2012phase}. Here, to estimate the phase error rate in the presence of a THA against the PM, we apply the method introduced in Ref.~\cite{tamaki2012phase} to our protocol.

\subsubsection{The Asymptotic Limit}\label{pma}
To simplify the analysis, let us first consider a scenario where Alice's and Bob's light sources are both ideal single-photon sources. Also, let us assume that Alice's and Bob's basis choices are random and do not depend on the IM or on previous emitted pulses. Let ${\left| {\Psi _{\rm{Z}}^i} \right\rangle _{\rm{A,E}}}$ and ${\left| {\Psi _{\rm{Z}}^i} \right\rangle _{\rm{B,E}}}$ (${\left| {\Psi _{\rm{X}}^i} \right\rangle _{\rm{A,E}}}$ and ${\left| {\Psi _{\rm{X}}^i} \right\rangle _{\rm{B,E}}}$) denote the states that Alice and Bob prepare (in an equivalent entanglement-based scenario) in the Z~(X) basis together with Eve's system from a THA in the $i$th trial of the protocol. The subscripts A, B and E denote the systems of Alice, Bob and Eve, respectively. As already mentioned, here we consider a virtual entanglement scenario where each of Alice and Bob prepares a bipartite entangled state and then measures one of the two systems to actually prepare the states that are sent to the relay. More precisely, the system A (B) above contains a virtual qubit ${\rm A}_{\rm q}$ (${\rm B}_{\rm q}$) indicating Alice's~(Bob's) bit value choice, and Alice's~(Bob's) photonic system ${\rm A}_{\rm p}$~(${\rm B}_{\rm p}$) that is sent to the relay via the quantum channel. In addition, the system A (B) could also contain an ancilla system ${\rm A}_{\rm a}$~(${\rm B}_{\rm a}$) stored in Alice's~(Bob's) lab to account for the loss in the transmitter. And Eve's system E corresponds to the back-reflected light from a THA. That is, the system $\rm A\equiv {\rm A}_{\rm q}{\rm A}_{\rm p}{\rm A}_{\rm a}$ and it is similar for B. All these quantum systems are listed in Table.~\ref{sys2}.

\begin{table}[tbp]
\caption{\footnotesize {Quantum systems defined in a THA against the PM}}
\centering \label{sys2}
\begin{tabular}{|m{3.5cm}<{\centering}|m{12cm}<{\centering}|}
\hline
  ${\left| {\Psi _{\rm{Z}}^i} \right\rangle _{\rm{A,E}}}$ (${\left| {\Psi _{\rm{Z}}^i} \right\rangle _{\rm{B,E}}}$)& The state that Alice (Bob) prepares in the Z basis together with Eve's system in the $i$th trial of the protocol\\
  \hline
  ${\left| {\Psi _{\rm{X}}^i} \right\rangle _{\rm{A,E}}}$ (${\left| {\Psi _{\rm{X}}^i} \right\rangle _{\rm{B,E}}}$) & The state that Alice (Bob) prepares in the X basis together with Eve's system in the $i$th trial of the protocol\\
  \hline
  ${\rm A}_{\rm q}$ (${\rm B}_{\rm q}$) & A virtual qubit that contains Alice's~(Bob's) bit value choice\\
  \hline
  ${\rm A}_{\rm p}$~(${\rm B}_{\rm p}$) & Alice's~(Bob's) photonic system that is sent to the relay via a quantum channel\\
  \hline
  ${\rm A}_{\rm a}$~(${\rm B}_{\rm a}$)  & An additional ancilla system stored in Alice's~(Bob's) lab to account for the loss in the transmitter\\
  \hline
  ${\rm E}$  & The back-reflected light from Eve's THA\\
  \hline
\end{tabular}
\end{table}

The phase error rate is the fictitious bit error rate that Alice and Bob would obtain if they would measure the systems $\rm A_{q}$ and $\rm B_{q}$ in the X basis, given that they prepared the states ${\left| {\Psi _{\rm{Z}}^i} \right\rangle _{\rm{A,E}}}$ and ${\left| {\Psi _{\rm{Z}}^i} \right\rangle _{\rm{B,E}}}$, respectively. In order to estimate the phase error rate in the presence of information leakage from the PM, we consider a fictitious protocol where we assume that Alice and Bob meet together~\cite{tamaki2012phase} and they decide the basis choices by  measuring a so-called quantum coin~\cite{gottesman2004security,koashi2009simple}. Particularly, we assume that in the $i$th trial of this protocol, Alice and Bob first prepare a joint state
\begin{equation}\label{xbstate}
\begin{array}{*{20}{l}}
{\left| {{\Psi ^i}} \right\rangle  \equiv }&{{p_{\rm{Z}}}{{\left| {{0_{\rm{Z}}}} \right\rangle }_{{{\rm{A}}_{{\rm{ba}}}}}}{{\left| {{0_{\rm{Z}}}} \right\rangle }_{{{\rm{A}}_{\rm{c}}}}}{{\left| {\Psi _{\rm{Z}}^i} \right\rangle }_{\rm{A,E}}}{{\left| {\Psi _{\rm{Z}}^i} \right\rangle }_{\rm{B,E}}} + {{p_{\rm{X}}}}{{\left| {{0_{\rm{Z}}}} \right\rangle }_{{{\rm{A}}_{{\rm{ba}}}}}}{{\left| {{1_{\rm{Z}}}} \right\rangle }_{{{\rm{A}}_{\rm{c}}}}}{{\left| {\Psi _{\rm{X}}^i} \right\rangle }_{\rm{A,{E}}}}{{\left| {\Psi _{\rm{X}}^i} \right\rangle }_{\rm{B,{E}}}}}\\
{}&{ + {\sqrt {{p_{\rm{Z}}}{p_{\rm{X}}}}} \left( {{{\left| {{1_{\rm{Z}}}} \right\rangle }_{{{\rm{A}}_{{\rm{ba}}}}}}{{\left| {{0_{\rm{Z}}}} \right\rangle }_{{{\rm{A}}_{\rm{c}}}}}{{\left| {\Psi _{\rm{Z}}^i} \right\rangle }_{\rm{A,{E}}}}{{\left| {\Psi _{\rm{X}}^i} \right\rangle }_{\rm{B,{E}}}} + {{\left| {{1_{\rm{Z}}}} \right\rangle }_{{{\rm{A}}_{{\rm{ba}}}}}}{{\left| {{1_{\rm{Z}}}} \right\rangle }_{{{\rm{A}}_{\rm{c}}}}}{{\left| {\Psi _{\rm{X}}^i} \right\rangle }_{\rm{A,{E}}}}{{\left| {\Psi _{\rm{Z}}^i} \right\rangle }_{\rm{B,{E}}}}} \right),}
\end{array}
\end{equation}
where
\begin{equation}\label{stateb}
\begin{array}{*{20}{l}}
{{{\left| {\Psi _{\rm{Z}}^i} \right\rangle }_{{\rm{A}},{\rm{E}}}} \equiv \frac{1}{{\sqrt 2 }}\left( {{{\left| {{0_{\rm{Z}}}} \right\rangle }_{{{\rm{A}}_{\rm{q}}}}}{{\left| {\Psi _{0,{\rm{Z}}}^i} \right\rangle }_{{\rm{A'}},{\rm{E}}}} + {{\left| {{1_{\rm{Z}}}} \right\rangle }_{{{\rm{A}}_{\rm{q}}}}}{{\left| {\Psi _{1,{\rm{Z}}}^i} \right\rangle }_{{\rm{A'}},{\rm{E}}}}} \right),}\\
{{{\left| {\Psi _{\rm{X}}^i} \right\rangle }_{{\rm{A}},{\rm{E}}}} \equiv \frac{1}{{\sqrt 2 }}\left( {{{\left| {{0_{\rm{Z}}}} \right\rangle }_{{{\rm{A}}_{\rm{q}}}}}{{\left| {\Psi _{0,{\rm{X}}}^i} \right\rangle }_{{\rm{A'}},{\rm{E}}}} + {{\left| {{1_{\rm{Z}}}} \right\rangle }_{{{\rm{A}}_{\rm{q}}}}}{{\left| {\Psi _{1,{\rm{X}}}^i} \right\rangle }_{{\rm{A'}},{\rm{E}}}}} \right),}\\
{{{\left| {\Psi _{\rm{Z}}^i} \right\rangle }_{{\rm{B}},{\rm{E}}}} \equiv \frac{1}{{\sqrt 2 }}\left( {{{\left| {{0_{\rm{Z}}}} \right\rangle }_{{{\rm{B}}_{\rm{q}}}}}{{\left| {\Psi _{0,{\rm{Z}}}^i} \right\rangle }_{{\rm{B'}},{\rm{E}}}} + {{\left| {{1_{\rm{Z}}}} \right\rangle }_{{{\rm{B}}_{\rm{q}}}}}{{\left| {\Psi _{1,{\rm{Z}}}^i} \right\rangle }_{{\rm{B'}},{\rm{E}}}}} \right),}\\
{{{\left| {\Psi _{\rm{X}}^i} \right\rangle }_{{\rm{B}},{\rm{E}}}} \equiv \frac{1}{{\sqrt 2 }}\left( {{{\left| {{0_{\rm{Z}}}} \right\rangle }_{{{\rm{B}}_{\rm{q}}}}}{{\left| {\Psi _{0,{\rm{X}}}^i} \right\rangle }_{{\rm{B'}},{\rm{E}}}} + {{\left| {{1_{\rm{Z}}}} \right\rangle }_{{{\rm{B}}_{\rm{q}}}}}{{\left| {\Psi _{1,{\rm{X}}}^i} \right\rangle }_{{\rm{B'}},{\rm{E}}}}} \right).}
\end{array}
\end{equation}
In Eq.~(\ref{xbstate}), the first system $\rm A_{ba}$ denotes a system in Alice's hands which decides whether or not Alice's and Bob's basis choices match by measuring it in the Z basis. More precisely, if she obtains the measurement outcome corresponding to $\left| {{0_{\rm{Z}}}} \right\rangle _{\rm{A_{ba}}}$~($\left| {{1_{\rm{Z}}}} \right\rangle _{\rm{A_{ba}}}$), then Alice's and Bob's basis choices (do not) match. The relay can perform any operation on each received signal pair from Alice and Bob to decide in which rounds there will be ``click'' events. For each click event, the relay then performs some measurement on the received signal pair and both Alice and Bob measure their systems $\rm {A_q}$ and $\rm B_q$ in the X basis. Besides, Alice selects the $\rm Z_{A_c}$ or $\rm X_{A_c}$ basis with probabilities $p_{\rm Z_{A_c}}$ and $p_{\rm X_{A_c}}$, respectively, to measure her quantum coin, denoted by the system $\rm A_c$ in Eq.~(\ref{xbstate}) in the selected basis. In Eq.~(\ref{stateb}), the system $\rm A'$ is defined as $\rm A'\equiv {\rm A}_{\rm p}{\rm A}_{\rm a}$ and the definition of the system $\rm{B}'$ is similar.

After applying the Bloch sphere bound~\cite{tamaki2003unconditionally} to this fictitious scenario, we obtain
\begin{equation}
\begin{array}{*{20}{l}}
&1 - 2{{\Pr }^i}\left( {{{{X}}_{{{\rm{A}}_{\rm{c}}}}} =  - \left| {{\rm{click}}{\rm{,sb}}{\rm{,X - error}},{{\rm{X}}_{{{\rm{A}}_{\rm{c}}}}}} \right.} \right)\\
 &\le2\sqrt {{{\Pr }^i}\left( {{{\rm{Z}}_{{{\rm{A}}_{\rm{c}}}}} = 1\left| {{\rm{click}}{\rm{,sb}}{\rm{,X - error}},{{\rm{Z}}_{{{\rm{A}}_{\rm{c}}}}}} \right.} \right)}\\
 &\times\sqrt{ {1 - {{\Pr }^i}\left( {{{\rm{Z}}_{{{\rm{A}}_{\rm{c}}}}} = 1\left| {{\rm{click}}{\rm{,sb}}{\rm{,X - error}},{{\rm{Z}}_{{{\rm{A}}_{\rm{c}}}}}} \right.} \right)} } ,
\end{array}\label{MDIX error}
\end{equation}
and
\begin{equation}
 \begin{array}{*{20}{l}}
&1- 2{{\Pr }^i}\left( {{{{X}}_{{{\rm{A}}_{\rm{c}}}}} =  - \left| {{\rm{click}}{\rm{,sb}}{\rm{,No}}\;{\rm{X - error}},{{\rm{X}}_{{{\rm{A}}_{\rm{c}}}}}} \right.} \right)\\
&\le2\sqrt {{{\Pr }^i}\left( {{{\rm{Z}}_{{{\rm{A}}_{\rm{c}}}}} = 1\left| {{\rm{click}}{\rm{,sb}}{\rm{,No}}\;{\rm{X - error}},{{\rm{Z}}_{{{\rm{A}}_{\rm{c}}}}}} \right.} \right)}\\
 &\times\sqrt{ {1 - {{\Pr }^i}\left( {{{\rm{Z}}_{{{\rm{A}}_{\rm{c}}}}} = 1\left| {{\rm{click}}{\rm{,sb}}{\rm{,No}}\;{\rm{X - error}},{{\rm{Z}}_{{{\rm{A}}_{\rm{c}}}}}} \right.} \right)} },
\end{array} \label{MDINo Xerror}
\end{equation}
where ${{\Pr }^i}\left( {{{{X}}_{{{\rm{A}}_{\rm{c}}}}} =  - \left| {{\rm{click}}{\rm{,sb}},\;{\rm{X - error}},{{\rm{X}}_{{{\rm{A}}_{\rm{c}}}}}} \right.} \right)$ is the conditional probability that in the $i$th trial Alice's X basis measurement result on the quantum coin is `$-$' given that the relay obtains a successful result in his measurement device, Alice and Bob select the same basis (sb) for the state preparation, Bob's X basis measurement outcome on $\rm B_q$ differs from that obtained by Alice when she measures her system ${\rm A}_{\rm q}$ in the X basis (which we call an `$\rm X-error$'), and Alice performs the $\rm X_{A_c}$ basis measurement on the quantum coin. The other conditional probabilities that appear in Eqs. (\ref{MDIX error}) and (\ref{MDINo Xerror}) are defined similarly.

Then, we multiply Eq.~(\ref{MDIX error}) by
\begin{equation}
\begin{array}{*{20}{l}}
{{\Pr} ^i}\left( {{{\rm{Z}}_{{{\rm{A}}_{\rm{c}}}}}\left| {{\rm{click}}} \right.} \right){{\Pr} ^i}\left( {{\rm{sb}},{\rm{X - error}}\left| {{\rm{click}},{{\rm{X}}_{{{\rm{A}}_{\rm{c}}}}}} \right.} \right) \\= {{\Pr} ^i}\left( {{{\rm{Z}}_{{{\rm{A}}_{\rm{c}}}}}\left| {{\rm{click}}} \right.} \right){{\Pr} ^i}\left( {{\rm{sb}},{\rm{X - error}}\left| {{\rm{click}},{{\rm{Z}}_{{{\rm{A}}_{\rm{c}}}}}} \right.} \right),
\end{array}\label{error}
\end{equation}
and we multiply Eq.~(\ref{MDINo Xerror}) by
\begin{equation}
\begin{array}{*{20}{l}}
{{\Pr} ^i}\left( {{{\rm{Z}}_{{{\rm{A}}_{\rm{c}}}}}\left| {{\rm{click}}} \right.} \right){{\Pr} ^i}\left( {{\rm{sb}},{\rm{No}}\;{\rm{X - error}}\left| {{\rm{click}},{{\rm{X}}_{{{\rm{A}}_{\rm{c}}}}}} \right.} \right)\\ = {{\Pr} ^i}\left( {{{\rm{Z}}_{{{\rm{A}}_{\rm{c}}}}}\left| {{\rm{click}}} \right.} \right){{\Pr} ^i}\left( {{\rm{sb}},{\rm{No}}\;{\rm{X - error}}\left| {{\rm{click}},{{\rm{Z}}_{{{\rm{A}}_{\rm{c}}}}}} \right.} \right).
\end{array}\label{no error}
\end{equation}
After adding both results together, we obtain
\begin{equation}
\begin{array}{*{20}{l}}
&{\Pr} ^i \left( {{{\rm{Z}}_{{{\rm{A}}_{\rm{c}}}}}\left| {{\rm{click}}} \right.} \right){\Pr} ^i \left( {{\rm{sb}}\left| {{\rm{click}},{{\rm{X}}_{{{\rm{A}}_{\rm{c}}}}}} \right.} \right) - 2{\Pr} ^i \left( {{{\rm{Z}}_{{{\rm{A}}_{\rm{c}}}}}\left| {{\rm{click}}} \right.} \right){\Pr} ^i \left( {{X_{{{\rm{A}}_{\rm{c}}}}} =  - ,{\rm{sb}}\left| {{\rm{click}},{{\rm{X}}_{{{\rm{A}}_{\rm{c}}}}}} \right.} \right)\\
 &\le{2\sqrt {{{\Pr }^i}\left( {{{\rm{Z}}_{{{\rm{A}}_{\rm{c}}}}} = 1,{\rm sb},{\rm{X - error}},{{\rm{Z}}_{{{\rm{A}}_{\rm{c}}}}}\left| {{\rm{click}}} \right.} \right) {{{\Pr }^i}\left( {{{\rm{Z}}_{{{\rm{A}}_{\rm{c}}}}} = 0,{\rm sb},{\rm{X - error}},{{\rm{Z}}_{{{\rm{A}}_{\rm{c}}}}}\left| {{\rm{click}}} \right.} \right)} } }\\
& +{2\sqrt {{{\Pr }^i}\left( {{{\rm{Z}}_{{{\rm{A}}_{\rm{c}}}}} = 1,{\rm sb},{\rm{No}}\;{\rm{X - error}},{{\rm{Z}}_{{{\rm{A}}_{\rm{c}}}}}\left| {{\rm{click}}} \right.} \right){{{\Pr }^i}\left( {{{\rm{Z}}_{{{\rm{A}}_{\rm{c}}}}} = 0,{\rm sb},{\rm{No}}\;{\rm{X - error}},{{\rm{Z}}_{{{\rm{A}}_{\rm{c}}}}}\left| {{\rm{click}}} \right.} \right)} } }.
\end{array}\label{MDIPZac}
\end{equation}

Note that ${{{\Pr }^i}\left( {{{\rm{Z}}_{{{\rm{A}}_{\rm{c}}}}} \left| {{\rm{click}}} \right.} \right)}=p_{{{\rm{Z}}_{{{\rm{A}}_{\rm{c}}}}}}$ for any round `$i$' and we also have ${\Pr }^i\left( {{\rm{sb}}\left| {{\rm{click}},{{\rm{X}}_{{{\rm{A}}_{\rm{c}}}}}} \right.} \right) = {\Pr }^i \left( {{\rm{sb}}\left| {{\rm{click}}} \right.} \right)$. To relate the probabilities in Eq.~(\ref{MDIPZac}) to the expected numbers of events, we take the sum over $i \in \{1,2,...,N_{\rm click}\}$, where $N_{\rm click}$ is the number of click events. Due to the concavity of the square root function, we have that
\begin{equation}
\begin{array}{*{20}{l}}
&{p_{{{\rm{Z}}_{{{\rm{A}}_{\rm{c}}}}}}}\sum\limits_{i = 1}^{{N_{{\rm{click}}}}} {{\Pr} ^i \left( {{\rm{sb}}\left| {{\rm{click}}} \right.} \right)}  - 2{p_{{{\rm{Z}}_{{{\rm{A}}_{\rm{c}}}}}}}\sum\limits_{i = 1}^{{N_{{\rm{click}}}}} {{\Pr} ^i \left( {{X_{{{\rm{A}}_{\rm{c}}}}} =  - ,{\rm{sb}}\left| {{\rm{click}},{{\rm{X}}_{{{\rm{A}}_{\rm{c}}}}}} \right.} \right)} \\
 &\le{2\sqrt {\sum\limits_{i = 1}^{{N_{{\rm{click}}}}} {{{\Pr }^i}\left( {{{\rm{Z}}_{{{\rm{A}}_{\rm{c}}}}} = 1,{\rm sb},{\rm{X - error}},{{\rm{Z}}_{{{\rm{A}}_{\rm{c}}}}}\left| {{\rm{click}}} \right.} \right)} \sum\limits_{i = 1}^{{N_{{\rm{click}}}}} {{{\Pr }^i}\left( {{{\rm{Z}}_{{{\rm{A}}_{\rm{c}}}}} = 0,{\rm sb},{\rm{X - error}},{{\rm{Z}}_{{{\rm{A}}_{\rm{c}}}}}\left| {{\rm{click}}} \right.} \right)} } }\\
& +{2\sqrt {\sum\limits_{i = 1}^{{N_{{\rm{click}}}}} {{{\Pr }^i}\left( {{{\rm{Z}}_{{{\rm{A}}_{\rm{c}}}}} = 1,{\rm sb},{\rm{No}}\;{\rm{X - error}},{{\rm{Z}}_{{{\rm{A}}_{\rm{c}}}}}\left| {{\rm{click}}} \right.} \right)}}}\\
 &\times \sqrt{{\sum\limits_{i = 1}^{{N_{{\rm{click}}}}} {{{\Pr }^i}\left( {{{\rm{Z}}_{{{\rm{A}}_{\rm{c}}}}} = 0,{\rm sb},{\rm{No}}\;{\rm{X - error}},{{\rm{Z}}_{{{\rm{A}}_{\rm{c}}}}}\left| {{\rm{click}}} \right.} \right)} } }.
\end{array}\label{MDIN ph}
\end{equation}
In Eq.~(\ref{MDIN ph}), we have that $\sum\limits_{i = 1}^{{N_{{\rm{click}}}}} {{\Pr} ^i \left( {{\rm{sb}}\left| {{\rm{click}}} \right.} \right)}=\mathcal{E}_{\rm sb|click}$, with $\mathcal{E}_{\rm sb|click}$ being the expected number of events where Alice and Bob select the same basis given that there is a click. Also, we have that
\begin{equation}
\begin{array}{*{20}{l}}
\sum\limits_{i = 1}^{{N_{{\rm{click}}}}} {{\Pr} ^i \left( {{X_{{{\rm{A}}_{\rm{c}}}}} =  - ,{\rm{sb}}\left| {{\rm{click}},{{\rm{X}}_{{{\rm{A}}_{\rm{c}}}}}} \right.} \right)} & = \sum\limits_{i = 1}^{{N_{{\rm{click}}}}} {{\Pr} ^i \left( {{\rm{sb}}\left| {{\rm{click}},{{\rm{X}}_{{{\rm{A}}_{\rm{c}}}}}} \right.} \right){\Pr} ^i \left( {{X_{{{\rm{A}}_{\rm{c}}}}} =  - \left| {{\rm{click}},{{\rm{X}}_{{{\rm{A}}_{\rm{c}}}}},{\rm{sb}}} \right.} \right)} \\
& = \sum\limits_{i = 1}^{{N_{{\rm{click}}}}} {{\Pr} ^i \left( {{\rm{sb}}\left| {{\rm{click}}} \right.} \right){\Pr} ^i \left( {{X_{{{\rm{A}}_{\rm{c}}}}} =  - \left| {{\rm{click}},{{\rm{X}}_{{{\rm{A}}_{\rm{c}}}}},{\rm{sb}}} \right.} \right)}.
\end{array}
\end{equation}

Note that, although in the actual protocol there is no data corresponding to the event `${{{{X}}_{{{\rm{A}}_{\rm{c}}}}} =  - }$', we can still upper bound the probability ${{{\Pr }^i}\left( {{X_{{{\rm{A}}_{\rm{c}}}}} =  - \left| {{\rm{click}},{{\rm{X}}_{{{\rm{A}}_{\rm{c}}}}}{\rm{,sb}}} \right.} \right)}$. For this, we assume a worst-case scenario where we take the maximum value of this probability in the total number $N$ of rounds. More precisely, we have that
\begin{equation}
\begin{array}{*{20}{l}}
{{{\Pr }^i}\left( {{X_{{{\rm{A}}_{\rm{c}}}}} =  - \left| {{\rm{click}},{{\rm{X}}_{{{\rm{A}}_{\rm{c}}}}}{\rm{,sb}}} \right.} \right)}&{ \le \mathop {\max }\limits_{j \in \left\{ {1,2,...N} \right\}} {{\Pr }^j}\left( {{X_{{{\rm{A}}_{\rm{c}}}}} =  - \left| {{{\rm{X}}_{{{\rm{A}}_{\rm{c}}}}}{\rm{,sb}}} \right.} \right)}\\
{}&{ = \frac{1}{2}\left\{ {1 - \frac{{2{p_{\rm{Z}}}{p_{\rm{X}}}}}{{p_{\rm{Z}}^2 + p_{\rm{X}}^2}}\mathop {\min }\limits_{j \in \left\{ {1,2,...N} \right\}} {\rm{Re}}\left( {\langle \Psi _{\rm{Z}}^j|{{\left. {\Psi _{\rm{X}}^j} \right\rangle }_{{\rm{A}},{\rm{E}}}}\langle \Psi _{\rm{Z}}^j|{{\left. {\Psi _{\rm{X}}^j} \right\rangle }_{{\rm{B}},{\rm{E}}}}} \right)} \right\}}\\
{}&\equiv \Delta_{{{X_{{{\rm{A}}_{\rm{c}}}}} =  -}}.
\end{array}\label{MDIN++--}
\end{equation}
The detailed calculation of the probability ${{\Pr }^j}\left( {{X_{{{\rm{A}}_{\rm{c}}}}} =  - \left| {{{\rm{X}}_{{{\rm{A}}_{\rm{c}}}}}{\rm{,sb}}} \right.} \right)$ can be found in Appendix B.

Let us denote
 \begin{equation}
\begin{array}{l}
\sum\limits_{i = 1}^{{N_{{\rm{click}}}}} {{{\Pr }^i}\left( {{{\rm{Z}}_{{{\rm{A}}_{\rm{c}}}}} = 1,{{\rm{sb}}},{\rm{X - error}},{{\rm{Z}}_{{{\rm{A}}_{\rm{c}}}}}\left| {{\rm{click}}} \right.} \right)}  = {\mathcal{E}_{{\rm{X}},{\rm{X - error}}}},\\
\sum\limits_{i = 1}^{{N_{{\rm{click}}}}} {{{\Pr }^i}\left( {{{\rm{Z}}_{{{\rm{A}}_{\rm{c}}}}} = 0,{{\rm{sb}}},{\rm{X - error}},{{\rm{Z}}_{{{\rm{A}}_{\rm{c}}}}}\left| {{\rm{click}}} \right.} \right)}  = {\mathcal{E}_{{\rm{Z}},{\rm{X - error}}}},\\
\sum\limits_{i = 1}^{{N_{{\rm{click}}}}} {{{\Pr }^i}\left( {{{\rm{Z}}_{{{\rm{A}}_{\rm{c}}}}} = 1,{{\rm{sb}}},{\rm{No}}\;{\rm{X - error}},{{\rm{Z}}_{{{\rm{A}}_{\rm{c}}}}}\left| {{\rm{click}}} \right.} \right)}  = {\mathcal{E}_{{\rm{X}},{\rm{No}}\;{\rm{X - error}}}},\\
\sum\limits_{i = 1}^{{N_{{\rm{click}}}}} {{{\Pr }^i}\left( {{{\rm{Z}}_{{{\rm{A}}_{\rm{c}}}}} = 0,{{\rm{sb}}},{\rm{No}}\;{\rm{X - error}},{{\rm{Z}}_{{{\rm{A}}_{\rm{c}}}}}\left| {{\rm{click}}} \right.} \right)}  = {\mathcal{E}_{{\rm{Z}},{\rm{No}}\;{\rm{X - error}}}}.
\end{array}
\end{equation}

Then Eq.~(\ref{MDIN ph}) can be written as
\begin{equation}
{{p_{{{\rm{Z}}_{{{\rm{A}}_{\rm{c}}}}}}}{{\cal E}_{{\rm{sb}}\left| {{\rm{click}}} \right.}}\left( {1 - 2{\Delta _{{X_{{{\rm{A}}_{\rm{c}}}}} =  - }}} \right) \le 2\sqrt {{{\cal E}_{{\rm{X}},{\rm{X - error}}}}{{\cal E}_{{\rm{Z}},{\rm{X - error}}}}}  + 2\sqrt {{{\cal E}_{{\rm{X}},{\rm{No}}\;{\rm{X - error}}}}{{\cal E}_{{\rm{Z}},{\rm{No}}\;{\rm{X - error}}}}} }.\label{MDIn phase}
\end{equation}

Eq.~(\ref{MDIn phase}) gives the mathematical relation between the expected number of events in the asymptotic limit. Next we will explain how to extend Eq.~(\ref{MDIn phase}) to the actual protocol in the finite-key regime by applying Azuma's inequality.

\subsubsection{The Finite-Key Regime}\label{pmf}
Here, we apply Azuma's inequality~\cite{azuma1967weighted} again to relate the expected numbers of events to the corresponding actual numbers of events. Let $N_{\lambda}$ denote the actual number of times that the event `$\lambda$' occurs in $N_{\rm click}$ trials.

Then Eq.~(\ref{MDIn phase}) can be rewritten as
\begin{equation}
\begin{array}{*{20}{l}}
{{p_{{{\rm{Z}}_{{{\rm{A}}_{\rm{c}}}}}}}\left( {N_{{\rm{sb}}|{\rm{click}}}} - {\delta _{{\rm{sb}}|{\rm{click}}}} \right)\left( {1 - 2{\Delta _{{X_{{{\rm{A}}_{\rm{c}}}}} =  - }}} \right)}\\
{ \le 2\sqrt {\left( {{N_{{\rm{X}},{\rm{X - error}}}} + {\delta _{{\rm{X}},{\rm{X - error}}}}} \right)\left( {{N_{{\rm{Z}},{\rm{X - error}}}} + {\delta _{{\rm{Z}},{\rm{X - error}}}}} \right)} }\\
{ + 2\sqrt {\left( {{N_{{\rm{click}},{\rm{X}}}} - {N_{{\rm{X}},{\rm{X - error}}}} + {\delta _{{\rm{X}},{\rm{No}}\;{\rm{X - error}}}}} \right)\left( {{N_{{\rm{click}},{\rm{Z}}}} - {N_{{\rm{Z}},{\rm{X - error}}}} + {\delta _{{\rm{Z}},{\rm{No}}\;{\rm{X - error}}}}} \right)} }
\end{array},\label{MDINphase}
\end{equation}
except for an exponentially small error probability $\sum\limits_\lambda  {\left( {{\varepsilon _\lambda } + {{\hat \varepsilon }_\lambda }} \right)}$, where $\lambda \in \{({\rm{sb}}|{\rm{click}}),\rm(X,X-error),(Z,X-error),(X,No~X-error),(Z,No~X-error)\}$. In Eq.~(\ref{MDINphase}), ${{N_{\rm{click,Z(X)}}}} $ denotes the actual number of events when the relay obtains a successful measurement result and Alice and Bob select the Z (X) basis, {\it i.e.}, ${{{N_{{\rm{click,Z(X)}}}}} = {{N_{{\rm{Z(X),X - error}}}}} + {{N_{{\rm Z(X),\rm{No}}\;{\rm{X - error}}}}} }$. Finally, $ {{N_{{\rm{Z}},{\rm{X - error}}}}} $ is the quantity to be estimated, {\it i.e.}, the actual number of phase errors.

So far we have considered that Alice and Bob have single-photon sources, however, it is straightforward to adapt the analysis above to the MDI-QKD protocol described in Sec.~\ref{ass} based on phase randomized WCPs. Since the final key is only distilled from the data associated with the signal intensity setting, we only need to consider that now all the actual numbers in Eq.~(\ref{MDINphase}) actually represent the single-photon contributions within $Z^{\rm ss}$. More precisely, $\left\{ N_{{\rm{sb|click}}},{{N_{{\rm{X}},{\rm{X - error}}}}} ,~ {{N_{{\rm{Z}},{\rm{X - error}}}}},~ {{N_{{\rm{click,X}}}}} ,~ {{N_{{\rm{click,Z}}}}} \right\}$ now refer to $\left\{  N_{{\rm{sb,11,ss|click}}},{{N_{{\rm{X - error}},{\rm{11,ss|X}}}}} ,~{{N_{{\rm{X - error}},{\rm{11,ss|Z}}}}} ,~ {{N_{{\rm{click,11,ss|X}}}}} ,~ {{N_{{\rm{click,11,ss|Z}}}}} \right\}$, where $N_{{\rm{sb,11,ss|click}}}$ denotes the actual number of events where Alice and Bob both select the same basis and the intensity setting $\gamma^{\rm s}$ and send a single photon state given that the relay obtains a successful measurement result. This replacement is allowed because in principle Alice and Bob can perform a quantum non-demolition measurement to know in which pulse each of them emits a single-photon, and we can apply the analysis above to such instances. As long as we do not use explicity in which instance Alice and Bob emit a single-photon, which is exactly the case in our analysis, the security follows.

Then we can rewrite Eq.~(\ref{MDINphase}) as follows:
\begin{equation}\label{MDISphase}
\begin{array}{*{20}{l}}
{{{p_{{{\rm{Z}}_{{{\rm{A}}_{\rm{c}}}}}}}\left( {N_{{\rm{sb,11,ss|click}}}} - {\delta _{{\rm{sb,11,ss|click}}}} \right)\left( {1 - 2{\Delta _{{X_{{{\rm{A}}_{\rm{c}}}}} =  - }}} \right)}}\\
{ \le 2\sqrt {\left( {{N_{{\rm{X-error}},{\rm 11,ss}\left| {\rm{X}} \right.}} + {\delta _{{\rm{X-error}},{\rm 11,ss}\left| {\rm{X}} \right.}}} \right)\left( {{N_{{\rm{X - error}},{\rm 11,ss}\left| {\rm{Z}} \right.}} + {\delta _{{\rm{X - error}},{\rm 11,ss}\left| {\rm{Z}} \right.}}} \right)} }\\
+ 2\sqrt {\left( {{N_{{\rm{click}},{\rm 11,ss}{\rm{|X}}}} - {N_{{\rm{X-error}},{\rm 11,ss}\left| {\rm{X}} \right.}} + {\delta _{{\rm{No}}\;{\rm{X-error}},{\rm 11,ss}\left| {\rm{X}} \right.}}} \right)}\\
\times \sqrt{\left( {{N_{{\rm{click}},{\rm 11,ss}{\rm{|Z}}}} - {N_{{\rm{X - error}},{\rm 11,ss}\left| {\rm{Z}} \right.}} + {\delta _{{\rm{No}}\;{\rm{X - error}},{\rm 11,ss}\left| {\rm{Z}} \right.}}} \right)} .
\end{array}
\end{equation}
Since we have that ${N_{{\rm{sb,11,ss|click}}}}={{N_{{\rm{click,11,ss|Z}}}}} + {{N_{{\rm{click,11,ss|X}}}}}\geq 2\sqrt{{{N_{{\rm{click,11,ss|Z}}}}}{{N_{{\rm{click,11,ss|X}}}}}}$, if we divide the LHS of Eq.~(\ref{MDISphase}) by ${N_{{\rm{sb,11,ss|click}}}}$ and RHS of Eq.~(\ref{MDISphase}) by $2\sqrt{{{N_{{\rm{click,11,ss|Z}}}}}{{N_{{\rm{click,11,ss|X}}}}}}$, respectively, then we obtain that
\begin{equation}
\begin{array}{*{20}{l}}\label{MDIPsphase}
{{{p_{{{\rm{Z}}_{{{\rm{A}}_{\rm{c}}}}}}}\left( 1 - \frac{{\delta _{{\rm{sb,11,ss|click}}}}}{{N_{{\rm{sb,11,ss|click}}}}}\right)\left( {1 - 2{\Delta _{{X_{{{\rm{A}}_{\rm{c}}}}} =  - }}} \right)}}\\
{ \le \sqrt {\frac{{\left( {{N_{{\rm{X-error}},{\rm 11,ss}\left| {\rm{X}} \right.}} + {\delta _{{\rm{X-error}},{\rm 11,ss}\left| {\rm{X}} \right.}}} \right)}}{{{N_{{\rm{click}},{\rm 11,ss}\left| {\rm{X}} \right.}}}}\frac{{\left( {{N_{{\rm{X - error}},{\rm 11,ss}\left| {\rm{Z}} \right.}} + {\delta _{{\rm{X - error}},{\rm 11,ss}\left| {\rm{Z}} \right.}}} \right)}}{{{N_{{\rm{click}},{\rm 11,ss}\left| {\rm{Z}} \right.}}}}} }\\
{ + \sqrt {\left( {1 - \frac{{{N_{{\rm{No~X-error}},{\rm 11,ss}\left| {\rm{X}} \right.}} - {\delta _{{\rm{No}}\;{\rm{X-error}},{\rm 11,ss}\left| {\rm{X}} \right.}}}}{{{N_{{\rm{click}},{\rm 11,ss}\left| {\rm{X}} \right.}}}}} \right)\left( {1 - \frac{{{N_{{\rm{No~X - error}},{\rm 11,ss}\left| {\rm{Z}} \right.}} + {\delta _{{\rm{No}}\;{\rm{X - error}},{\rm 11,ss}\left| {\rm{Z}} \right.}}}}{{{N_{{\rm{click}},{\rm 11,ss}\left| {\rm{Z}} \right.}}}}} \right)} ,}
\end{array}
\end{equation}
also holds except for a small failure probability.

To solve Eq.~(\ref{MDIPsphase}) one can use the same procedure based on the linear optimization method that we employed in the previous section to first estimate the quantities $\left\{{{{N_{{\rm{sb,11,ss|click}}}}}} ,~ {{N_{{\rm{X - error}},{\rm{11,ss|X}}}}},~ {{N_{{\rm{click,11,ss|X}}}}} ,~ {{N_{{\rm{click,11,ss|Z}}}}} \right\}$. Note that, although the relation ${{{N_{{\rm{sb,11,ss|click}}}}}}={{N_{{\rm{click,11,ss|Z}}}}} + {{N_{{\rm{click,11,ss|X}}}}}$ holds in terms of the actual numbers, to estimate an upper bound on the number of phase errors by using Eq.~(\ref{MDIPsphase}), we need to obtain either a lower bound or an upper bound on each of these three parameters. This means that, we need to estimate these three parameters independently. Also, we set $p_{\rm Z_{A_c}}$ as input and search for its optimal value by using a Monte Carlo method. For each given value of $p_{\rm Z_{A_c}}$, the only unknown variable in Eq.~(\ref{MDIPsphase}) is ${{N_{{\rm{X - error}},{\rm{11,ss|Z}}}}}$, which can be numerically estimated by using the optimization toolbox of Matlab.

\section{Simulation of the secret key rate}\label{sim}
\begin{figure}[!t]
\includegraphics*[scale=0.6]{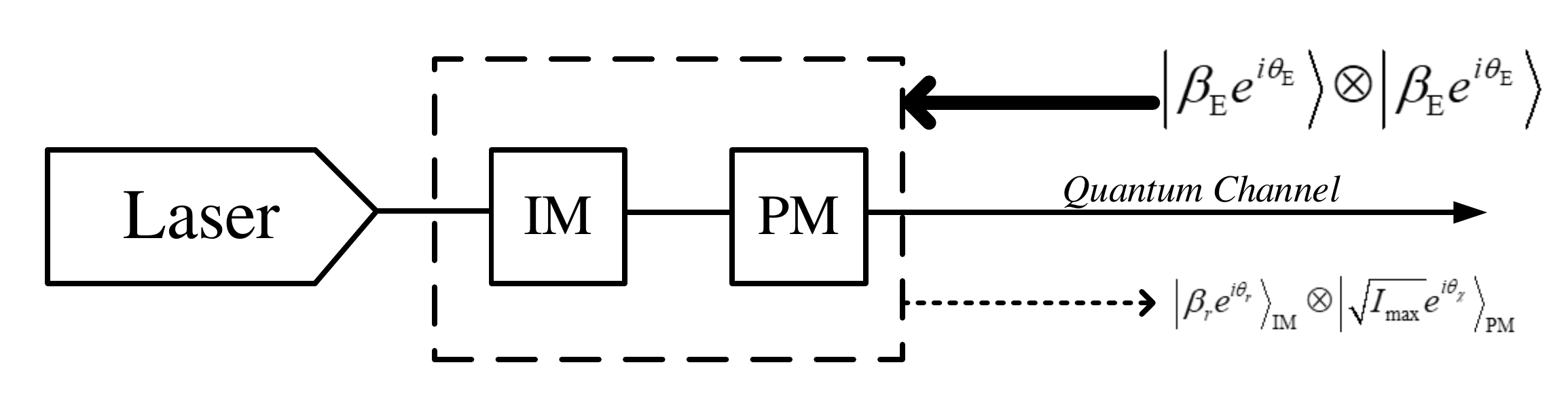}
\caption{ \footnotesize { Example of a THA against the IM and the PM of Alice (Bob). Eve sends Alice (Bob) two high intensity single-mode coherent pulses, each of which is denoted by $\left| {\beta _{\rm E}{e^{i\theta_{\rm E}}}} \right\rangle$. One of them targets the IM and the other one targets the PM. We further assume for simplicity that the back-reflected light from the IM and the PM to Eve is in a product state of two coherent states. One comes from the IM, which we denote by $\left| {{\beta _{r}}{e^{i{\theta_{r}}}}} \right\rangle $, and the other comes from the PM, which has the form $\left| {{\sqrt{I_{\rm max}} }{e^{i{\theta_{\chi}}}}} \right\rangle $, where $r$ and $\chi$ refer to the intensity setting and basis choice, respectively, with $r \in \{\rm s,v,w\}$ and $\chi \in \{\rm Z,X\}$}. Eve can learn partial information about the intensity settings and the basis choices by separately measuring the states $\left| {{\beta _{r}}{e^{i{\theta_{r}}}}} \right\rangle $ and $\left| {{\sqrt{I_{\rm max}} }{e^{i{\theta_{\chi}}}}} \right\rangle $.
\label{THAmodel}}
\end{figure}

In the simulation, only for illustration purposes we assume a particular example of THA, which is shown in Fig.~\ref{THAmodel}. Eve sends Alice (Bob) two high intensity single-mode coherent pulses, each of which is denoted by $\left| {\beta _{\rm E}{e^{{\rm i}\theta_{\rm E}}}} \right\rangle$, with $\beta_{\rm E}$ representing the amplitude and $\theta_{\rm E}$ the phase of the coherent state. One of them targets the IM and the other one targets the PM. For simplicity and due to the lack of experimental data, we shall also assume that the back-reflected light from both the IM and the PM to Eve is still in a coherent state. We further assume that the back-reflected light from the IM has the form $\left| {{\beta _{r}}{e^{{\rm i}{\theta_{r}}}}} \right\rangle $, where the values of the parameters $\beta_{r}$ and $\theta _{r}$ depend on Alice's and Bob's intensity settings each given time with $r\in \{\rm s,v,w\}$, and the back-reflected light from the PM is given by $\left| {{\sqrt{I_{\rm max}} }{e^{i{\theta_{\chi}}}}} \right\rangle $, where $I_{\rm max}$ is the maximum intensity of the back-reflected light and $\chi \in \{\rm Z,~X\}$ refers to the basis choice. Note that, here for simplicity and in order to compare our simulation results with those in~\cite{Weilong2018Finite}, we assume that Eve's back-reflected light from the PM only contains the basis information. That is, we assume that $|{\Psi_{0,\rm{Z}}^i}\rangle_{\rm A',E}=|{\Psi_{0,\rm{Z}}^i}\rangle_{\rm A'}\otimes|{\phi_{\rm{Z}}}\rangle_{\rm E}$ and $|{\Psi_{1,\rm{Z}}^i}\rangle_{\rm A',E}=|{\Psi_{1,\rm{Z}}^i}\rangle_{\rm A'}\otimes|{\phi_{\rm{Z}}}\rangle_{\rm E}$, where the state $|{\phi _{\rm{Z}}}{\rangle _{\rm{E}}}{\rm{ = }}|\sqrt {{I_{{\rm{max}}}}} {e^{i{\theta _{\rm{Z}}}}}\rangle $ of Eve's back-reflected light is the same for both bit values (and similarly for the X basis). To learn partial information about the intensity settings, Eve can measure the state $\left| {{\beta _{r}}{e^{i{\theta_{r}}}}} \right\rangle $, and to learn partial information about the basis choices, Eve can measure the state $\left| {{\sqrt{I_{\rm max}} }{e^{i{\theta_{\chi}}}}} \right\rangle $. We emphasize, however, that this is just a particular model of a THA that we use to evaluate the secret key rate. Our security analysis can be applied to any THA.

We remark that incorporating the information leakage about the bit value is straightforward and our results remain exactly the same even if we include the information leakage about the bit value as long as the fidelities are the same, {\it i.e.}, as long as the value of $\Delta_{\rm X_{A_c}=-}$ is unchanged.

In the presence of information leakage, the actual secret key length, $\ell'$, is bounded by
\begin{equation} \label{MDIkey'}
\ell' \geq \mathop {\max }\limits_{{\Gamma _{\rm AB}}} \mathop {\min }\limits_{{\Gamma _{\rm E}}} \ell,
\end{equation}
where $\ell$ is given by Eq.~(\ref{MDIkey}). Here, $\Gamma_{\rm AB}$ and $\Gamma _{\rm E}$ denote the spaces of the parameters controlled by Alice and Bob, and by Eve, respectively. In the simulation, we assume a practically reasonable value for the weakest decoy state, $\gamma^{\rm w}=5\times10^{-4}$, and, without loss of generality, we assume that $\theta_{\rm s}=0$. The experimental parameters used in the simulations are listed in Table~\ref{MDIparas}.

\begin{table}[tbp]
\caption{\footnotesize {Experimental parameters used in the simulations. The parameter $e_{\rm d}$ is the intrinsic error rate due to the misalignment of the MDI-QKD system; $p_{\rm d}$ is the dark count rate of the relay's detectors, which we assume is equal for all of them; $\eta_{\rm det}$ is the overall detection efficiency of the relay's receiver; $\alpha$ is the loss coefficient of the channel measured in dB/km; and $f_{\rm EC}$ is the efficiency of the error correction code.}}
\centering \label{MDIparas}
\begin{tabular}{|p{2cm}<{\centering}|p{2cm}<{\centering}|p{2cm}<{\centering}|p{2cm}<{\centering}|p{2cm}<{\centering}|p{2cm}r}
\hline
$e_{\rm d}$ & $p_{\rm d}$ & $\eta_{\rm det}$ & $\alpha$  & $f_{\rm EC}$\\ \hline
 $1\%$& $5\times 10^{-6}$ & $0.25$ & $0.2$ & $1.2$\\
 \hline
\end{tabular}
\end{table}

Below we present the simulation results of the secret key rate in three practical cases within the framework of the THA described above. Each case corresponds to a particular model for the back-reflected light.

\subsection{Case 1}
In the framework of the THA considered, it is clear that the higher the intensity of the back-reflected light is, the more information Eve can extract. In this first example, we evaluate a worst-case scenario, where Alice and Bob may overestimate the intensity of the back-reflected light leaked to Eve. In particular, we suppose that the intensity $ {{\beta_{r}}} ^2$ is always upper bounded by a certain value $I_{\rm max}$ for all $r$ and we conservatively assume that
\begin{equation}
{ {{\beta_{{\rm s}}}} ^2} = { {{\beta_{{\rm v}}}} ^2} = { {{\beta_{{\rm w}}}}^2} ={I_{\rm max }}. \label{MDICase1}
\end{equation}

\begin{figure}[!t]
\includegraphics*[width=8.1cm,height=5.5cm]{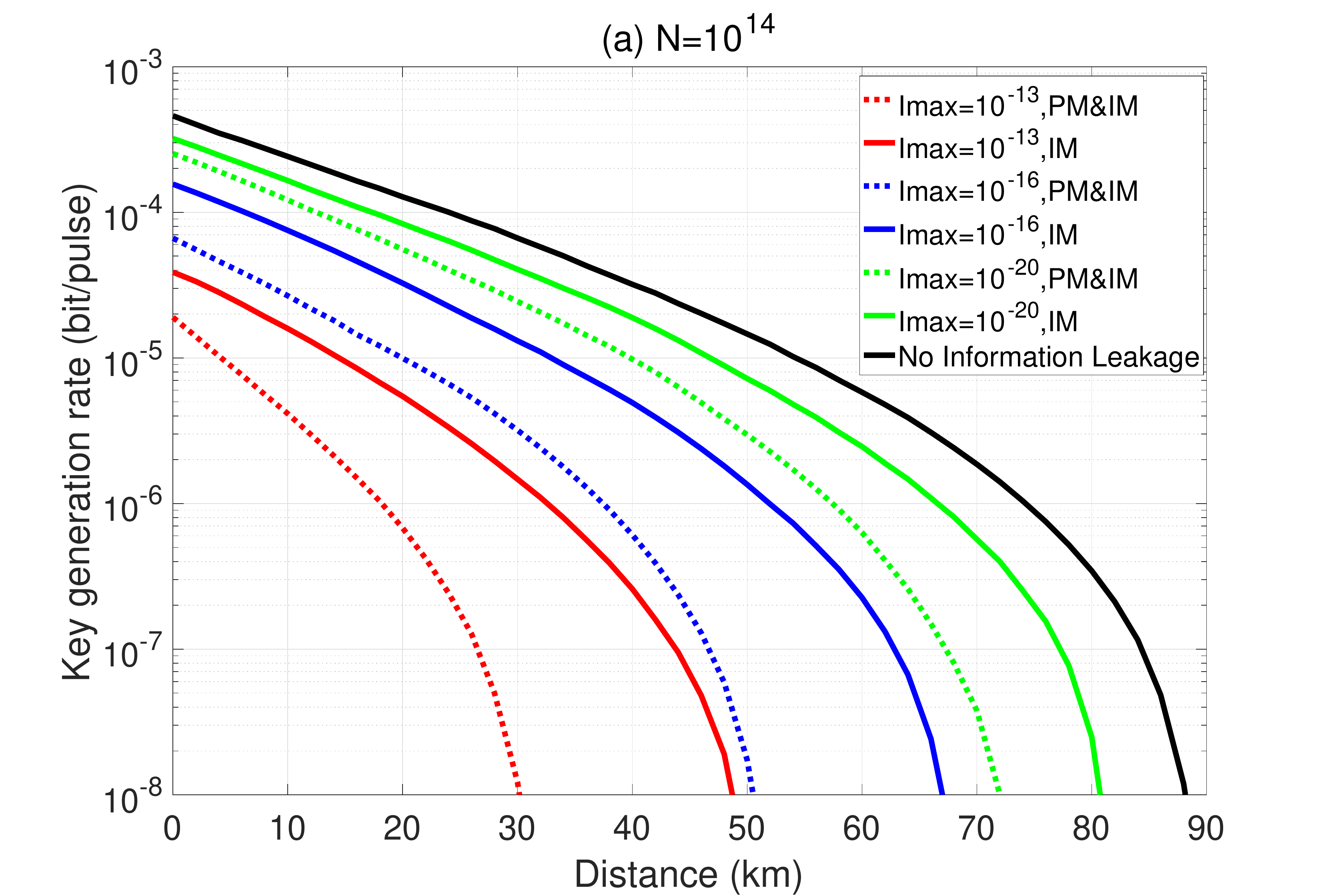}
\includegraphics*[width=8.1cm,height=5.5cm]{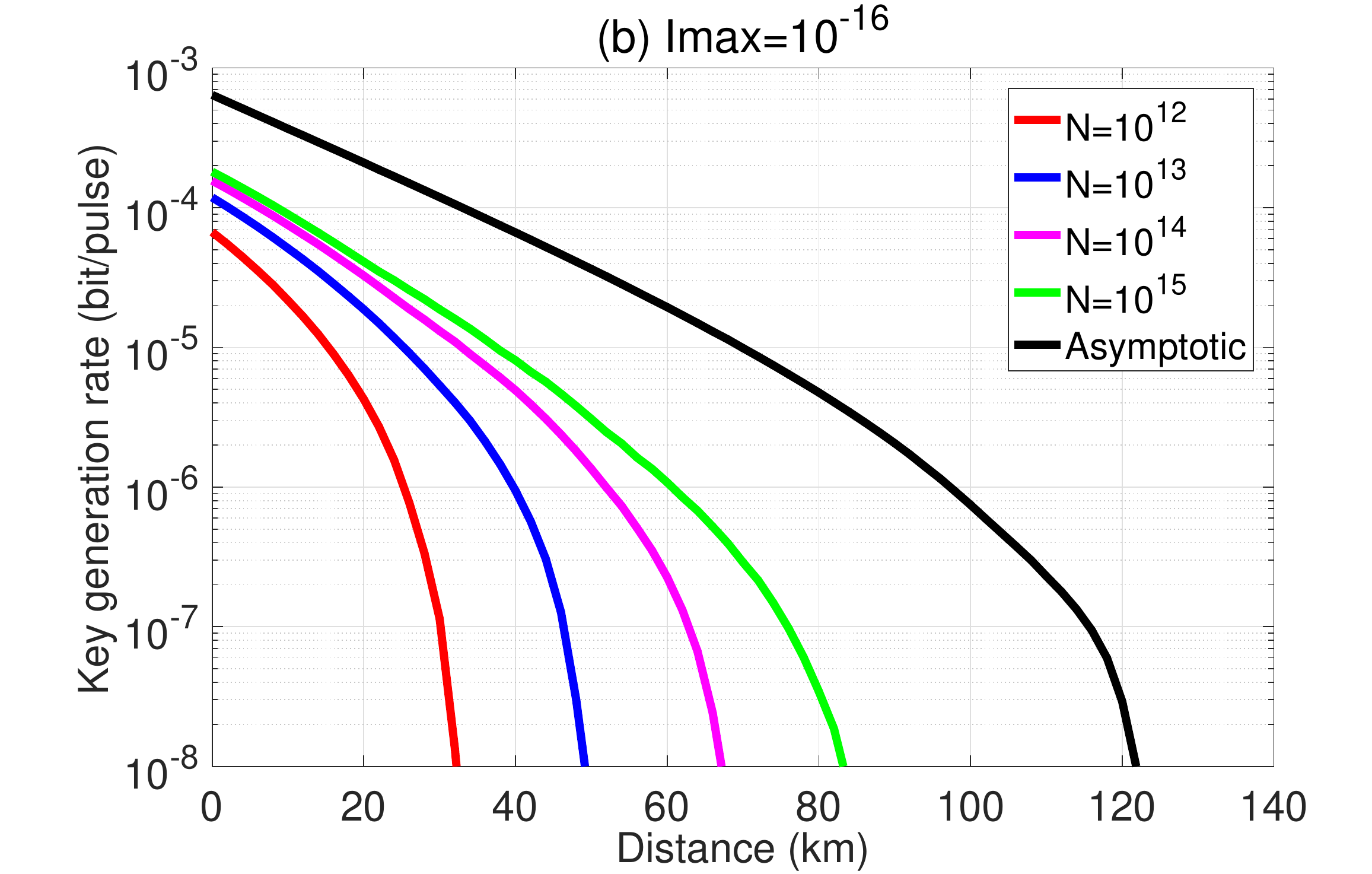}
\caption{ \footnotesize { Case 1. (a) The secret key rate in logarithmic scale as a function of the distance for a fixed value of the total number of transmitted pulses, $N=10^{14}$. The black solid line
represents the perfectly isolated situation where there is no information leakage ({\it i.e.}, $I_{\rm max}=0$) and the different colored lines correspond to different amounts of information leakage. More precisely, the colored solid (dotted) lines represent the secret key rates in the presence of a THA against the IM (both the IM and the PM). (b) The secret key rate in logarithmic scale as a function of the distance for a fixed value of information leakage, $I_{\rm max}=10^{-16}$,  from only the IM. Different colored lines correspond to different values of the number of transmitted pulses. In our simulations, for each value of the distance we maximize the secret key rate over the amplitudes $\gamma^{\rm s}$ and $\gamma^{\rm v}$, and the probabilities $p_{\rm Z_{A_c}}$, $p_{\rm s}$, $p_{\rm v}$ and $p_{\rm Z}$ which are controlled by Alice and Bob, and we minimize it over the angles $\theta_r$ and $\theta_{\chi}$} controlled by Eve, respectively. That is, we consider the worst-case scenario where the phases $\theta_r$ and $\theta_{\chi}$ are selected such that they provide maximal information to Eve.
} \label{MDIfigcase1}
  \end{figure}

The simulation result of the secret key rate, $\ell'/N$, as a function of the transmission distance between Alice and Bob in this case is shown in Fig.~\ref{MDIfigcase1}~(a) for a fixed value of the total number of transmitted pulses, $N=10^{14}$. In this figure, the black solid line represents the key rate in the situation where there is no information leakage, namely $I_{\rm max}=0$, and the different colored lines correspond to different amounts of information leakage. More precisely, the colored solid lines represent the key rates in the presence of a THA against only the IM. If we compare these results with the longest achievable distance without information leakage, which is about 88 km, we find that now the secret key rate vanishes at about 48 km even when $I_{\rm max}$ is as small as $10^{-13}$. The colored dotted lines represent the secret key rates in the presence of a THA against both the IM and the PM. Now the secret key rates are obviously lower than the ones corresponding to a THA against only the IM. For example, when $I_{\rm max}=10^{-13}$ the secret key rate now vanishes at only 30 km.

As already observed in the finite-key analysis for decoy-state QKD~\cite{Weilong2018Finite}, here we also find that in MDI-QKD Alice and Bob need to discard part of their data (on average about $Np_{\rm X_{A_c}}$ pulse pairs) to estimate the phase error rate when there is information leakage from the PM. In our simulation, we find that the optimal value of $p_{\rm Z_{A_c}}$ typically lies in the interval $\left[0.65,0.9 \right]$. Note that, compared to the simulation result in~\cite{Weilong2018Finite}, we have that the value of $p_{\rm Z_{A_c}}$ is typically smaller in the MDI-QKD protocol, which means that MDI-QKD has to sacrifice a bigger proportion of data than in the case of the standard decoy-state QKD protocol to estimate the phase error rate.

Also, we find that MDI-QKD seems to be more sensitive to information leakage. In order to obtain a certain performance, the value of $I_{\rm max}$ in MDI-QKD is much smaller than that in standard decoy-state QKD (indeed, the value of $I_{\rm max}$ in MDI-QKD is roughly the square of that in standard decoy-state QKD). The main reason for this behavior is that in MDI-QKD there are two leaky sources (Alice and Bob) instead of only one leaky source as is the case in standard decoy-state QKD. Thus, to implement the MDI-QKD protocol, both Alice and Bob need to carefully isolate their devices from the external environment to guarantee the security of the system.

In Fig.~\ref{MDIfigcase1}~(b), the different colored lines show the secret key rate as a function of the distance for a fixed value $I_{\rm max}=10^{-16}$ and for different total numbers of transmitted pulses. Here, for simplicity, we only plot the key rates against information leakage from the IM and omit the results when there is also information leakage from the PM as they are similar to those shown in Fig.~\ref{MDIfigcase1}~(b). That is, in this figure we can see the effect of the information leakage as a function of the number of transmitted pulses. For example, when $I_{\rm max}=10^{-16}$, the longest achievable distance is about 84 km when the total number of transmitted pulses is $N=10^{15}$. However, when $N=10^{12}$, this distance decreases to 32 km. Our results indicate that the finite-key effect has a much bigger impact on the secret key rate in the presence of information leakage~\cite{curty2014finite}. The reason for this is mainly that, in order to estimate the statistical fluctuations for a finite sampling size in the presence of information leakage from the IM, our methodology relies on applying Azuma's inequality~\cite{azuma1967weighted} to the total number of \emph{transmitted pulses}. In contrast, when there is no information leakage from the IM, one can apply Azuma's inequality to the number of \emph{pulses detected}. This is so because in this latter case, one can assume a counterfactual scenario where Alice and Bob select their intensity settings {\it a posteriori}, {\it i.e.}, after the relay has detected the successful events. As a consequence, the performance of MDI-QKD in the finite-key regime is comparatively worse in the presence of information leakage from the IM. Note that for the case of information leakage from the PM, we actually apply Azuma's inequality to the number of the detected events.

\begin{figure}[!t]
 \centering
 \includegraphics*[scale=0.46]{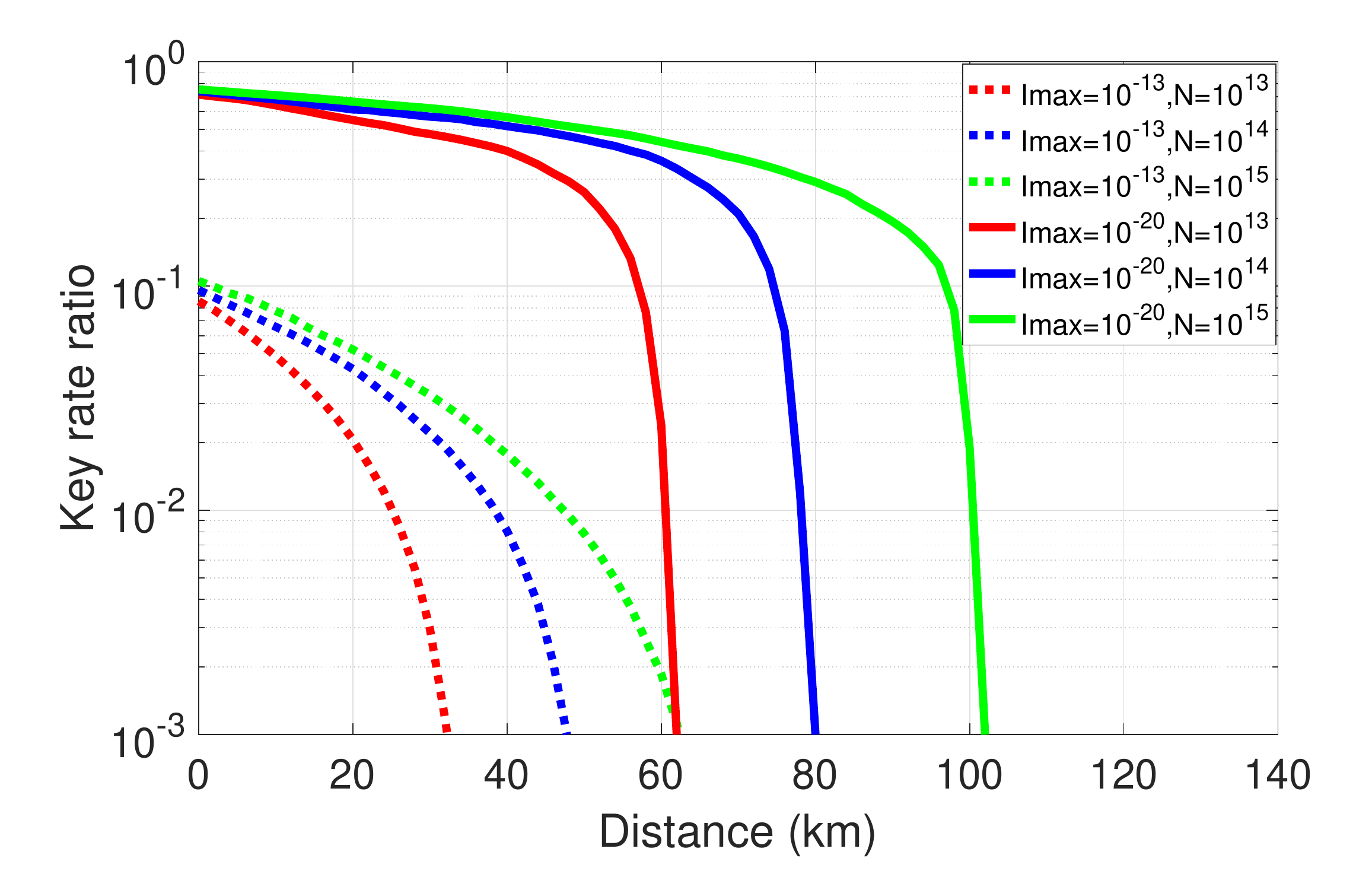}
\caption{ \footnotesize {The ratio ($\ell'_{I_{\rm max}>0}/\ell'_{I_{\rm max}=0}$) between the secret key rates in logarithmic scale with and without information leakage as a function of the distance for two fixed positive values of $I_{\rm max}=\{10^{-13},~10^{-20}\}$. The solid (dotted) lines represent the case $I_{\rm max}=10^{-20}$ ($I_{\rm max}=10^{-13}$). Different colored lines correspond to different values of $N$. }} \label{ratio}
\end{figure}

To further illustrate how the information leakage affects the secret key rate as a function of the number of transmitted pulses, in Fig.~\ref{ratio} we plot the ratio ($\ell'_{I_{\rm max}>0}/\ell'_{I_{\rm max}=0}$) between the secret key rates for two fixed positive values of information leakage, $I_{\rm max}=\{10^{-13},~10^{-20}\}$ and those when $I_{\rm max}=0$ ({\it i.e.}, when there is no information leakage) for different values of $N$. Here, for simplicity, we disregard again the information leakage from the PM. From Fig.~\ref{ratio} one can see that given a fixed distance and a fixed value of $N$, the ratio when $I_{\rm max}=10^{-13}$ is at least one order of magnitude lower than that when $I_{\rm max}=10^{-20}$. And the ratio when $I_{\rm max}=10^{-13}$ drops faster as the distance increases than that when $I_{\rm max}=10^{-20}$. For instance, if we focus on the red lines, from 0 km to 30 km, the ratio drops from about $10^{-1}$ to $10^{-3}$ when $I_{\rm max}=10^{-13}$ ({\it i.e.}, two orders of magnitude) while the ratio drops only from 0.71 to 0.49 ({\it i.e.}, of the same order of magnitude) when $I_{\rm max}=10^{-20}$. This suggests that the effect of information leakage increases when $N$ decreases, and the finite-size effect is amplified when the amount of information leakage increases. We remark that the simulation results for the other two cases that we consider next are analogous to those of Fig.~\ref{ratio} and thus we omit them in the next two subsections.

\subsection{Case 2}
\begin{figure}[!t]
\includegraphics*[width=8.1cm,height=5.5cm]{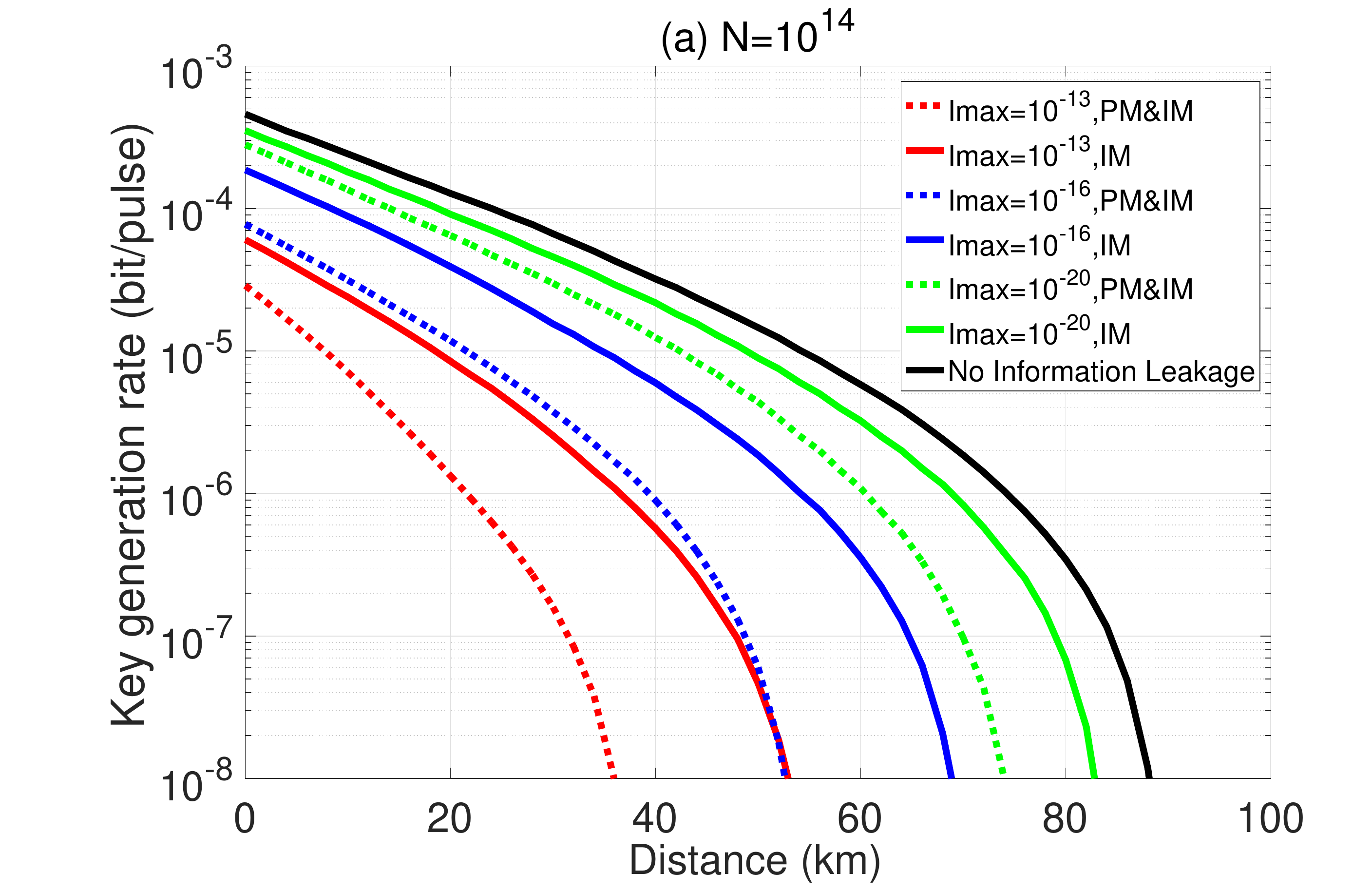}
\includegraphics*[width=8.1cm,height=5.5cm]{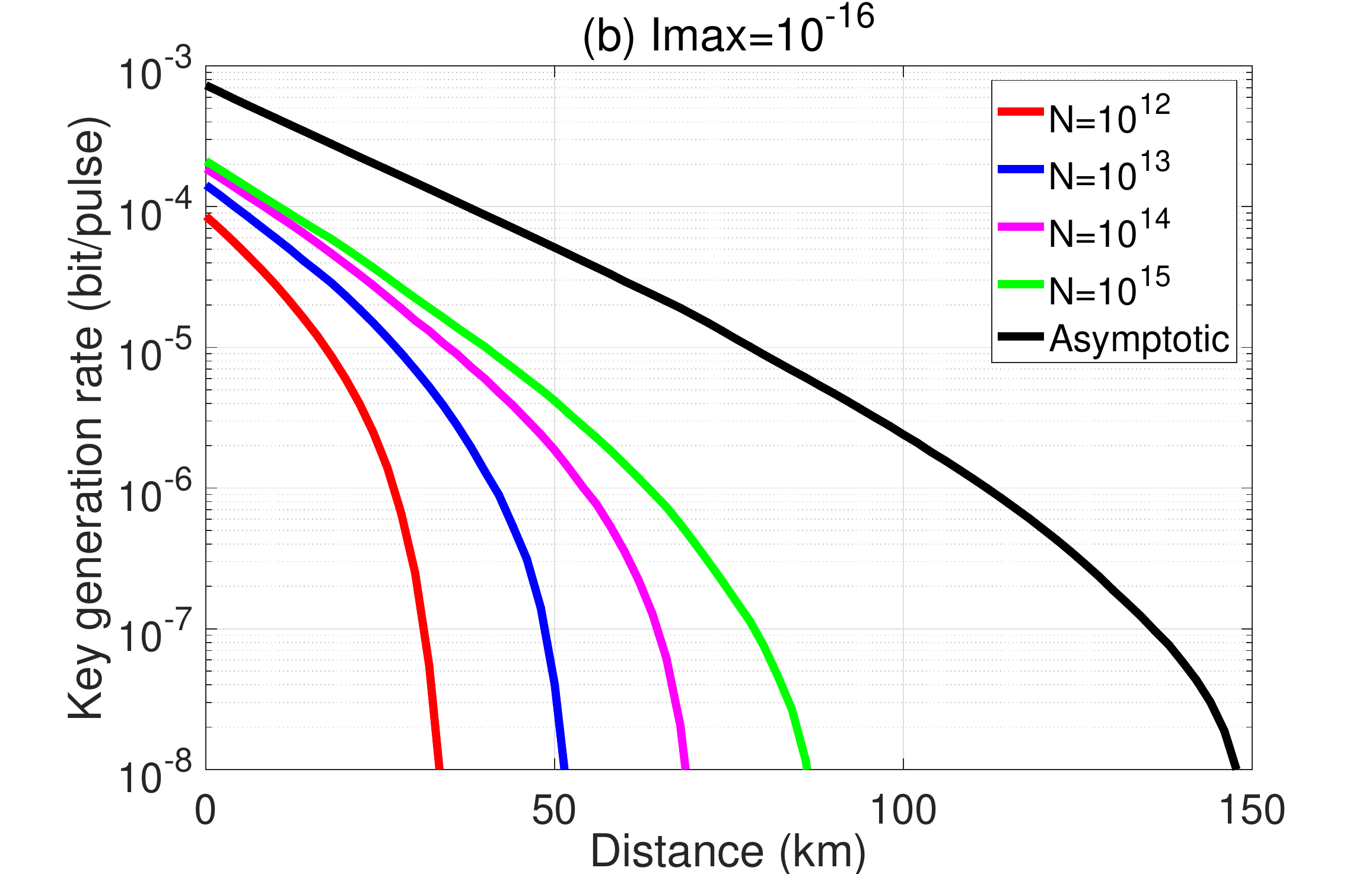}
\caption{ \footnotesize { Case 2. (a) The secret key rate in logarithmic scale as a function of the distance for a fixed value of the total number of transmitted pulses, $N=10^{14}$. The black solid line
represents the perfectly isolated situation where there is no information leakage ({\it i.e.}, $I_{\rm max}=0$) and the different colored lines correspond to different amounts of information leakage. More precisely, the colored solid (dotted) lines represent the secret key rates in the presence of a THA against the IM (both the IM and the PM). (b) The secret key rate in logarithmic scale as a function of the distance for a fixed value of information leakage, $I_{\rm max}=10^{-16}$,  from only the IM. Different colored lines correspond to different values of the number of transmitted pulses. In our simulations, for each value of the distance we maximize the secret key rate over the amplitudes $\gamma^{\rm s}$ and $\gamma^{\rm v}$, and the probabilities $p_{\rm Z_{A_c}}$, $p_{\rm s}$, $p_{\rm v}$ and $p_{\rm Z}$ which are controlled by Alice and Bob, and we minimize it over the angles $\theta_r$ and $\theta_{\chi}$ controlled by Eve, respectively. That is, we consider the worst-case scenario where the phases $\theta_r$ and $\theta_{\chi}$ are selected such that they provide maximal information to Eve.}} \label{MDIfigcase2}
  \end{figure}
In the previous case, we considered a conservative scenario for Alice and Bob, where the intensity of the back-reflected light is maximal and independent of the settings of the IM. Thus, the amount of information leaked might be overestimated, which results in a relatively pessimistic lower bound on the secret key rate. However, in practice, the input light of Eve may also go through the IM. As a consequence, the back-reflected light could be modulated in the same manner as the senders' pulses during the state preparation process. In this case, we have that
\begin{equation}
{ { {{\beta_{{\rm s}}}} ^2} = \frac{{{\gamma ^{\rm s}}}}{{{\gamma ^{\rm v}}}}{ {{\beta_{{\rm v}}}} ^2} = \frac{{{\gamma ^{\rm s}}}}{{{\gamma ^{\rm w}}}}{ {{\beta_{{\rm w}}}} ^2= I_{\max }}}. \label{MDICase2}
\end{equation}
That is, here we assume that the maximum amount of information leakage comes from the largest intensity setting of the senders, namely ${I_{\max }} = { {{\beta_{{\rm s}}}} ^2}$. The intensity of the back-reflected light corresponding to the other intensity settings fulfills the conditions: ${{\beta_{{\rm s}}}} ^2/{{\beta_{{\rm v}}}} ^2={\gamma ^{\rm s}}/{\gamma ^{\rm v}}$ and ${{\beta_{{\rm s}}}} ^2/{{\beta_{{\rm w}}}} ^2={\gamma ^{\rm s}}/{\gamma ^{\rm w}}$.

The simulation result of the secret key rate as a function of the transmission distance between Alice and Bob when $N=10^{14}$ and for different values of $I_{\rm max}$ is shown in Fig.~\ref{MDIfigcase2}~(a). Fig.~\ref{MDIfigcase2}~(b) shows the secret key rate as a function of the distance in the presence of information leakage (with a fixed value of $I_{\rm max}=10^{-16}$) from only the IM and for different values of $N$. The behavior of the curves is very similar to those in Case 1, and in the simulation we find that the optimized value of $p_{\rm Z_{A_c}}$ is similar as well. One main difference is that with the same experimental parameters the secret key rate is now a little bit higher and can go a bit further than that in Case 1. For example, when the total number of transmitted pulses is $10^{14}$ and $I_{\rm max}=10^{-13}$, we find that the secret key is positive up to about 54 km while in Case 1 this distance is 48 km in the presence of information leakage only from the IM.

\subsection{Case 3}
In this case we consider a more favorable situation for Alice and Bob where they implement an additional step to randomize the phase of each signal going out of their transmitters including the back-reflected light to Eve. Moreover, we optimistically assume that there is no information leakage from this phase randomization step. Furthermore, we suppose that the amplitudes $\beta_{k}$ still satisfy Eq.~(\ref{MDICase2}) like in the previous case. Then, we have that the state of Eve's back-reflected light from the IM and the PM are given by:
\begin{equation}
\begin{array}{*{20}{l}}
{\rho _{{\gamma ^k}}} = {e^{ -(\beta_{k})^2}}\sum\limits_{n = 0}^\infty {\frac{{(\beta_{k})^2}}{{n!}}} |n\rangle\langle n|,\\
{{\rho _{{I_{\max }}}} = {e^{ - {I_{\max }}}}\sum\limits_{n = 0}^\infty  {\frac{{{I_{\max }}}}{{n!}}} |n\rangle \langle n|,
}\end{array}\label{MDICase3}
\end{equation}respectively.

This means that the information about Alice's and Bob's inner settings can only be leaked via the amplitudes of the back-reflected light but Eve cannot obtain any information from the phases. We remark that here we consider a model which is different from the ones considered in previous works~\cite{tamaki2016decoy,Weilong2018Finite}. To be precise, in Refs.~\cite{tamaki2016decoy,Weilong2018Finite} the authors consider that the phase randomization step is only applied to the back-reflected light from the IM. However, here we consider that this step is applied to the back-reflected light from both the IM and the PM. This means that, now Eve cannot exploit any information leakage from the PM, but only information leakage from the IM as the state $\rho_{I_{\rm max}}$ does not depend on the basis choice. Thus, in this situation, the resulting secret key rate when Eve only attacks the IM and that corresponding to the case where Eve attacks both the IM and the PM coincide.

\begin{figure}[!t]
\includegraphics*[width=8.1cm,height=5.5cm]{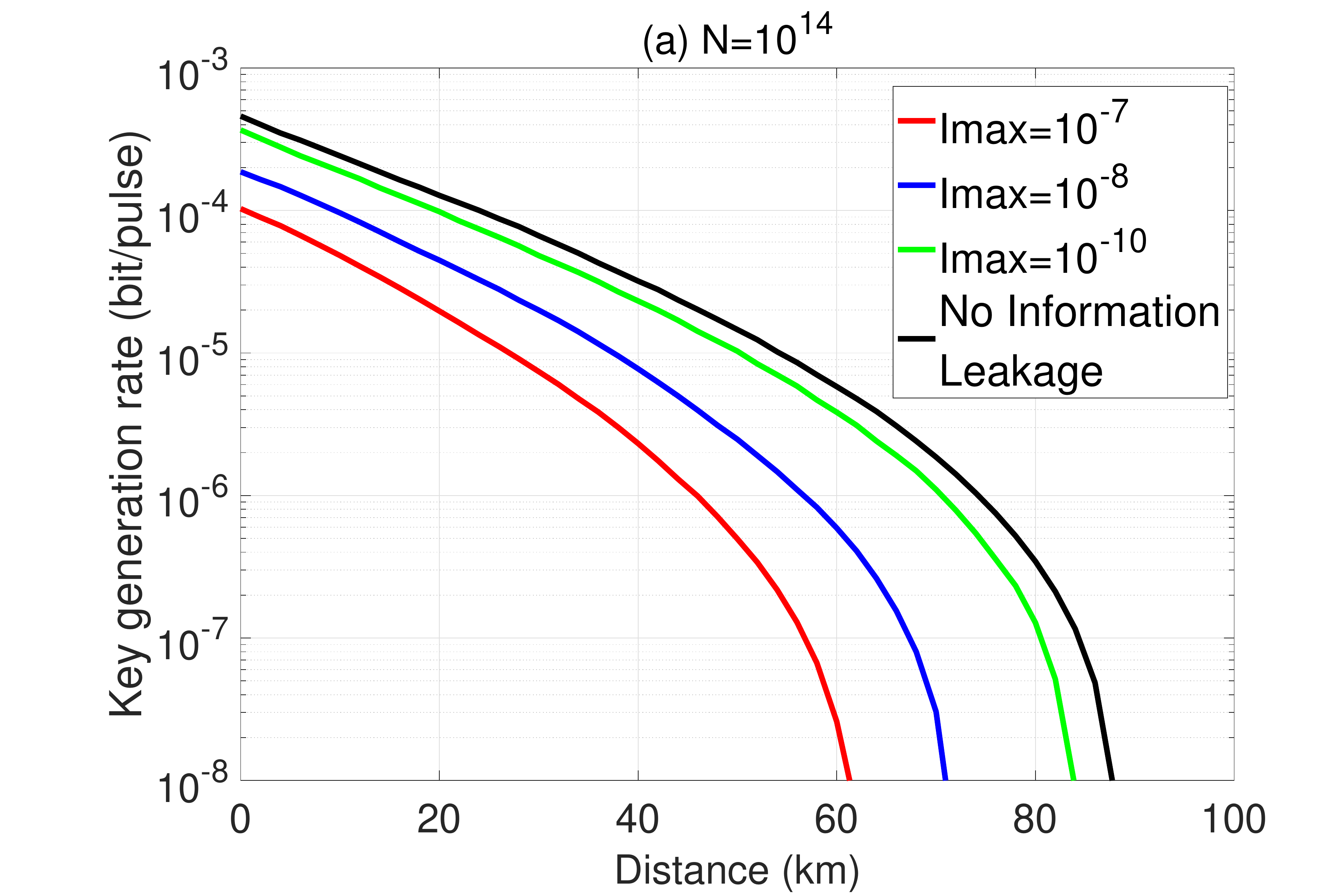}
\includegraphics*[width=8.1cm,height=5.5cm]{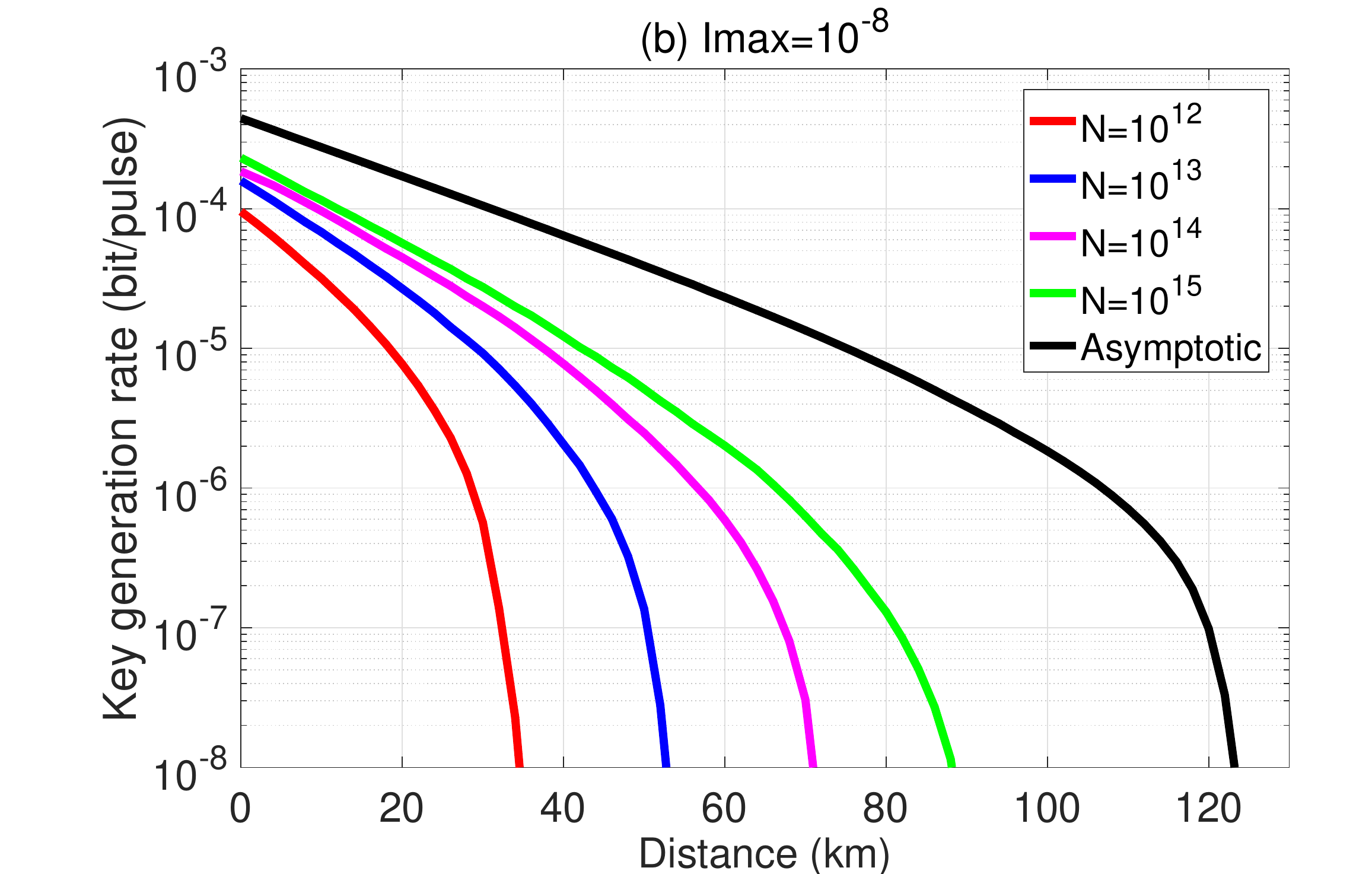}
\caption{ \footnotesize { Case 3. (a) The secret key rate in logarithmic scale as a function of the distance for a fixed value of the total number of transmitted pulses, $N=10^{14}$. The black solid line represents the perfectly isolated situation where there is no information leakage ({\it i.e.}, $I_{\rm max}=0$) and the different colored lines correspond to different amounts of information leakage. Here, the case where Eve attacks only the IM and the case where Eve attacks both the IM and the PM coincide}. (b) The secret key rate in logarithmic scale as a function of the distance for a fixed value of information leakage, $I_{\rm max}=10^{-8}$. Different colored lines correspond to different values of the number of transmitted pulses. In our simulations, for each value of the distance we maximize the secret key rate over the amplitudes $\gamma^{\rm s}$ and $\gamma^{\rm v}$, and the probabilities $p_{\rm Z_{A_c}}$, $p_{\rm s}$, $p_{\rm v}$ and $p_{\rm Z}$ which are controlled by Alice and Bob.} \label{MDIfigcase3}
  \end{figure}

The simulation result of the secret key rate as a function of the transmission distance between Alice and Bob when $N=10^{14}$ and for different values of $I_{\rm max}$ is shown in Fig.~\ref{MDIfigcase3}~(a). Fig.~\ref{MDIfigcase3}~(b) shows the finite-key effect on the secret key rate as a function of the distance for a fixed value of $I_{\rm max}=10^{-8}$ and for different values of $N$. Here, we find that the typical interval where $p_{\rm Z_{A_c}}$ lies is $\left[0.71,0.93 \right]$. Compared to the secret key rate shown in Figs~\ref{MDIfigcase1} and \ref{MDIfigcase2}, now the secret key rate is obviously improved. For example, when the total number of transmitted pulses is $N=10^{14}$ and $I_{\rm max}=10^{-7}$, the secret key rate remains positive up to about 62 km. In comparison, the maximum achievable distance with the same number of transmitted pulses and assuming an $I_{\rm max}$ as low as $10^{-13}$ is only about 36 km in Case 2, and it is even worse in Case 1.

In practice, however, Eve might also perform a THA against the phase randomization step to obtain some information about the random phase applied by Alice and Bob each given time. This will reduce the benefit of the phase randomization step. One can also analyze this last scenario with the techniques in this paper, but for simplicity we omit it here.

\section{The four-intensity decoy-state MDI-QKD protocol}\label{XBP}
In this section, we now consider the MDI-QKD protocol introduced in~\cite{zhou2016making}, which has been recently implemented over a record distance of 404 km~\cite{yin2016measurement} (we note that the current record is 421 km~\cite{boaron2018secure}). In that protocol, each of Alice and Bob uses one intensity setting $\gamma^{\rm s}$ for the Z basis states, and three intensity settings $\gamma^{\rm v}$, $\gamma^{\rm w}$ and $\gamma^{\rm 0}=0$ for the X basis states. This is motivated by the fact that in order to increase the number of single-photon pulses emitted in the Z basis used for key generation, the intensity of the signal states, $\gamma_{\rm s}$, needs to be close to one, while in order to have a tight estimation of the relevant parameters, the intensities in the X basis used for parameter estimation need to be weak. With the four-intensity decoy-state MDI-QKD protocol, one can optimize the intensities for key generation and parameter estimation independently. The probabilities to select the corresponding intensities are $p_{\rm s}$, $p_{\rm v}$, $p_{\rm w}$ and $p_{\rm 0}$, respectively, with $p_{\rm s}+p_{\rm v}+p_{\rm w}+p_{\rm 0}=1$. Note that the probability to choose the Z basis is now $p_{\rm{Z}}=p_{\rm s}$ and the probability to choose the X basis is given by $p_{\rm{X}}=p_{\rm v}+p_{\rm w}+p_{\rm 0}$.

The security analysis of this protocol against information leakage from the IM and the PM is slightly different from that in the previous section. This is because of the following. Since now the intensity setting in the Z basis is unique and it is typically different from the intensity settings in the X basis, by analyzing the information leakage from the IM Eve can also learn partial information about the users' basis choices. Similarly, by analyzing the information leakage from the PM Eve can learn partial information about the users' intensity settings as well. That is, the information leakage from the IM and the PM of each user is now correlated. Fortunately, a general procedure to estimate the relevant parameters in this situation has already been briefly introduced in Ref.~\cite{tamaki2016decoy}. It applies the quantum coin idea to estimate the phase error rate, together with a similar ``\emph{post-selection} step" like that employed in~\cite{Weilong2018Finite} where Alice post-selects part of her data (with probability $p_{\rm Z_{A_c}}$) and discards the other part.

\begin{figure}[!t]
\includegraphics*[scale=0.6]{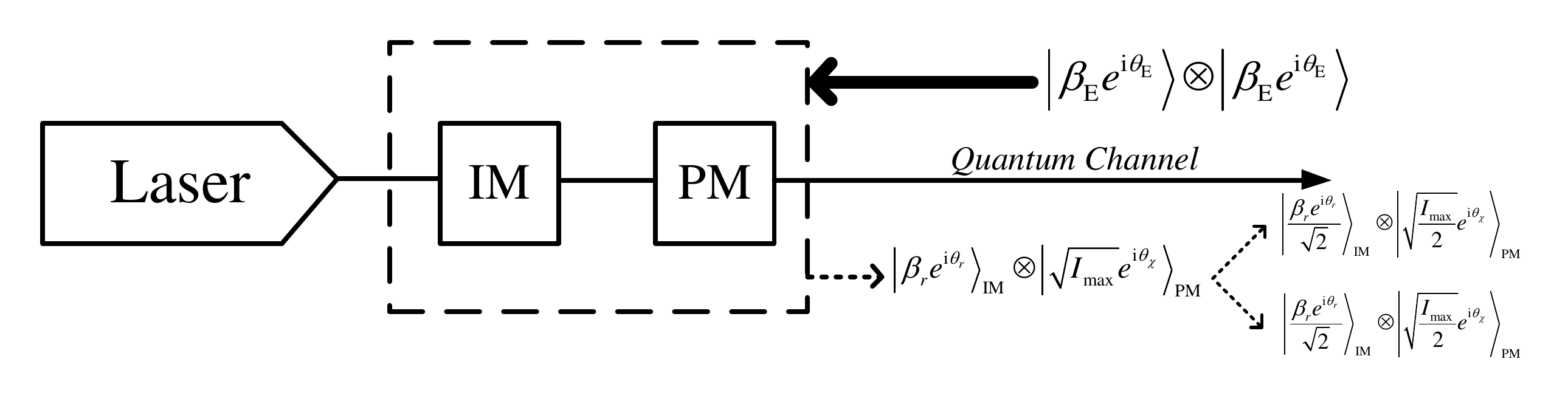}
\caption{ \footnotesize { Example of a THA against correlated IM and PM of Alice (Bob). Eve sends Alice (Bob) two high intensity single-mode coherent pulses, each of which is denoted by $\left| {\beta _{\rm E}{e^{i\theta_{\rm E}}}} \right\rangle$. One of them targets the IM and the other one targets the PM. We further assume for simplicity that the back-reflected light from the IM and the PM to Eve is in a product state of two coherent states. One comes from the IM, which we denote by $\left| {{\beta _{r}}{e^{i{\theta_{r}}}}} \right\rangle $, and the other comes from the PM, which has the form $\left| {{\sqrt{I_{\rm max}} }{e^{i{\theta_{\chi}}}}} \right\rangle $, where $r$ and $\chi$ refer to the intensity setting and basis choice, respectively, with $r \in \{\rm s,v,w,0\}$ and $\chi \in \{\rm Z,X\}$} for the MDI-QKD protocol introduced in~\cite{zhou2016making}. In addition, for simplicity, we assume that Eve splits the joint back-reflected light (from both the IM and the PM together) into two parts by means of a 50:50 beamsplitter, one part is used to learn partial information about the intensity settings and the other part is used to learn partial information about the basis choices.}
\label{THAXiangbin}
\end{figure}

In what follows, for illustration purposes we consider a particular example of a THA against the correlated IM and PM performed by Eve, which is shown in Fig.~\ref{THAXiangbin}. For comparison purposes, we use the same assumptions in this THA like those in Sec.~\ref{sim}. That is, we suppose that Eve sends Alice (Bob) two high intensity single-mode coherent pulses, each of which is denoted by $\left| {\beta _{\rm E}{e^{i\theta_{\rm E}}}} \right\rangle$. One of them targets the IM and the other one targets the PM. And the state of the back-reflected light from the IM has the form $\left| {{\beta _{r}}{e^{i{\theta_{r}}}}} \right\rangle $ and the one from the PM is $\left| {{\sqrt{I_{\rm max}} }{e^{i{\theta_{\chi}}}}} \right\rangle $, where $r$ and $\chi$ refer to the intensity setting and basis choice, respectively, with $r \in \{\rm s,v,w,0\}$ and $\chi \in \{\rm Z,X\}$. Now, since the IM and the PM are correlated, in this THA Eve can jointly measure the states $\left| {{\beta _{r}}{e^{i{\theta_{r}}}}} \right\rangle $ and $\left| {{\sqrt{I_{\rm max}} }{e^{i{\theta_{\chi}}}}} \right\rangle $ to extract partial information about both the intensity settings and the basis choices. Importantly, to have a fair comparison with the simulation results shown in Sec.~\ref{sim}, we assume that the amount of information leaked to Eve in both protocols is the same. That is, we assume that the intensity of the back-reflected light is equal in both cases. In addition, in this THA we assume, for example, that Eve splits the joint back-reflected light (from both the IM and the PM together, which can be described with the product state $\left| {{\beta _{r}}{e^{i{\theta_{r}}}}} \right\rangle \otimes \left| {{\sqrt{I_{\rm max}} }{e^{i{\theta_{\chi}}}}} \right\rangle $) into two parts by means of a 50:50 beamsplitter, one part is used to learn partial information about the intensity settings and the other part is used to learn partial information about the basis choices. We remark, however, that our method to estimate the phase error rate could be applied to any strategy applied by Eve.

Note that, in general, when the IM and the PM are correlated, the yields associated with different photon number states can also depend on the bit value~\cite{tamaki2016decoy}. However, for simplicity, in the model above we assume that the back-reflected light does not carry information about the bit value. This latter case is briefly discussed in Appendix C.

\subsection{Estimation of the parameters $N^{L}_{{\rm click,00,\rm {ss}}}$ and $N^{L}_{{\rm click,11,\rm {ss}}}$}
In this section, we present the procedure to estimate the parameters $N^{L}_{{\rm click,00,\rm {ss}}|\rm Z}$ and $N^{L}_{{\rm click,11,\rm {ss}}|\rm Z}$ which are needed to evaluate the secret key rate formula given by Eq.~(\ref{MDIkey}). Due to the fact that in this protocol the intensity $\gamma^{\rm s}$ is only used for the data in the Z basis, we have that $N^{L}_{{\rm click,00,\rm {ss}}|\rm Z}\equiv N^{L}_{{\rm click,00,\rm {ss}}}$ and $N^{L}_{{\rm click,11,\rm {ss}}|\rm Z}\equiv N^{L}_{{\rm click,11,\rm {ss}}}$, where $N^{L}_{{\rm click,00,\rm {ss}}}$ ($N^{L}_{{\rm click,11,\rm {ss}}}$) is a lower bound on the number of events where Alice and Bob both select the intensity $\gamma^{\rm s}$ and send a vacuum (single-photon) pulse, and the relay obtains a successful measurement result. In the MDI-QKD protocol introduced in~\cite{zhou2016making}, the data in the Z basis is used for key distillation and the data in the X basis is used for parameter estimation, and we first estimate the quantities $N^{L}_{{\rm click,00,\rm {vv}}}$ and $N^{L}_{{\rm click,11,\rm {vv}}}$, which correspond to the intensity $\gamma^{\rm v}$ only in the X basis. For this we use the same techniques, like in the previous section, which rely on the trace distance argument among events only in the X basis. More precisely, the quantities $N^{L}_{{\rm click,00,\rm {vv}}}$ and $N^{L}_{{\rm click,11,\rm {vv}}}$ can be estimated by applying similar linear programming techniques with all the sifted data in the X basis as what has been done in Sec.~\ref{THAIM}. This analysis is valid because we can imagine a fictitious delayed measurement scenario in which Alice and Bob determine the basis first, and then for the events where both of them use the X basis, Alice and Bob start to choose the intensity settings. Note that since this estimation only involves events in the X basis, it is not affected by the information leakage from the PM. Next, we can relate these two quantities to the parameters $N^{L}_{{\rm click,00,\rm {ss}}}$ and $N^{L}_{{\rm click,11,\rm {ss}}}$, respectively, by using the trace distance argument between the X and Z basis states. For example, in Eqs.~(\ref{MDITHA})-(\ref{MDIYD}) if we focus on the total number of events $N$ instead of on the $N_{\chi}$ events where both Alice and Bob select the $\chi$ basis and take $n=m=0$, $j_{\rm A}=j_{\rm B}={\rm s}$ and $k_{\rm A}=k_{\rm B}={\rm v}$, we have that
\begin{equation}\label{XBMDIYDel}
{{\mathcal E}_{{\rm{click}},00,{\rm{ss}}}} = \frac{{p_{\rm{s}}^2{{(p_0^{\rm{s}})}^2}}}{{p_{\rm{v}}^2{{(p_0^{\rm{v}})}^2}}}{{\mathcal E}_{{\rm{click}},00,{\rm{vv}}}} + \Delta _{00}^{{\rm{ssvv}}},
\end{equation}
where $\mathcal{E}_{{\rm{click}},00,{\rm vv}}$ is the expected number of events where both Alice and Bob select the intensity setting $\gamma^{{\rm v}}$ and send a vacuum pulse and the relay obtains a successful event. Note that, $\mathcal{E}_{{\rm{click}},00,{\rm vv}}\equiv\mathcal{E}_{{\rm{click}},00,{\rm vv}|\rm X}$ as the intensity setting $\gamma^{\rm v}$ is only selected in the X basis. The quantity $\mathcal{E}_{{\rm{click}},00,{{\rm s}}{{\rm s}}}$ is defined in an analogous way and it is equal to $\mathcal{E}_{{\rm{click}},00,{{\rm s}}{{\rm s}}|\rm Z}$. The deviation term ${\Delta_{00} ^{{{\rm s}}{{\rm s}}{{\rm v}}{{\rm v}}}}\in \left[  - p^2_{{{\rm s}}}(p_0^{{\rm{s}}})^2ND^{{\rm s}{\rm s} ,{\rm v}{\rm v}}_{00 },~p^2_{{{\rm s}}}(p_0^{{\rm{s}}})^2ND^{{\rm s}{\rm s} ,{\rm v}{\rm v}}_{00 } \right]$, where
\begin{equation}D^{{\rm s}{\rm s} ,{\rm v}{\rm v}}_{00 } =\frac{1}{N}\sum\limits_{i = 1}^{{N }} {D_{00}^{{{\rm{s}}}{{\rm{s}}},{{\rm{v}}}{{\rm{v}}},i}}:=\frac{1}{{{N }}}\sum\limits_{i = 1}^{{N }} {{\rm{Tr}}} \left[ \sqrt{\left({\rho _{00}^{\gamma^{{\rm s}}\gamma^{{\rm s}},i}}-\rho _{00}^{\gamma^{{\rm v}}\gamma^{{\rm v}},i}\right)^2} \right]
\end{equation}
with ${\rho _{00}^{\gamma^{{\rm v}}\gamma^{{\rm v}},i}}$ (${\rho _{00}^{\gamma^{{\rm s}}\gamma^{{\rm s}},i}}$) being the normalized joint state of both Alice's and Bob's vacuum pulses when they both select the intensity setting $\gamma^{\rm v}$ ($\gamma^{\rm s}$) together with the systems $E_{\rm a}^{\rm A}$, $E_{\rm a}^{\rm B}$, $E_{\rm p}^{\rm A'}$, $E_{\rm p}^{\rm B'}$ in the $i$th trial. The definitions of $E_{\rm a}^{\rm A}$, $E_{\rm a}^{\rm B}$, $E_{\rm p}^{\rm A'}$ and $E_{\rm p}^{\rm B'}$ are the same as those of Sec.~\ref{THAIM}. Next we apply Azuma's inequality to Eq.~(\ref{XBMDIYDel}) and obtain
\begin{equation}\label{XBYDel}
{{N_{{\rm{click}},00,{\rm{ss}}}} = \frac{{p_{\rm{s}}^2{{(p_0^{\rm{s}})}^2}}}{{p_{\rm{v}}^2{{(p_0^{\rm{v}})}^2}}}({N_{{\rm{click}},00,{\rm{vv}}}} + \delta _{00}^{{\rm{vv}}}) - \delta _{00}^{{\rm{ss}}} + \Delta _{00}^{{\rm{ssvv}}}},
\end{equation}
where the parameter $\delta^{\rm ss}_{00}$ denotes the deviation term between $\mathcal{E}_{{\rm{click}},00,{\rm ss}}$ and $N_{{\rm{click}},00,{\rm ss}}$ and it can be bounded by $[  - \Delta _{00}^{\rm ss},\widehat{\Delta} _{00}^{\rm ss} ]$ except for a small error probability $\varepsilon_{00}^{\rm ss}+\widehat{\varepsilon}_{00}^{\rm ss}$. The bounds are given by $ \Delta _{00}^{\rm ss}=f(N,\varepsilon_{00}^{\rm ss})$ and $\widehat{\Delta} _{00}^{\rm ss}=  f(N,\widehat{\varepsilon}_{00}^{\rm ss})$, and the parameter $\delta^{\rm vv}_{00}$ denotes the deviation term between $\mathcal{E}_{{\rm{click}},00,{\rm vv}}$ and $N_{{\rm{click}},00,{\rm vv}}$ and it can be bounded by $[  - \Delta _{00}^{\rm vv},\widehat{\Delta} _{00}^{\rm vv} ]$ except for a small error probability $\varepsilon_{00}^{\rm vv}+\widehat{\varepsilon}_{00}^{\rm vv}$. The bounds are given by $ \Delta _{00}^{\rm vv}=f(N,\varepsilon_{00}^{\rm vv})$ and $\widehat{\Delta} _{00}^{\rm vv}=  f(N,\widehat{\varepsilon}_{00}^{\rm vv})$. In this way, given the quantity $N^{L}_{{\rm click,00,\rm {vv}}}$, we can estimate $N^{\rm L}_{{\rm{click}},00,{{\rm s}}{{\rm s}}}$ according to Eq.~(\ref{XBYDel}).

Note that, in general, one could alternatively first estimate all $N_{{\rm{click}},{\rm{00}},{{kl}}}^{L}$ for $k,l\in \{\rm v,w,0\}$ by using the linear programming method. Then by performing the trace distance argument, one obtains
\begin{equation}\label{sv}
  \left|\mathcal{E}_{{\rm{click}},00,{{\rm s}}{{\rm s}}} - \sum\limits_{k,l} {{q_{00,kl}}\mathcal{E}_{{\rm{click}},00,kl}} \right| \le {D^{{\rm{ss}},\sum kl}_{{\rm{00}}}},
\end{equation}
with $k,l\in \{\rm v,w,0\}$, where $q_{00,kl}$ is the normalization probability given by ${q_{00,kl}} = \frac{{{p_k}p^k_0{p_l}p^l_0}}{{\sum\limits_{k',l' \in \{ {\rm{v}}{\rm{,w}}{\rm{,0}}\} } {{p_{k{'}}}p^{k'}_0{p_{l{'}}}p^{l'}_0} }}$ and ${D^{{\rm{ss}},\sum kl}_{{\rm{00}}}}=\frac{1}{N}\sum\limits_{i = 1}^{{N }} {D_{00}^{{{\rm{s}}}{{\rm{s}}},\sum kl,i}}:=\frac{1}{{{N }}}\sum\limits_{i = 1}^{{N }} {{\rm{Tr}}} \left[ \sqrt{(\sum\limits_{k,l} {{q_{00,kl}}{\rho _{00}^{\gamma^k\gamma^l,i}}-\rho _{00}^{\gamma^{{\rm s}}\gamma^{{\rm s}},i}})^2} \right]$. In so doing, and after applying Azuma's inequality in Eq.~(\ref{sv}), one can relate $N_{{\rm{click}},{\rm{00}},{\rm{ss}}}^{L}$ to all $N_{{\rm{click}},{\rm{00}},{{kl}}}^{L}$ for $k,l\in \{\rm v,w,0\}$ in a similar way like Eq.~(\ref{XBYDel}). Finally, one can derive $N_{{\rm{click}},{\rm{00}},{\rm{ss}}}^{L}$ from the estimations of all the quantities $N_{{\rm{click}},{\rm{00}},{{kl}}}^{L}$. However, our simulations suggest that the improvement obtained when estimating $N_{{\rm{click}},{\rm{00}},{\rm{ss}}}^{L}$ with this general method is negligible compared to the simpler method given by Eq.~(\ref{XBYDel}). Thus, in the following we use Eq.~(\ref{XBYDel}) to obtain $N_{{\rm{click}},{\rm{00}},{\rm{ss}}}^{L}$ from $N_{{\rm{click}},{\rm{00}},{\rm{vv}}}^{L}$.

Similarly, if we focus on the total number of events, $N$, and take $n=m=1$, $j_{\rm A}=j_{\rm B}={\rm s}$ and $k_{\rm A}=k_{\rm B}={\rm v}$ in Eqs.~(\ref{MDITHA})-(\ref{MDIYD}), we can estimate the quantity $N^{\rm L}_{{\rm{click}},11,{{\rm s}}{{\rm s}}}$ by following the same procedure explained above. We omit the explicit calculations here for simplicity.

\subsection{Estimation of the parameter ${e_{{\rm{ph}}}^{{\rm{U}}}}$}\label{pmxb}

To apply the idea of the quantum coin, we need to determine the form of the joint state prepared by Alice and Bob together with Eve's systems in a virtual scenario. To be precise, we shall consider a virtual single-photon scenario, where we assume that Alice and Bob meet together and follow the procedure introduced in~\cite{tamaki2012phase}. That is, they prepare a joint state in each round of the protocol, which has the form~\cite{tamaki2012phase}:
\begin{equation}\label{MDIstatexb}
\begin{array}{*{20}{l}}
{\left| \Psi  \right\rangle  \equiv }&{{p_{\rm{Z}}}{{\left| {{0_{\rm{Z}}}} \right\rangle }_{{{\rm{A}}_{{\rm{ba}}}}}}{{\left| {{0_{\rm{Z}}}} \right\rangle }_{{{\rm{A}}_{\rm{c}}}}}{{\left| {{\Psi _{\rm{Z}}}} \right\rangle }_{{{\rm{A}}_{\rm{q}}},{{\rm{A}}_{\rm{p}}},{{\rm{A}}_{\rm{a}}},{{\rm{A}}_{\rm int}},{{\rm{E}}_{{\rm{IM}}}},{{\rm{E}}_{{\rm{PM}}}}}}{{\left| {{\Psi _{\rm{Z}}}} \right\rangle }_{{{\rm{B}}_{\rm{q}}},{{\rm{B}}_{\rm{p}}},{{\rm{B}}_{\rm{a}}},{{\rm{B}}_{\rm int}},{{\rm{E}}_{{\rm{IM}}}},{{\rm{E}}_{{\rm{PM}}}}}}}\\
{}&{ + {p_{\rm{X}}}{{\left| {{0_{\rm{Z}}}} \right\rangle }_{{{\rm{A}}_{{\rm{ba}}}}}}{{\left| {{1_{\rm{Z}}}} \right\rangle }_{{{\rm{A}}_{\rm{c}}}}}{{\left| {{\Psi _{\rm{X}}}} \right\rangle }_{{{\rm{A}}_{\rm{q}}},{{\rm{A}}_{\rm{p}}},{{\rm{A}}_{\rm{a}}},{{\rm{A}}_{\rm int}},{{\rm{E}}_{{\rm{IM}}}},{{\rm{E}}_{{\rm{PM}}}}}}{{\left| {{\Psi _{\rm{X}}}} \right\rangle }_{{{\rm{B}}_{\rm{q}}},{{\rm{B}}_{\rm{p}}},{{\rm{B}}_{\rm{a}}},{{\rm{B}}_{\rm int}},{{\rm{E}}_{{\rm{IM}}}},{{\rm{E}}_{{\rm{PM}}}}}}}\\
{}&{ + \sqrt {{p_{\rm{Z}}}{p_{\rm{X}}}} ({{\left| {{1_{\rm{Z}}}} \right\rangle }_{{{\rm{A}}_{{\rm{ba}}}}}}{{\left| {{0_{\rm{Z}}}} \right\rangle }_{{{\rm{A}}_{\rm{c}}}}}{{\left| {{\Psi _{\rm{Z}}}} \right\rangle }_{{{\rm{A}}_{\rm{q}}},{{\rm{A}}_{\rm{p}}},{{\rm{A}}_{\rm{a}}},{{\rm{A}}_{\rm int}},{{\rm{E}}_{{\rm{IM}}}},{{\rm{E}}_{{\rm{PM}}}}}}{{\left| {{\Psi _{\rm{X}}}} \right\rangle }_{{{\rm{B}}_{\rm{q}}},{{\rm{B}}_{\rm{p}}},{{\rm{B}}_{\rm{a}}},{{\rm{B}}_{\rm int}},{{\rm{E}}_{{\rm{IM}}}},{{\rm{E}}_{{\rm{PM}}}}}}}\\
{}&{ + {{\left| {{1_{\rm{Z}}}} \right\rangle }_{{{\rm{A}}_{{\rm{ba}}}}}}{{\left| {{1_{\rm{Z}}}} \right\rangle }_{{{\rm{A}}_{\rm{c}}}}}{{\left| {{\Psi _{\rm{X}}}} \right\rangle }_{{{\rm{A}}_{\rm{q}}},{{\rm{A}}_{\rm{p}}},{{\rm{A}}_{\rm{a}}},{{\rm{A}}_{\rm int}},{{\rm{E}}_{{\rm{IM}}}},{{\rm{E}}_{{\rm{PM}}}}}}{{\left| {{\Psi _{\rm{Z}}}} \right\rangle }_{{{\rm{B}}_{\rm{q}}},{{\rm{B}}_{\rm{p}}},{{\rm{B}}_{\rm{a}}},{{\rm{B}}_{\rm int}},{{\rm{E}}_{{\rm{IM}}}},{{\rm{E}}_{{\rm{PM}}}}}}),}
\end{array}
\end{equation}where the system $\rm A_{ba}$ denotes a system in Alice's hands which decides whether or not Alice's and Bob's basis choices match by measuring it in the Z basis. The system $\rm A_c$ denotes Alice's quantum coin and this system determines her basis choice by measuring it in the Z basis. The system ${\rm A}_{\rm q}$ (${\rm B}_{\rm q}$) denotes a virtual qubit, which contains Alice's (Bob's) bit value choice, the system ${\rm A}_{\rm p}$ (${\rm B}_{\rm p}$) represents Alice's (Bob's) single-photon system that she (he) sends to the relay via the quantum channel, the system ${\rm A}_{\rm a}$~(${\rm B}_{\rm a}$) denotes an ancilla system stored in Alice's~(Bob's) lab to account for the loss in the transmitter, and the system ${\rm{A}}_{\rm int}$ (${\rm B}_{\rm int}$) denotes a virtual system which decides Alice's (Bob's) intensity setting choice. ${{\rm{E}_{\rm IM}}}$ and ${{\rm{E}_{\rm PM}}}$, on the other hand, denote the systems corresponding to the back-reflected light to Eve from the IM and the PM, respectively, together with the systems ${\rm E_a^A}$ and ${\rm E_a^B}$ in Eve's hands.

In this virtual protocol, the relay can perform any operation on each received signal pair from Alice and Bob to decide in which rounds there will be `click' events ({\it i.e.}, successful measurement results with the notation used in this paper). For each click event, the relay then performs some measurement on the received signal pair and both Alice and Bob measure their systems $\rm {A_q}$ and $\rm B_q$ in the X basis. Besides, Alice selects the $\rm Z_{A_c}$ or $\rm X_{A_c}$ basis with probabilities $p_{\rm Z_{A_c}}$ and $p_{\rm X_{A_c}}$, respectively, to measure her quantum coin in the selected basis.

In general, we have that the states prepared by Alice for the Z and X bases are given by
\begin{equation}\label{ZXb}
\begin{array}{*{20}{l}}
{\left| {{\Psi _{\rm{Z}}}} \right\rangle _{{{\rm{A}}_{\rm{q}}},{{\rm{A}}_{\rm{p}}},{{\rm{A}}_{{\rm{int}}}},{{\rm{E}}_{{\rm{IM}}}},{{\rm{E}}_{{\rm{PM}}}}}}&  = \frac{1}{{\sqrt 2 }}\left( {{{\left| {{0_{\rm{Z}}}} \right\rangle }_{{{\rm{A}}_{\rm{q}}}}}{{\left| {{0_{\rm{Z}}}} \right\rangle }_{{{\rm{A}}_{\rm{p}}}}}{{\left| {\sqrt {\frac{{{I_{\max }}}}{{{\rm{2}}}}} {e^{i{\theta _{0,{\rm{Z}}}}}}} \right\rangle }_{{\rm{PM}}}} + {{\left| {{1_{\rm{Z}}}} \right\rangle }_{{{\rm{A}}_{\rm{q}}}}}{{\left| {{1_{\rm{Z}}}} \right\rangle }_{{{\rm{A}}_{\rm{p}}}}}{{\left| {\sqrt {\frac{{{I_{\max }}}}{{{\rm{2}}}}} {e^{i{\theta _{1,{\rm{Z}}}}}}} \right\rangle }_{{\rm{PM}}}}} \right)\\
& {\kern 11pt} \otimes {\left| {{\Phi _{{\gamma ^{\rm{s}}}}}} \right\rangle _{{{\rm{A}}_{{\rm{int}}}}}}{\left| {\frac{{{\beta _{\rm{s}}}}}{{{\sqrt 2} }}{e^{i{\theta _{\rm{s}}}}}} \right\rangle _{{\rm{IM}}}}\\
{\left| {{\Psi _{\rm{X}}}} \right\rangle_{{{\rm{A}}_{\rm{q}}},{{\rm{A}}_{\rm{p}}},{{\rm{A}}_{\rm{int}}},{{\rm{E}_{\rm IM}}},{{\rm{E}_{\rm PM}}}} }& = \frac{1}{{\sqrt {2\left( {{p_{\rm{v}}} + {p_{\rm{w}}}} \right)} }}\big[{\left| {{0_{\rm{X}}}} \right\rangle_{\rm A_{\rm{q}}}{{{\left| {{0_{\rm{X}}}} \right\rangle }}}_{\rm A_{\rm{p}}} }({e^{i{\theta _1}}}\sqrt {{p_{\rm{v}}}} {\left| \Phi_{{\gamma ^{\rm{v}}}} \right\rangle_{{\rm{A}}_{\rm int}} }{\left| {\frac{\beta _{\rm{v}}}{{\sqrt 2}}{e^{i{\theta _{\rm{v}}}}}} \right\rangle _{{\rm{IM}}}} \\
&{\kern 60pt} +{e^{i{\theta _2}}}\sqrt {{p_{\rm{w}}}} {\left| \Phi_{{\gamma ^{\rm{w}}}} \right\rangle_{{\rm{A}}_{\rm int}} }{\left| {{\frac{\beta _{\rm{w}}}{{\sqrt 2}}}{e^{i{\theta _{\rm{w}}}}}} \right\rangle _{{\rm{IM}}}}){\left| {\sqrt {\frac{I_{\max }}{2}} {e^{i{\theta _{0,\rm{X}}}}}} \right\rangle _{{\rm{PM}}}} \\
&{\kern 60pt}+ {\left| {{1_{\rm{X}}}} \right\rangle_{\rm A_{\rm{q}}}{{{\left| {{1_{\rm{X}}}} \right\rangle }}}_{\rm A_{\rm{p}}} }( {e^{i{\theta _3}}}\sqrt {{p_{\rm{v}}}} {\left| \Phi_{{\gamma ^{\rm{v}}}} \right\rangle_{{\rm{A}}_{\rm int}} }{\left| {{\frac{\beta _{\rm{v}}}{\sqrt{2}}}{e^{i{\theta _{\rm{v}}}}}} \right\rangle _{{\rm{IM}}}}\\
&{\kern 60pt}+{e^{i{\theta _4}}}\sqrt {{p_{\rm{w}}}} {\left| \Phi_{{\gamma ^{\rm{w}}}} \right\rangle_{{\rm{A}}_{\rm int}} }{\left| {{\frac{\beta _{\rm{w}}}{\sqrt{2}}}{e^{i{\theta _{\rm{w}}}}}} \right\rangle _{{\rm{IM}}}}){\left| {\sqrt {\frac{I_{\max }}{2}} {e^{i{\theta _{1,\rm{X}}}}}} \right\rangle _{{\rm{PM}}}}\big],
\end{array}
\end{equation}
where ${{{\left| {{0_{\rm{Z}}}} \right\rangle }}}_{\rm A_{\rm{q}}}$, ${{{\left| {{1_{\rm{Z}}}} \right\rangle }}}_{\rm A_{\rm{q}}}$ and ${{{\left| {{0_{\rm{X}}}} \right\rangle }}}_{\rm A_{\rm{q}}}$, ${{{\left| {{1_{\rm{X}}}} \right\rangle }}}_{\rm A_{\rm{q}}}$ are the states of the virtual qubits which Alice measures in the Z or X basis, respectively, to prepare a state with a particular bit value. $\left| \Phi_{{\gamma ^{k}}} \right\rangle_{{\rm{A}}_{\rm int}} $ is the virtual state that determines the intensity setting choice $k$ with $k \in \{\rm s,v,w\}$. It holds that $\left\langle {{\Phi _{{\gamma ^{\rm{v}}}}}} \right.{\left| {{\Phi _{{\gamma ^{\rm{w}}}}}} \right\rangle _{{{\rm{A}}_{\rm int}}}}=0$. This is so because in the X basis there are various intensity settings and Alice knows which one she uses each given time. Note that, these states could also include information leakage from the bit value choice as shown in Eq.~(\ref{ZXb}).
Since Alice always sends a vacuum state when she selects the intensity setting $\gamma^{0}$, there is no single-photon contribution to the state $\left| {{\Psi _{\rm{X}}}} \right\rangle_{{{\rm{A}}_{\rm{q}}},{{\rm{A}}_{\rm{p}}},{{\rm{A}}_{\rm{a}}},{{\rm{A}}_{\rm int}},{{\rm{E}_{\rm IM}}},{{\rm{E}_{\rm PM}}}}$ from the intensity setting $\gamma^{0}$. Likewise, Bob's states $\left| {{\Psi _{\rm{Z}}}} \right\rangle_{{{\rm{B}}_{\rm{q}}},{{\rm{B}}_{\rm{p}}},{{\rm{B}}_{\rm{a}}},{{\rm{B}}_{\rm int}},{{\rm{E}_{\rm IM}}},{{\rm{E}_{\rm PM}}}}$ and $\left| {{\Psi _{\rm{X}}}} \right\rangle_{{{\rm{B}}_{\rm{q}}},{{\rm{B}}_{\rm{p}}},{{\rm{B}}_{\rm{a}}},{{\rm{B}}_{\rm int}},{{\rm{E}_{\rm IM}}},{{\rm{E}_{\rm PM}}}}$ are defined in a similar way like Eq.~(\ref{ZXb}) by changing the subscript `A' with `B'. In what follows, and in order to simplify the notation, we omit the subscripts `${{{\rm{A}}_{\rm{q}}},{{\rm{A}}_{\rm{p}}},{{\rm{A}}_{\rm{a}}},{{\rm{A}}_{\rm int}},{{\rm{E}_{\rm IM}}},{{\rm{E}_{\rm PM}}}}$' of the states prepared by Alice as well as those of the states prepared by Bob; instead, we use the subscript `A,E' for Alice and similarly for Bob.

Now we are ready to apply the quantum coin idea to this virtual scenario. By following the same procedure applied to the virtual single-photon scenario explained in Sec.~\ref{PM3}, we obtain a similar expression like Eq.~(\ref{MDIPsphase}). In particular, we find that
{\begin{equation}
\begin{array}{*{20}{l}}\label{XBphase}
{{{p_{{{\rm{Z}}_{{{\rm{A}}_{\rm{c}}}}}}}\left( {1 - \frac{{{\delta _{{\rm{sb}},{\rm{11}},{\rm{ss + vv|click}}}}}}{{{N_{{\rm{sb}},{\rm{11}},{\rm{ss + vv|click}}}}}}} \right)\left( {1 - 2{\Delta _{{X_{{{\rm{A}}_{\rm{c}}}}} =  - }}} \right)}}\\
{ \le \sqrt {\frac{{\left( {{N_{{\rm{X-error}},{\rm 11,vv}\left| {\rm{X}} \right.}} + {\delta _{{\rm{X-error}},{\rm 11,vv}\left| {\rm{X}} \right.}}} \right)}}{{{N_{{\rm{click}},{\rm 11,vv}\left| {\rm{X}} \right.}}}}\frac{{\left( {{N_{{\rm{X - error}},{\rm 11,ss}\left| {\rm{Z}} \right.}} + {\delta _{{\rm{X - error}},{\rm 11,ss}\left| {\rm{Z}} \right.}}} \right)}}{{{N_{{\rm{click}},{\rm 11,ss}\left| {\rm{Z}} \right.}}}}} }\\
{ + \sqrt {\left( {1 - \frac{{{N_{{\rm{No~X-error}},{\rm 11,vv}\left| {\rm{X}} \right.}} - {\delta _{{\rm{No}}\;{\rm{X-error}},{\rm 11,vv}\left| {\rm{X}} \right.}}}}{{{N_{{\rm{click}},{\rm 11,vv}\left| {\rm{X}} \right.}}}}} \right)\left( {1 - \frac{{{N_{{\rm{No~X - error}},{\rm 11,ss}\left| {\rm{Z}} \right.}} + {\delta _{{\rm{No}}\;{\rm{X - error}},{\rm 11,ss}\left| {\rm{Z}} \right.}}}}{{{N_{{\rm{click}},{\rm 11,ss}\left| {\rm{Z}} \right.}}}}} \right)} ,}
\end{array}
\end{equation}
where ${{N_{{\rm{11}},{{\rm{ss+vv}}}|{\rm{click}}}}}={N_{{\rm{click}},{\rm{11}},{{\rm{ss}}}}}+{N_{{\rm{click}},{\rm{11}},{{\rm{vv}}}}}$. The quantity ${{N_{{\rm{X - error}},{\rm{11,ss|Z}}}}}$ is the actual number of phase errors, which can be numerically estimated by using a similar method as that explained in Sec.~\ref{pmf} and $\delta{_{{\rm{X - error}},{\rm{11,ss|Z}}}}$ is the deviation term when using Azuma's inequality to estimate it.

Note, however, that the value of $\Delta _{{X_{{{\rm{A}}_{\rm{c}}}}} =  - }$ is now different. To calculate $\Delta _{{X_{{{\rm{A}}_{\rm{c}}}}} =  - }$ we apply Eq.~(\ref{MDIN++--}) to the states defined by Eq.~(\ref{ZX}). From Alice and Bob's point of view, the bigger $\rm{Re}\left(\left\langle {{\Psi _{\rm{Z}}}} \right.\left| {{\Psi _{\rm{X}}}} \right\rangle_{\rm A,E}\left\langle {{\Psi _{\rm{Z}}}} \right.\left| {{\Psi _{\rm{X}}}} \right\rangle_{\rm B,E} \right)$ is, the better. Indeed, in the ideal scenario without information leakage we have that $\rm{Re}\left(\left\langle {{\Psi _{\rm{Z}}}} \right.\left| {{\Psi _{\rm{X}}}} \right\rangle_{\rm A,E}\left\langle {{\Psi _{\rm{Z}}}} \right.\left| {{\Psi _{\rm{X}}}} \right\rangle_{\rm B,E} \right) =1$. In general, Alice and Bob can choose the states $\left| \Phi_{{\gamma ^{\rm{s}}}} \right\rangle$, $\left| \Phi_{{\gamma ^{\rm{v}}}} \right\rangle$ and $\left| \Phi_{{\gamma ^{\rm{w}}}} \right\rangle$ such that $\left\langle {{\Phi _{{\gamma ^{\rm{v}}}}}} \right.{\left| {{\Phi _{{\gamma ^{\rm{w}}}}}} \right\rangle }=0$ and the phases $\theta_1$, $\theta_2$, $\theta_3$ and $\theta_4$ in order to maximize the quantity $\rm{Re}\left(\left\langle {{\Psi _{\rm{Z}}}} \right.\left| {{\Psi _{\rm{X}}}} \right\rangle_{\rm A,E}\left\langle {{\Psi _{\rm{Z}}}} \right.\left| {{\Psi _{\rm{X}}}} \right\rangle_{\rm B,E} \right)$. On the other hand, Eve can choose the values of $\theta_{\rm s}$, $\theta_{\rm v}$, $\theta_{\rm w}$, $\theta_{\rm Z}$ and $\theta_{\rm X}$ to minimize $\rm{Re}\left(\left\langle {{\Psi _{\rm{Z}}}} \right.\left| {{\Psi _{\rm{X}}}} \right\rangle_{\rm A,E}\left\langle {{\Psi _{\rm{Z}}}} \right.\left| {{\Psi _{\rm{X}}}} \right\rangle_{\rm B,E} \right)$. To maximize the quantity $\rm{Re}\left(\left\langle {{\Psi _{\rm{Z}}}} \right.\left| {{\Psi _{\rm{X}}}} \right\rangle_{\rm A,E}\left\langle {{\Psi _{\rm{Z}}}} \right.\left| {{\Psi _{\rm{X}}}} \right\rangle_{\rm B,E} \right)$, without loss of generality, we can choose $\theta_1=\theta_2=\theta_3=\theta_4=0$ and obtain that
\begin{equation}
\begin{array}{*{20}{l}}
&\rm{Re}\left(\left\langle {{\Psi _{\rm{Z}}}} \right.\left| {{\Psi _{\rm{X}}}} \right\rangle_{\rm A,E}\left\langle {{\Psi _{\rm{Z}}}} \right.\left| {{\Psi _{\rm{X}}}} \right\rangle_{\rm B,E} \right)\\
&= \exp \left( {{\frac{I_{\max }}{2}}} \right)\cos \left( { - {\frac{I_{\max }}{2}}} \right)\frac{{\sqrt {{p_{\rm{v}}}} {\rm{Re}}\left(\left\langle {{\Phi _{{\gamma ^{\rm{s}}}}}} \right.\left| {{\Phi _{{\gamma ^{\rm{v}}}}}} \right\rangle \left\langle \frac{{{\beta _{\rm{s}}}{e^{i{\theta _{\rm{s}}}}}}}{\sqrt{2}} \right.{{\left| \frac{{{\beta _{\rm{v}}}{e^{i{\theta _{\rm{v}}}}}}}{\sqrt{2}} \right\rangle }_{{\rm{IM}}}}\right) + \sqrt {p_{\rm{w}}}\rm{Re}\left( \left\langle {{\Phi _{{\gamma ^{\rm{s}}}}}} \right.\left| {{\Phi _{{\gamma ^{\rm{w}}}}}} \right\rangle \left\langle \frac{{{\beta _{\rm{s}}}{e^{i{\theta _{\rm{s}}}}}}}{\sqrt{2}} \right.{{\left| \frac{{{\beta _{\rm{w}}}{e^{i{\theta _{\rm{w}}}}}}}{\sqrt{2}} \right\rangle }_{{\rm{IM}}}}\right)}}{{\sqrt {{p_{\rm{v}}} + {p_{\rm{w}}}} }}\\
 &= \exp\left( {{\frac{I_{\max }}{2}}} \right)\cos \left( { -{\frac{I_{\max }}{2}}} \right)\frac{{{p_{\rm{v}}}{\rm{Re}}\left(\left\langle {{\beta _{\rm{s}}}{e^{i{\theta _{\rm{s}}}}}} \right.{{\left| {{\beta _{\rm{v}}}{e^{i{\theta _{\rm{v}}}}}} \right\rangle }_{{\rm{IM}}}}\right) + {p_{\rm{w}}}\rm{Re}\left(\left\langle {{\beta _{\rm{s}}}{e^{i{\theta _{\rm{s}}}}}} \right.{{\left| {{\beta _{\rm{w}}}{e^{i{\theta _{\rm{w}}}}}} \right\rangle }_{{\rm{IM}}}}\right)}}{2({{p_{\rm{v}}} + {p_{\rm{w}}}})},
\end{array}
\end{equation}where we have chosen ${\left\langle {{\Phi _{{\gamma ^{\rm{s}}}}}} \right.\left| {{\Phi _{{\gamma ^{\rm{v}}}}}} \right\rangle }=\frac{{\sqrt {{p_{\rm{v}}}} }}{{\sqrt {{p_{\rm{v}}} + {p_{\rm{w}}}} }}$ and ${\left\langle {{\Phi _{{\gamma ^{\rm{s}}}}}} \right.\left| {{\Phi _{{\gamma ^{\rm{w}}}}}} \right\rangle }=\frac{{\sqrt {{p_{\rm{w}}}} }}{{\sqrt {{p_{\rm{v}}} + {p_{\rm{w}}}} }}$.

\subsection{Simulation}
The lower bound on the length of the secret key is given by Eq.~(\ref{MDIkey}). Let $\Gamma_{\rm AB}$ and $\Gamma _{\rm E}$ denote the spaces of the parameters controlled by Alice and Bob, and by Eve, respectively. In the simulation, we assume, without loss of generality, that $\theta_{\rm 0}=0$. Also, we assume that Eve's back-reflected light from the PM only depends on the basis information like we did in Sec.~\ref{sim}. In this case, we use the following equation to replace Eq.~(\ref{ZXb}) to calculate the quantity $\Delta_{{{X_{{{\rm{A}}_{\rm{c}}}}} =  -}}$ in the simulation:
\begin{equation}\label{ZX}
\begin{array}{*{20}{l}}
{\left| {{\Psi _{\rm{Z}}}} \right\rangle_{{{\rm{A}}_{\rm{q}}},{{\rm{A}}_{\rm{p}}},{{\rm{A}}_{\rm int}},{{\rm{E}_{\rm IM}}},{{\rm{E}_{\rm PM}}}} }& = \frac{1}{{\sqrt 2 }}\left( {{{\left| {{0_{\rm{Z}}}} \right\rangle }}_{\rm A_{\rm{q}}}{{{\left| {{0_{\rm{Z}}}} \right\rangle }}}_{\rm A_{\rm{p}}} + {{\left| {{1_{\rm{Z}}}} \right\rangle }}}_{\rm A_{\rm{q}}}{{{\left| {{1_{\rm{Z}}}} \right\rangle }}}_{\rm A_{\rm{p}}} \right)\\
&{\kern 11pt}\otimes{\left| \Phi_{{\gamma ^{\rm{s}}}} \right\rangle_{{\rm{A}}_{\rm int}} }{\left| {{\frac{{\beta _{\rm{s}}}{e^{i{\theta _{\rm{s}}}}}}{\sqrt{2}}}} \right\rangle _{{\rm{IM}}}}{\left| {\sqrt {{\frac{{I_{\max }}}{2}}} {e^{i{\theta _{\rm{Z}}}}}} \right\rangle _{{\rm{PM}}}},\\
{\left| {{\Psi _{\rm{X}}}} \right\rangle_{{{\rm{A}}_{\rm{q}}},{{\rm{A}}_{\rm{p}}},{{\rm{A}}_{\rm int}},{{\rm{E}_{\rm IM}}},{{\rm{E}_{\rm PM}}}} }& = \frac{1}{{\sqrt {2\left( {{p_{\rm{v}}} + {p_{\rm{w}}}} \right)} }}\times\big[{\left| {{0_{\rm{X}}}} \right\rangle_{\rm A_{\rm{q}}}{{{\left| {{0_{\rm{X}}}} \right\rangle }}}_{\rm A_{\rm{p}}} }({e^{i{\theta _1}}}\sqrt {{p_{\rm{v}}}} {\left| \Phi_{{\gamma ^{\rm{v}}}} \right\rangle_{{\rm{A}}_{\rm int}} }{\left| {\frac{{{\beta _{\rm{v}}}{e^{i{\theta _{\rm{v}}}}}}}{\sqrt{2}}} \right\rangle _{{\rm{IM}}}} \\
&{\kern 74pt} +{e^{i{\theta _2}}}\sqrt {{p_{\rm{w}}}} {\left| \Phi_{{\gamma ^{\rm{w}}}} \right\rangle_{{\rm{A}}_{\rm int}} }{\left| {\frac{{{\beta _{\rm{w}}}{e^{i{\theta _{\rm{w}}}}}}}{\sqrt{2}} }\right\rangle _{{\rm{IM}}}}) \\
&{\kern 74pt}+ {\left| {{1_{\rm{X}}}} \right\rangle_{\rm A_{\rm{q}}}{{{\left| {{1_{\rm{X}}}} \right\rangle }}}_{\rm A_{\rm{p}}} }( {e^{i{\theta _3}}}\sqrt {{p_{\rm{v}}}} {\left| \Phi_{{\gamma ^{\rm{v}}}} \right\rangle_{{\rm{A}}_{\rm int}} }{\left| {\frac{{{\beta _{\rm{v}}}{e^{i{\theta _{\rm{v}}}}}}}{\sqrt{2}}} \right\rangle _{{\rm{IM}}}}\\
&{\kern 74pt}+{e^{i{\theta _4}}}\sqrt {{p_{\rm{w}}}} {\left| \Phi_{{\gamma ^{\rm{w}}}} \right\rangle_{{\rm{A}}_{\rm int}} }{\left| {\frac{{{\beta _{\rm{w}}}{e^{i{\theta _{\rm{w}}}}}}}{\sqrt{2}}} \right\rangle _{{\rm{IM}}}})\big]{\left| {\sqrt {{\frac{{I_{\max }}}{2}}} {e^{i{\theta _{\rm{X}}}}}} \right\rangle _{{\rm{PM}}}}.
\end{array}
\end{equation}The states described by Eq.~(\ref{ZX}) correspond to the case illustrated in Fig.~\ref{THAXiangbin}.

Note that since the information leakage from the IM and the PM is correlated, in the following figures, we plot the secret key rates in the presence of information leakage from  both the IM and the PM. Moreover, in this section, we compare the simulation results of this protocol with those presented in Sec.~\ref{sim} when there is information leakage from  both the IM and the PM.

\subsubsection{Case 1}
\begin{figure}[!t]
\includegraphics*[width=8.1cm,height=5.5cm]{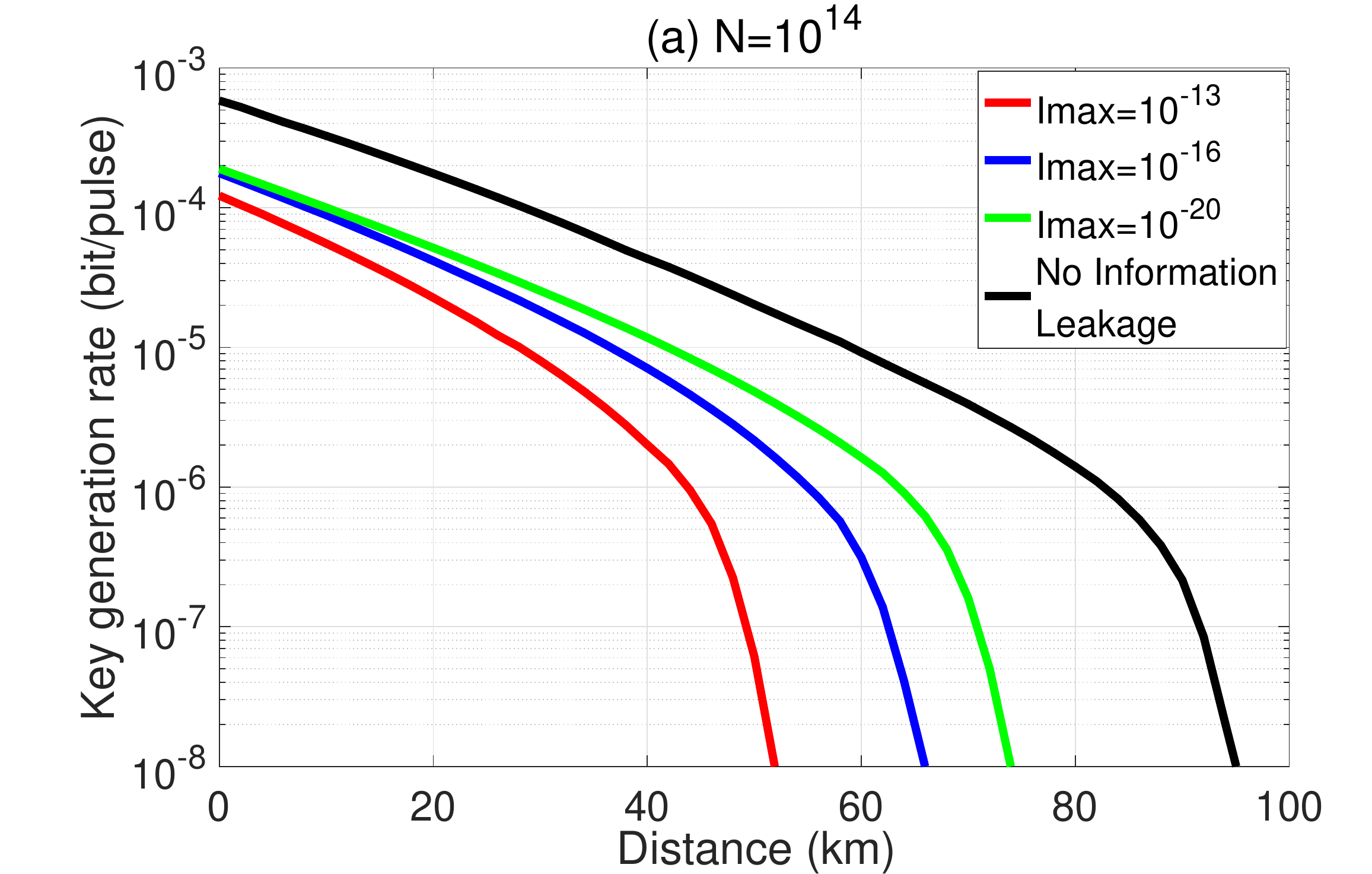}
\includegraphics*[width=8.1cm,height=5.5cm]{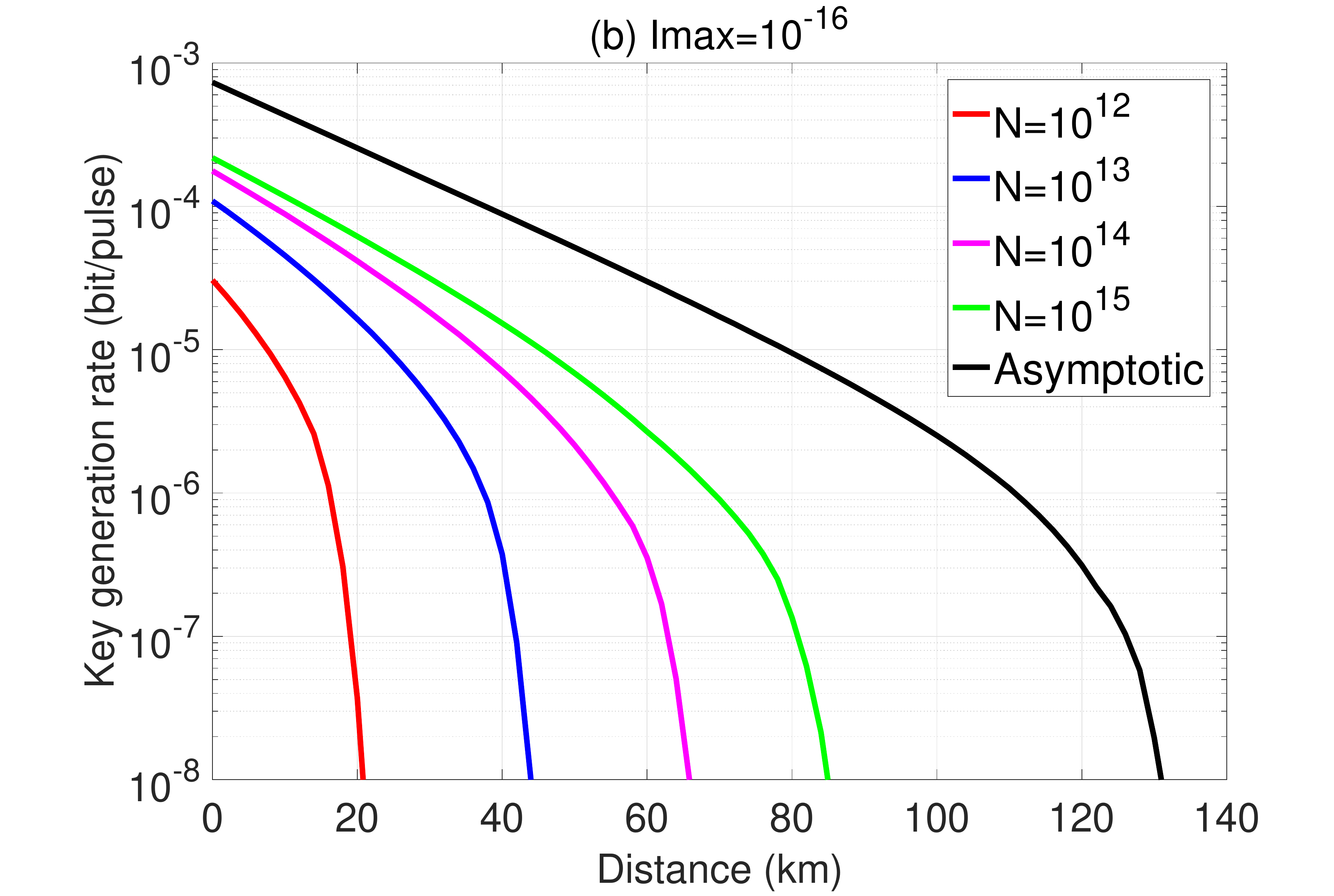}
\caption{ \footnotesize { Case 1. (a) The secret key rate in logarithmic scale as a function of the distance for a fixed value of the total number of transmitted pulses, $N=10^{14}$. The black solid line
represents the perfectly isolated situation where there is no information leakage ({\it i.e.}, $I_{\rm max}=0$) and the different colored lines correspond to different amounts of information leakage. (b) The secret key rate in logarithmic scale as a function of the distance for a fixed value of information leakage, $I_{\rm max}=10^{-16}$. Different colored lines correspond to different values of the number of transmitted pulses. In our simulations, for each value of the distance we maximize the secret key rate over the amplitudes $\gamma^{\rm s}$, $\gamma^{\rm v}$ and $\gamma^{\rm w}$, and the probabilities $p_{\rm Z_{A_c}}$, $p_{\rm s}$, $p_{\rm v}$ and $p_{\rm w}$ which are controlled by Alice and Bob, and we minimize it over the angles $\theta_r$, $\theta_{\rm Z}$ and $\theta_{\rm X}$ controlled by Eve, respectively.}} \label{Xiangfigcase1}
  \end{figure}
The simulation result of the secret key rate, $\ell'/N$, as a function of the transmission distance between Alice and Bob in this case is shown in Fig.~\ref{Xiangfigcase1}~(a) for a fixed value of the total number of transmitted pulses, $N=10^{14}$. The black solid line represents the key rate in the situation where there is no information leakage, and the different colored lines correspond to different amounts of information leakage. The longest achievable distance without information leakage is about 96 km. When $I_{\rm max}=10^{-13}$, the secret key rate vanishes at about 52 km. In the simulation, we find that in this case the optimized value of $p_{\rm Z_{A_c}}$ typically lies in the interval $\left[0.75,0.94 \right]$. That is, in this protocol Alice and Bob can sacrifice a smaller proportion of the data than that in the symmetric three-intensity decoy-state MDI-QKD protocol (where, as we have shown in the previous section, the typical interval of the optimized value of $p_{\rm Z_{A_c}}$ is $\left[0.65,0.9 \right]$).

Fig.~\ref{Xiangfigcase1} (b) shows the secret key rates as a function of the distance for a fixed value $I_{\rm max}=10^{-16}$ for different total numbers of transmitted pulses. That is, this figure illustrates the effect of the information leakage as a function of the number of transmitted pulses. For example, when $I_{\rm max}=10^{-16}$, the longest achievable distance is about 84 km when the total number of transmitted pulses is $N=10^{15}$. However, when $N=10^{12}$, this distance decreases to 21 km.

\begin{figure}[!t]
\includegraphics*[scale=0.3]{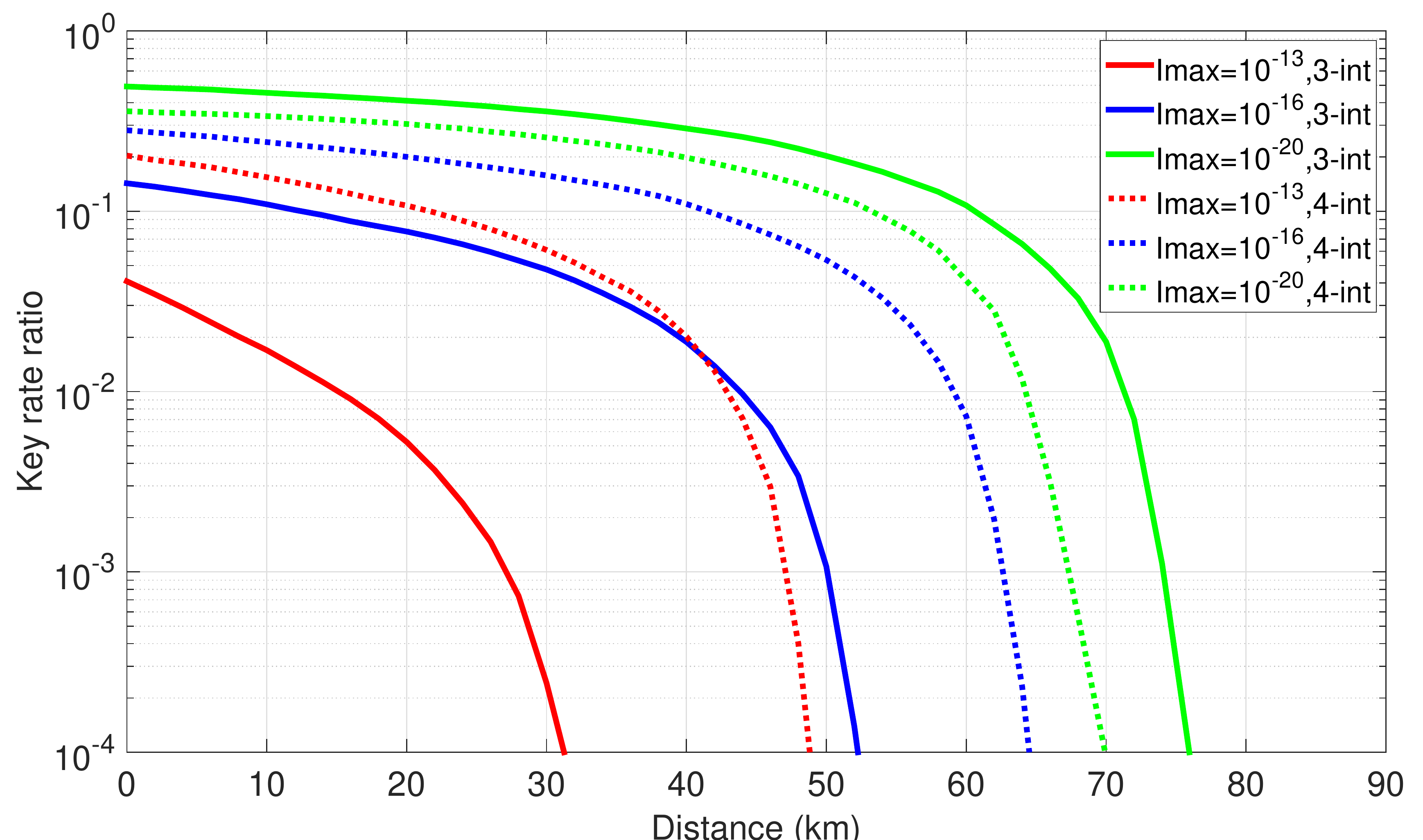}
\caption{ \footnotesize {The ratio ($\ell'_{I_{\rm max}>0}/\ell'_{I_{\rm max}=0}$) between the secret key rates in logarithmic scale with and without information leakage as a function of the distance for a fixed number, $N=10^{14}$, of transmitted pulses in the two protocols (3-int and 4-int represent the three-intensity decoy state MDI-QKD protocol and the four-intensity decoy-state MDI protocol that we consider, respectively). The solid (dotted) lines represent the 3-int protocol (the 4-int protocol). Different colored lines correspond to different values of $I_{\rm max}$. }}
\label{Ratiocase1}
  \end{figure}
To further compare the effect of the information leakage on the secret key rate in the two MDI-QKD protocols that we consider, we plot the the ratio ($\ell'_{I_{\rm max}>0}/\ell'_{I_{\rm max}=0}$) between the secret key rates for different positive values of information leakage, $I_{\rm max}$, and the secret key rate when there is no information leakage, {\it i.e.}, $I_{\rm max}=0$, given a fixed total number of transmitted pulses, say, $N=10^{14}$ in Fig.~\ref{Ratiocase1}. The solid and dotted lines represent the ratios in the symmetric three-intensity decoy-state MDI-QKD protocol and in the four-intensity decoy-state MDI-QKD protocol, respectively. In the following, for simplicity, let us denote these two protocols by `3-int' and `4-int', respectively. The result in Fig.~\ref{Ratiocase1} indicates that when the amount of information leakage is small enough, for instance, $I_{\rm max}=10^{-20}$, the impact of the information leakage on the 3-int protocol is smaller than that on the 4-int protocol as the green solid line is always above the green dotted line. However, the key rate ratio drops much faster as the amount of information leakage increases in the 3-int protocol than that in the 4-int protocol. From Fig.~\ref{Ratiocase1}, we find that when $I_{\rm max}=10^{-16}$  and $I_{\rm max}=10^{-13}$, the ratio in the 4-int protocol is bigger than that in the 3-int protocol. That is, when $I_{\rm max}$ increases, the effect of information leakage becomes more relevant  on the 3-int protocol than that on the 4-int protocol given a fixed total number of transmitted pulses.

The intuition for this behaviour could be the following: From Figs.~\ref{THAmodel} and~\ref{THAXiangbin}, we can see that the back-reflected light from the PM is the same for the 3-int and 4-int protocol. Now suppose that in the 4-int protocol Eve measures the back-reflected light from the IM and the PM independently instead of splitting the back-reflected light with a 50:50 BS. Then she learns the same information from the PM in both protocols. However, it may be more difficult for Eve to learn the information from the IM in the 4-int protocol than in the 3-int protocol because she needs to distinguish between four states in the former but she only needs to distinguish between three states in the latter. In this case, the 4-int protocol is more robust against information leakage than the 3-int protocol for all values of $I_{\rm max}$. Nevertheless, if Eve exploits the correlations between the back-reflected light from the IM and the PM, then which protocol is more robust seems to depend on the value of $I_{\rm max}$. In addition, note that the results illustrated in Fig.~\ref{Ratiocase1} consider the case where Eve splits the back-reflected light with a 50:50 BS, which might not be the optimal option for the example of THAs evaluated.

\subsubsection{Case 2}
\begin{figure}[!t]
\includegraphics*[width=8.1cm,height=5.5cm]{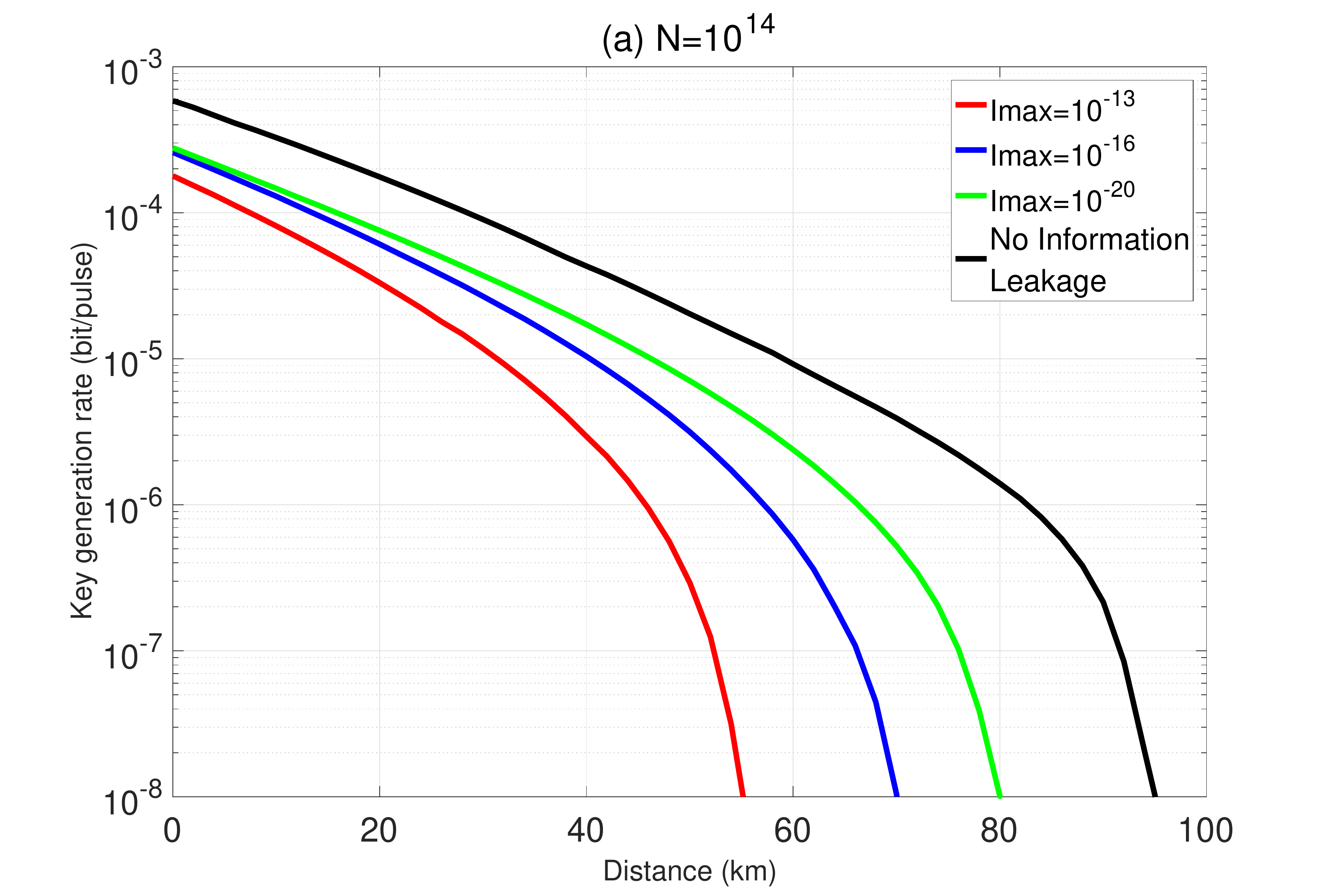}
\includegraphics*[width=8.1cm,height=5.5cm]{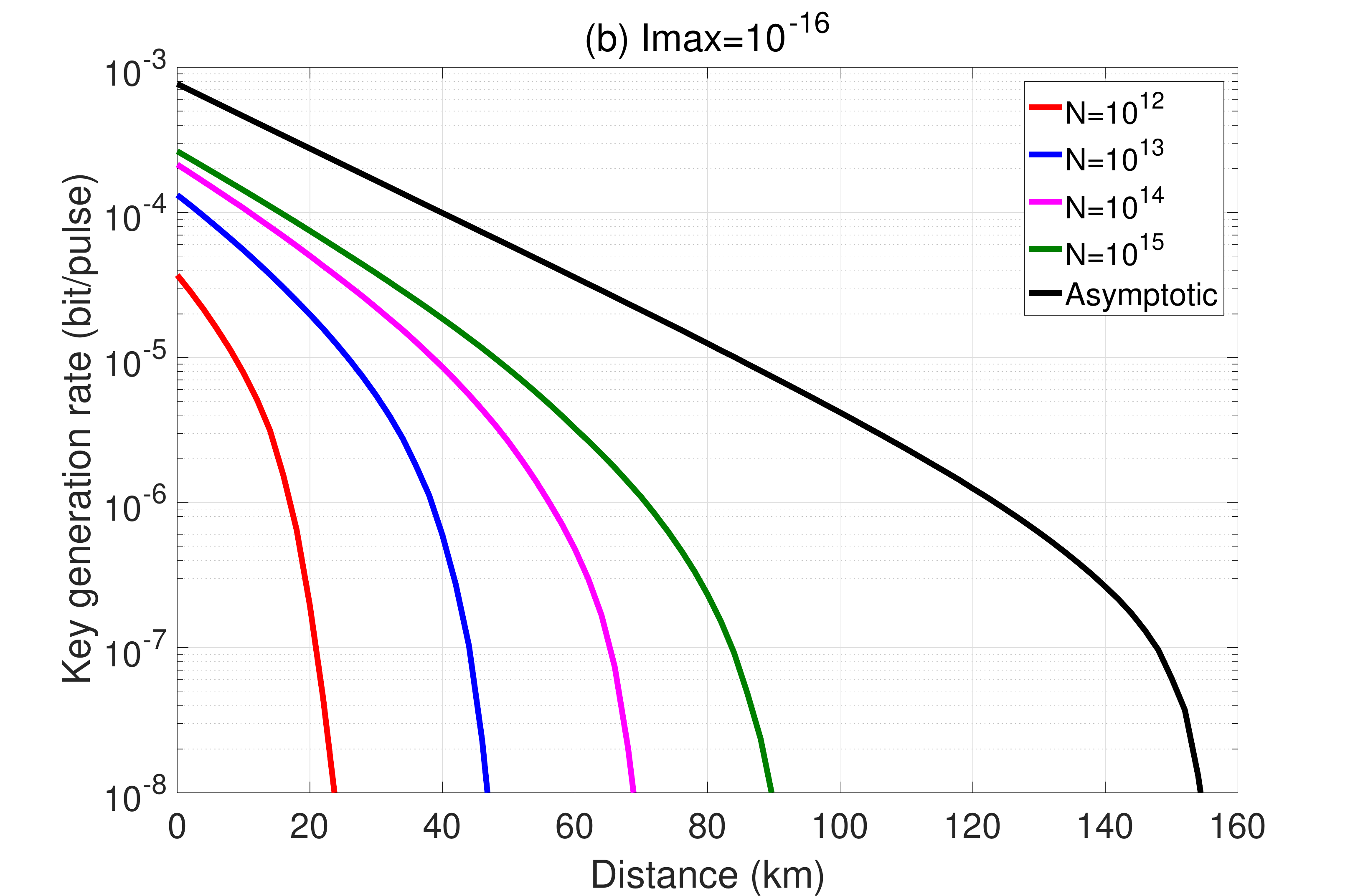}
\caption{ \footnotesize { Case 2. (a) The secret key rate in logarithmic scale as a function of the distance for a fixed value of the total number of transmitted pulses, $N=10^{14}$. The black solid line
represents the perfectly isolated situation where there is no information leakage ({\it i.e.}, $I_{\rm max}=0$) and the different colored lines correspond to different amounts of information leakage. (b) The secret key rate in logarithmic scale as a function of the distance for a fixed value of information leakage, $I_{\rm max}=10^{-16}$. Different colored lines correspond to different values of the number of transmitted pulses. In our simulations, for each value of the distance we maximize the secret key rate over the amplitudes $\gamma^{\rm s}$, $\gamma^{\rm v}$ and $\gamma^{\rm w}$, and the probabilities $p_{\rm Z_{A_c}}$, $p_{\rm s}$, $p_{\rm v}$ and $p_{\rm w}$ which are controlled by Alice and Bob, and we minimize it over the angles $\theta_r$, $\theta_{\rm Z}$ and $\theta_{\rm X}$ controlled by Eve, respectively.}} \label{Xiangfigcase2}
  \end{figure}

The simulation result of the secret key rate as a function of the transmission distance between Alice and Bob when $N=10^{14}$ and for different values of $I_{\rm max}$ is shown in Fig.~\ref{Xiangfigcase2}~(a). Fig.~\ref{Xiangfigcase2}~(b) shows the secret key rate as a function of the distance for a fixed value of $I_{\rm max}=10^{-16}$ and different values of $N$. The behavior of the curves is very similar to those in case 1, and in the simulation we find that the optimized value of $p_{\rm Z_{A_c}}$ is also similar. One main difference is that with the same experimental parameters (see Table.~\ref{para}) the secret key rate is a little higher and the achievable distance is a little longer than those in Case 1. For example, when the total number of transmitted pulses is $N=10^{14}$ and $I_{\rm max}=10^{-13}$, now we find that the secret key is positive up to about 57 km while in Case 1 this distance is 52  km.

Here we omit the comparison of the key rate ratios between the two protocols as the result in this case is similar to that shown in Fig.~(\ref{Ratiocase1}). And for the same reason, we omit such comparison in Case 3 as well.

\subsubsection{Case 3}
\begin{figure}[!t]
\includegraphics*[width=8.1cm,height=5.5cm]{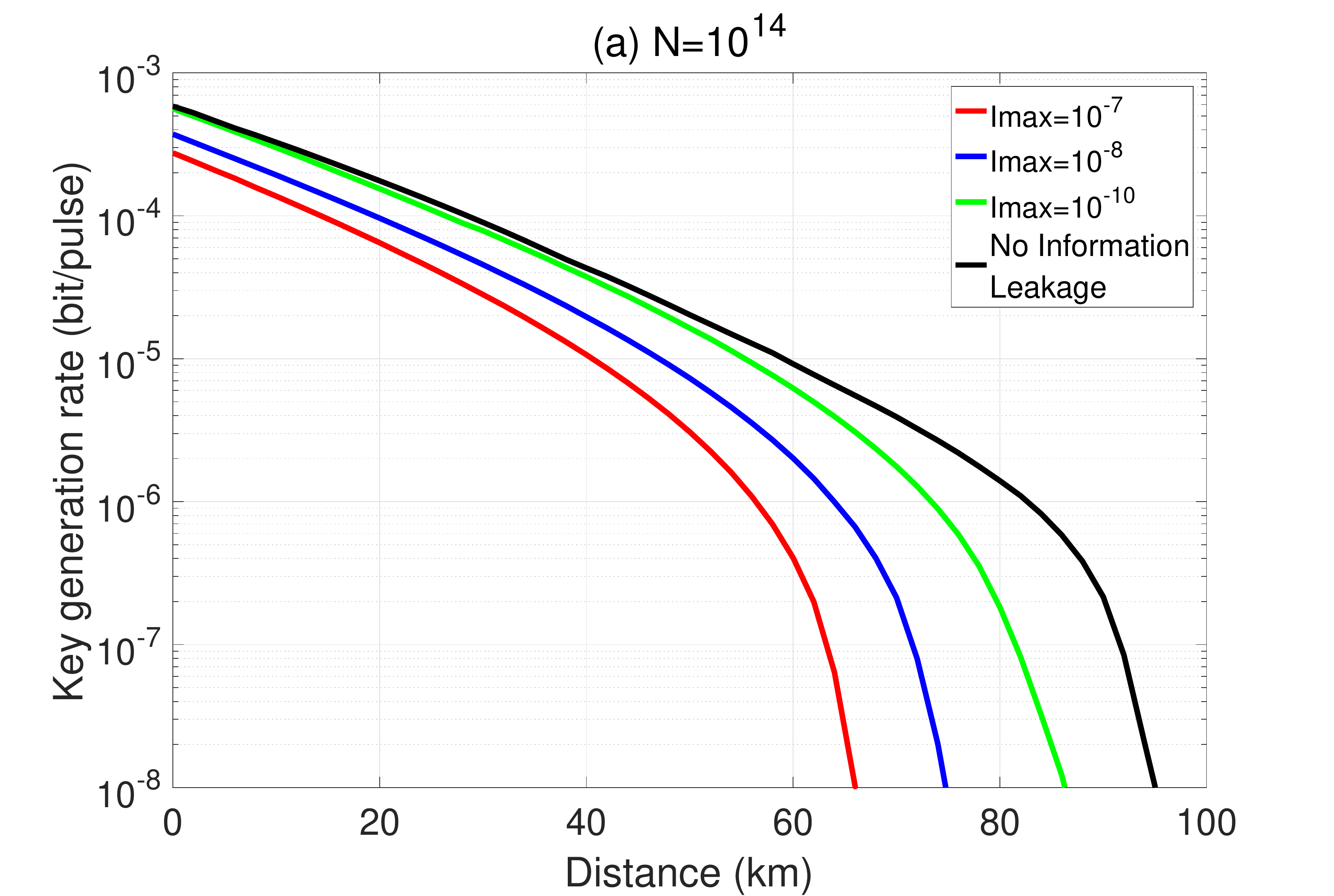}
\includegraphics*[width=8.1cm,height=5.5cm]{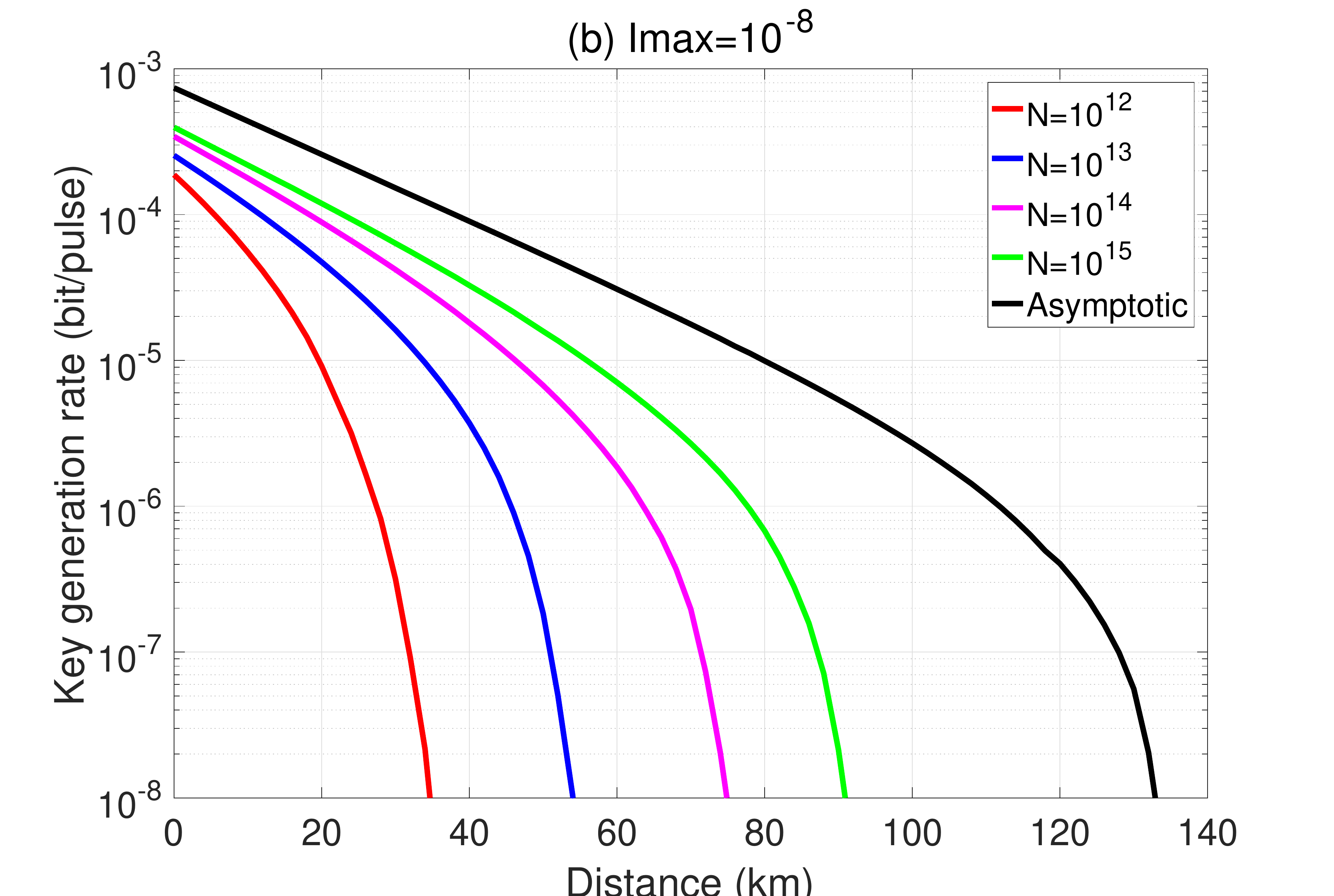}
\caption{ \footnotesize { Case 3. (a) The secret key rate in logarithmic scale as a function of the distance for a fixed value of the total number of transmitted pulses, $N=10^{14}$. The black solid line
represents the perfectly isolated situation where there is no information leakage ({\it i.e.}, $I_{\rm max}=0$) and the different colored lines correspond to different amounts of information leakage. (b) The secret key rate in logarithmic scale as a function of the distance for a fixed value of information leakage, $I_{\rm max}=10^{-8}$. Different colored lines correspond to different values of the number of transmitted pulses. In our simulations, for each value of the distance we maximize the secret key rate over the amplitudes $\gamma^{\rm s}$, $\gamma^{\rm v}$ and $\gamma^{\rm w}$, and the probabilities $p_{\rm Z_{A_c}}$, $p_{\rm s}$, $p_{\rm v}$ and $p_{\rm w}$ which are controlled by Alice and Bob.}} \label{Xiangfigcase3}
  \end{figure}
The simulation result of the secret key rate as a function of the transmission distance between Alice and Bob when $N=10^{14}$ and for different values of $I_{\rm max}$ is shown in Fig.~\ref{Xiangfigcase3}~(a). Fig.~\ref{Xiangfigcase3}~(b) shows the finite-key effect on the secret key rate as a function of the distance for a fixed value of $I_{\rm max}=10^{-8}$ and for different values of $N$. Here, we find that the typical interval that $p_{\rm Z_{A_c}}$ lies in is $\left[0.86,0.99 \right]$. Compared to the secret key rates shown in Figs~\ref{Xiangfigcase1} and \ref{Xiangfigcase2}, now it is obviously improved. For example, when the total number of transmitted pulses is $N=10^{14}$ and $I_{\rm max}=10^{-7}$, the secret key rate remains positive up to about 66 km. In comparison, the maximal achievable distance with the same number of transmitted pulses and assuming an $I_{\rm max}$ as low as $10^{-13}$ is only about 57 km (52 km) in Case 2 (Case 1). As discussed previously, this is because the phase randomization step removes the information leaked in the phase of the output states to Eve.

\section{Conclusion and Discussion}\label{CON}
In this paper, we have quantitatively analyzed the security of two decoy-state MDI-QKD protocols with leaky sources in the finite-key regime. Specially, we have simulated the secret key rate under three particular examples of THA, where Eve sends coherent pulses of light to probe the intensity modulators and phase modulators of the legitimate parties. Similar to the analysis presented in~\cite{Weilong2018Finite}, we have introduced an additional post-processing step in the actual protocol where Alice and Bob sacrifice part of their data. This step is necessary for the security proof to go through. Our simulation results suggest that MDI-QKD is more sensitive to information leakage than standard decoy-state QKD, but is possible to distill secure keys from leaky sources within a reasonable time frame of signal transmission given that Alice's and Bob's sources are sufficiently isolated. Furthermore, we have found that when the amount of information leakage is small enough, the effect of information leakage has a bigger impact on the four-intensity decoy-state MDI-QKD protocol than on the symmetric three-intensity decoy-state MDI-QKD protocol. However, when the amount of information leakage increases, the four-intensity MDI-QKD protocol becomes more robust against information leakage than the symmetric three-intensity MDI-QKD protocol.

In this context it would be interesting to consider a stronger THA, where Eve sends entangled probe states to Alice's and Bob's sources instead of sending them independent bright pulses. In this scenario, by performing a joint measurement on the outgoing states as well as on her ancilla system, Eve might be able to extract more information about Alice's and Bob's internal settings than what has been presented in this paper. This scenario, however, is beyond the scope of this work but could be evaluated with the techniques that have been introduced in this paper.

\emph{Acknowledgments}---This work was supported by the Galician Regional Government, consolidation of Research Units: AtlantTIC), the Spanish Ministry of Economy and Competitiveness (MINECO), the Fondo Europeo de Desarrollo Regional (FEDER) through grant TEC2017-88243-R and the European Union's Horizon 2020 research and innovation programme under the Marie Sklodowska-Currie grant agreement No. 675662 (Project QCALL). W. W gratefully acknowledges support from the National Natural Science Foundation of China (Grants No. 61701539, 61701539, 61901525) and the National Cryptography Development Fund (mmjj20180107, mmjj20180212). K.T. acknowledges support from JSPS KAKENHI Grant Numbers JP18H05237 18H05237 and JST-CREST JPMJCR 1671.

\section{Appendixes}
\subsection{Trace distance parameters}
\subsubsection{Three-intensity MDI-QKD protocol}
For simplicity, we assume that there is no quantum correlation between Alice's (Bob's) and Eve's systems~\cite{Weilong2018Finite}. As a result, the parameter $D^{{j_{\rm{A}}}{j_{\rm{B}}},{\rm{ss}}}_{{\rm Z},nm}$ does not depend on the basis Z nor on the photon number $n,m$. Therefore, we shall denote it by $D^{{j_{\rm{A}}}{j_{\rm{B}}},{\rm{ss}}}$. Below we show the values of $D^{{j_{\rm{A}}}{j_{\rm{B}}},{\rm{ss}}}$ for the three different cases considered in the simulation. The detailed calculations are similar to those in~\cite{tamaki2016decoy}.

\textbf{Case 1}

In this case, $D^{{j_{\rm{A}}}{j_{\rm{B}}},{\rm{ss}}}$ is given by
\begin{equation}
\begin{array}{*{20}{l}}
D^{{j_{\rm{A}}}{j_{\rm{B}}},{\rm{ss}}}&{ = \sqrt {1 - {{\left| {\left\langle {{\beta _{{{{j_{\rm{A}}}}}}}{e^{{\rm i}\theta_{{{j_{\rm{A}}}}}}}\left| {{\beta _{{{\rm{s}}}}}{e^{{\rm i}{\theta _{{{\rm{s}}}}}}}} \right.} \right\rangle  \times \left\langle {{\beta _{{{{j_{\rm{B}}}}}}}{e^{{\rm i}\theta_{{{j_{\rm{B}}}}}}}\left| {{\beta _{{{\rm{s}}}}}{e^{{\rm i}{\theta _{{{\rm{s}}}}}}}} \right.} \right\rangle } \right|}^2}} }\\
{}&{ = \sqrt {1 - \exp \left\{ {2{I_{\max }}\left[ {\cos \left( {{\theta _{{{\rm{s}}}}} - {\theta _{{{{j_{\rm{A}}}}}}}} \right) + \cos \left( {{\theta _{{{\rm{s}}}}} - {\theta _{{{{j_{\rm{B}}}}}}}} \right) - 2} \right]} \right\}} .}
\end{array}
\end{equation}

\textbf{Case 2}

In this case, we have that
\begin{equation}
\begin{array}{*{20}{l}}
D^{{j_{\rm{A}}}{j_{\rm{B}}},{\rm{ss}}}& =  \sqrt {1 - {{\left| {\left\langle {{\beta _{{{{j_{\rm{A}}}}}}}{e^{{\rm i}\theta_{{{j_{\rm{A}}}}}}}\left| {{\beta _{{{\rm{s}}}}}{e^{{\rm i}{\theta _{{{\rm{s}}}}}}}} \right.} \right\rangle  \times \left\langle {{\beta _{{{{j_{\rm{B}}}}}}}{e^{{\rm i}\theta_{{{j_{\rm{B}}}}}}}\left| {{\beta _{{{\rm{s}}}}}{e^{{\rm i}{\theta _{{{\rm{s}}}}}}}} \right.} \right\rangle } \right|}^2}}\\
&=\{\sqrt {1 - \exp \left\{ {\frac{{{I_{\max }}}}{{{\gamma ^{\rm{s}}}}}\left[ {2\sqrt {{\gamma ^{\rm{s}}}{\gamma ^{{j_{\rm{A}}}}}} \cos \left( {{\theta _{{{\rm{s}}}}} - {\theta _{{{{j_{\rm{A}}}}}}}} \right) + 2\sqrt {{\gamma ^{\rm{s}}}{\gamma ^{{j_{\rm{B}}}}}} \cos \left( {{\theta _{{{\rm{s}}}}} - {\theta _{{{{j_{\rm{B}}}}}}}} \right) - {\gamma ^{{j_{\rm{A}}}}} - {\gamma ^{{j_{\rm{B}}}}}} \right]} \right\}} .
\end{array}
\end{equation}

\textbf{Case 3}

In this last case, we further assume that the following constraints hold: $I_{\rm max}\leq \rm{log}2$ and $\gamma^{\rm w}\leq \gamma^{\rm v}\leq \gamma^{\rm s}$ for any $P_{\rm{cut}}\ge 1$. According to the definition of the states given by Eq.~(\ref{MDICase3}), it turns out that $D^{{j_{\rm{A}}}{j_{\rm{B}}},{\rm{ss}}}$ is given by
\begin{equation}
\begin{array}{*{20}{l}}
{{D^{{j_{\rm{A}}}{j_{\rm{B}}},{\rm{ss}}}}}&=  \sqrt {1 - {{\left| {\left\langle {{\beta _{{{{j_{\rm{A}}}}}}}{e^{{\rm i}\theta_{{{j_{\rm{A}}}}}}}\left| {{\beta _{{{\rm{s}}}}}{e^{{\rm i}{\theta _{{{\rm{s}}}}}}}} \right.} \right\rangle  \times \left\langle {{\beta _{{{{j_{\rm{B}}}}}}}{e^{{\rm i}\theta_{{{j_{\rm{B}}}}}}}\left| {{\beta _{{{\rm{s}}}}}{e^{{\rm i}{\theta _{{{\rm{s}}}}}}}} \right.} \right\rangle } \right|}^2}}
\\&{ = \frac{1}{2}\sum\limits_{n,m = 0}^\infty  \bigg| \left\{ {\exp \left( { - \beta _{{j_{\rm{A}}}}^2} \right)\frac{{{{\left( {{\beta _{{j_{\rm{A}}}}}} \right)}^{2n}}}}{{n!}} - \exp \left( { - \beta _{\rm{s}}^2} \right)\frac{{{{\left( {{\beta _{\rm{s}}}} \right)}^{2n}}}}{{n!}}} \right\}}\\
{}&{{{\kern 12pt}}\left\{ {\exp \left( { - \beta _{{j_{\rm{B}}}}^2} \right)\frac{{{{\left( {{\beta _{{j_{\rm{B}}}}}} \right)}^{2m}}}}{{m!}} - \exp \left( { - \beta _{\rm{s}}^2} \right)\frac{{{{\left( {{\beta _{\rm{s}}}} \right)}^{2m}}}}{{m!}}} \right\}\bigg|}\\
{}& = \frac{{\exp \left( { - {I_{{\rm{max}}}}} \right)}}{2}\sum\limits_{n,m = 0}^\infty  \frac{{{{\left( {{I_{{\rm{max}}}}} \right)}^{n + m}}}}{{n!m!}}\bigg| \left\{ {1 - \exp \left[ {{I_{{\rm{max}}}}\left( {1 - {\gamma ^{{j_{\rm{A}}}}}/{\gamma ^{\rm{s}}}} \right)} \right]\frac{{{{\left( {{\gamma ^{{j_{\rm{A}}}}}} \right)}^n}}}{{{{\left( {{\gamma ^{\rm{s}}}} \right)}^n}}}} \right\}\\
&{\kern 12pt}\times\left\{ {1 - \exp \left[ {{I_{{\rm{max}}}}\left( {1 - {\gamma ^{{j_{\rm{B}}}}}/{\gamma ^{\rm{s}}}} \right)} \right]\frac{{{{\left( {{\gamma ^{{j_{\rm{B}}}}}} \right)}^m}}}{{{{\left( {{\gamma ^{\rm{s}}}} \right)}^m}}}} \right\} \bigg| \\
&{ \le \frac{1}{2} - \frac{{\exp \left( { - {I_{{\rm{max}}}}} \right)}}{2}\;\;\;\sum\limits_{n,m = 0}^{{P_{{\rm{cut}}}}} {\frac{{{{\left( {{I_{{\rm{max}}}}} \right)}^{n + m}}}}{{n!m!}}\left[ {1 - \left| {1 - \exp \left[ {{I_{{\rm{max}}}}\left( {1 - \frac{{{\gamma ^{{j_{\rm{A}}}}}}}{{{\gamma ^{\rm{s}}}}}} \right)} \right]\frac{{{{\left( {{\gamma ^{{j_{\rm{A}}}}}} \right)}^n}}}{{{{\left( {{\gamma ^{\rm{s}}}} \right)}^n}}}} \right|} \right]} }\\
{}&{{{\kern 12pt}} \times \left[ {1 - \left| {1 - \exp \left[ {{I_{{\rm{max}}}}\left( {1 - \frac{{{\gamma ^{{j_{\rm{B}}}}}}}{{{\gamma ^{\rm{s}}}}}} \right)} \right]\frac{{{{\left( {{\gamma ^{{j_{\rm{B}}}}}} \right)}^m}}}{{{{\left( {{\gamma ^{\rm{s}}}} \right)}^m}}}} \right|} \right],}
\end{array}
\end{equation}
for any $P_{\rm cut} \geq0 $.

\subsubsection{Four-intensity MDI-QKD protocol}
In this section, we present the trace distance parameters for the four-intensity MDI-QKD protocol. We make the same assumptions as those in the previous section. The only difference comes from the fact that now the back-reflected light from the PM also contributes to the trace distance parameters. In particular, we obtain the following results in the three cases.

\textbf{Case 1}

In this case, $D^{{j_{\rm{A}}}{j_{\rm{B}}},{\rm{ss}}}$ is given by
\begin{equation}
\begin{array}{*{20}{l}}
{{D^{{j_{\rm{A}}}{j_{\rm{B}}},{\rm{ss}}}}}&{ = \sqrt {1 - {{\left| {\left\langle {\frac{{{\beta _{{j_{\rm{A}}}}}{e^{{\rm{i}}{\theta _{{j_{\rm{A}}}}}}}}}{{\sqrt 2 }}\left| {\frac{{{\beta _{\rm{s}}}{e^{{\rm{i}}{\theta _{\rm{s}}}}}}}{{\sqrt 2 }}} \right.} \right\rangle \left\langle {\frac{{{\beta _{{j_{\rm{B}}}}}{e^{{\rm{i}}{\theta _{{j_{\rm{B}}}}}}}}}{{\sqrt 2 }}\left| {\frac{{{\beta _{\rm{s}}}{e^{{\rm{i}}{\theta _{\rm{s}}}}}}}{{\sqrt 2 }}} \right.} \right\rangle {{\left( {\left\langle {\sqrt {\frac{{{I_{\max }}}}{2}} {e^{{\rm{i}}{\theta _{\rm{X}}}}}\left| {\sqrt {\frac{{{I_{\max }}}}{2}} {e^{{\rm{i}}{\theta _{\rm{Z}}}}}} \right.} \right\rangle } \right)}^2}} \right|}^2}} }\\
{}&{ = \sqrt {1 - \exp \left\{ {\frac{1}{8}{I_{\max }}\left[ {\cos \left( {{\theta _{\rm{s}}} - {\theta _{{j_{\rm{A}}}}}} \right) + \cos \left( {{\theta _{\rm{s}}} - {\theta _{{j_{\rm{B}}}}}} \right) + 2\cos \left( {{\theta _{\rm{Z}}} - {\theta _{\rm{X}}}} \right) - 4} \right]} \right\}} .}
\end{array}
\end{equation}

\textbf{Case 2}

In this case, we have that
\begin{equation}
\begin{array}{*{20}{l}}
{D^{{j_{\rm{A}}}{j_{\rm{B}}},{\rm{ss}}}} &{{ = \sqrt {1 - {{\left| {\left\langle {\frac{{{\beta _{{j_{\rm{A}}}}}{e^{{\rm{i}}{\theta _{{j_{\rm{A}}}}}}}}}{{\sqrt 2 }}\left| {\frac{{{\beta _{\rm{s}}}{e^{{\rm{i}}{\theta _{\rm{s}}}}}}}{{\sqrt 2 }}} \right.} \right\rangle \left\langle {\frac{{{\beta _{{j_{\rm{B}}}}}{e^{{\rm{i}}{\theta _{{j_{\rm{B}}}}}}}}}{{\sqrt 2 }}\left| {\frac{{{\beta _{\rm{s}}}{e^{{\rm{i}}{\theta _{\rm{s}}}}}}}{{\sqrt 2 }}} \right.} \right\rangle {{\left( {\left\langle {\sqrt {\frac{{{I_{\max }}}}{2}} {e^{{\rm{i}}{\theta _{\rm{X}}}}}\left| {\sqrt {\frac{{{I_{\max }}}}{2}} {e^{{\rm{i}}{\theta _{\rm{Z}}}}}} \right.} \right\rangle } \right)}^2}} \right|}^2}} }}\\
&=\sqrt {1 - \exp \left\{ {\frac{{{I_{\max }}}}{{8{\gamma ^{\rm{s}}}}}\Theta(\gamma ^{\rm{s}},\gamma ^{j_{\rm{A}}},\gamma ^{j_{\rm{B}}},\theta_{\rm Z},\theta_{\rm X})} \right\}},
\end{array}
\end{equation}
where
\begin{equation}
\begin{array}{l}
\Theta(\gamma ^{\rm{s}},\gamma ^{j_{\rm{A}}},\gamma ^{j_{\rm{B}}},\theta_{\rm Z},\theta_{\rm X})=\\
{\sqrt {{\gamma ^{\rm{s}}}{\gamma ^{{j_{\rm{A}}}}}} \cos \left( {{\theta _{{{\rm{s}}}}} - {\theta _{{{{j_{\rm{A}}}}}}}} \right) + \sqrt {{\gamma ^{\rm{s}}}{\gamma ^{{j_{\rm{B}}}}}} \cos \left( {{\theta _{{{\rm{s}}}}} - {\theta _{{{{j_{\rm{B}}}}}}}} \right) + 2{\gamma ^{\rm{s}}}\cos \left( {{\theta _{\rm{Z}}} - {\theta _{\rm{X}}}} \right) - {\gamma ^{{j_{\rm{A}}}}} - {\gamma ^{{j_{\rm{B}}}}} - 2}.
\end{array}
\end{equation}

\textbf{Case 3}

In this last case, we further assume that the following constraints hold: $I_{\rm max}\leq \rm{log}2$ and $0<\gamma^{\rm w}\leq \gamma^{\rm v}\leq \gamma^{\rm s}$ for any $P_{\rm{cut}}\ge 1$. According to the definition of the states given by Eq.~(\ref{MDICase3}), it turns out that $D^{{j_{\rm{A}}}{j_{\rm{B}}},{\rm{ss}}}$ is given by
\begin{equation}
\begin{array}{*{20}{l}}
{{D^{{j_{\rm{A}}}{j_{\rm{B}}},{\rm{ss}}}}}&{{ = \sqrt {1 - {{\left| {\left\langle {\frac{{{\beta _{{j_{\rm{A}}}}}{e^{{\rm{i}}{\theta _{{j_{\rm{A}}}}}}}}}{{\sqrt 2 }}\left| {\frac{{{\beta _{\rm{s}}}{e^{{\rm{i}}{\theta _{\rm{s}}}}}}}{{\sqrt 2 }}} \right.} \right\rangle \left\langle {\frac{{{\beta _{{j_{\rm{B}}}}}{e^{{\rm{i}}{\theta _{{j_{\rm{B}}}}}}}}}{{\sqrt 2 }}\left| {\frac{{{\beta _{\rm{s}}}{e^{{\rm{i}}{\theta _{\rm{s}}}}}}}{{\sqrt 2 }}} \right.} \right\rangle {{\left( {\left\langle {\sqrt {\frac{{{I_{\max }}}}{2}} {e^{{\rm{i}}{\theta _{\rm{X}}}}}\left| {\sqrt {\frac{{{I_{\max }}}}{2}} {e^{{\rm{i}}{\theta _{\rm{Z}}}}}} \right.} \right\rangle } \right)}^2}} \right|}^2}}}}\\
&{ = \frac{1}{2}\sum\limits_{n,m = 0}^\infty  \bigg| \left\{ {\exp \left[ { - \frac{1}{2}\left( {\beta _{{j_{\rm{A}}}}^2 + {I_{{\rm{max}}}}} \right)} \right]\frac{{{{\left( {\beta _{{j_{\rm{A}}}}^2 + {I_{{\rm{max}}}}} \right)}^n}}}{{n!}} - \exp \left( { - \beta _{\rm{s}}^2} \right)\frac{{{{\left( {{\beta _{\rm{s}}}} \right)}^{2n}}}}{{n!}}} \right\}}\\
{}&{{{\kern 12pt}}\left\{ {\exp \left[ { - \frac{1}{2}\left( {\beta _{{j_{\rm{B}}}}^2 + {I_{{\rm{max}}}}} \right)} \right]\frac{{{{\left( {\beta _{{j_{\rm{B}}}}^2 + {I_{{\rm{max}}}}} \right)}^m}}}{{m!}} - \exp \left( { - \beta _{\rm{s}}^2} \right)\frac{{{{\left( {{\beta _{\rm{s}}}} \right)}^{2m}}}}{{m!}}} \right\}\bigg|}\\
{}&{ = \frac{{\exp \left( { - {I_{{\rm{max}}}}/2} \right)}}{2}\sum\limits_{n,m = 0}^\infty  {\frac{{{{\left( {{I_{{\rm{max}}}}} \right)}^{n + m}}}}{{n!m!}}} \bigg|\left\{ {1 - \exp \left[ {{I_{{\rm{max}}}}\left( {1 - \frac{{{\gamma ^{{j_{\rm{A}}}}} + {\gamma ^{\rm{s}}}}}{{2{\gamma ^{\rm{s}}}}}} \right)} \right]\frac{{{{\left( {{\gamma ^{{j_{\rm{A}}}}} + {\gamma ^{\rm{s}}}} \right)}^n}}}{{{{\left( {2{\gamma ^{\rm{s}}}} \right)}^n}}}} \right\}}\\
{}&{{{\kern 12pt}} \times \left\{ {1 - \exp \left[ {{I_{{\rm{max}}}}\left( {1 - \frac{{{\gamma ^{{j_{\rm{B}}}}} + {\gamma ^{\rm{s}}}}}{{2{\gamma ^{\rm{s}}}}}} \right)} \right]\frac{{{{\left( {{\gamma ^{{j_{\rm{B}}}}} + {\gamma ^{\rm{s}}}} \right)}^m}}}{{{{\left( {2{\gamma ^{\rm{s}}}} \right)}^m}}}} \right\}\bigg|}\\
{}&{ \le \frac{1}{2} - \frac{{\exp \left( { - {I_{{\rm{max}}}}/2} \right)}}{2}\;\;\;\sum\limits_{n,m = 0}^{{P_{{\rm{cut}}}}} {\frac{{{{\left( {{I_{{\rm{max}}}}} \right)}^{n + m}}}}{{n!m!}}\left[ {1 - \left| {1 - \exp \left[ {{I_{{\rm{max}}}}\left( {1 - \frac{{{\gamma ^{{j_{\rm{A}}}}} + {\gamma ^{\rm{s}}}}}{{2{\gamma ^{\rm{s}}}}}} \right)} \right]\frac{{{{\left( {{\gamma ^{{j_{\rm{A}}}}} + {\gamma ^{\rm{s}}}} \right)}^n}}}{{{{\left( {2{\gamma ^{\rm{s}}}} \right)}^n}}}} \right|} \right]} }\\
{}&{{{\kern 12pt}} \times \left[ {1 - \left| {1 - \exp \left[ {{I_{{\rm{max}}}}\left( {1 - \frac{{{\gamma ^{{j_{\rm{B}}}}} + {\gamma ^{\rm{s}}}}}{{2{\gamma ^{\rm{s}}}}}} \right)} \right]\frac{{{{\left( {{\gamma ^{{j_{\rm{B}}}}} + {\gamma ^{\rm{s}}}} \right)}^m}}}{{{{\left( {2{\gamma ^{\rm{s}}}} \right)}^m}}}} \right|} \right],}
\end{array}
\end{equation}
for any $P_{\rm cut} \geq0 $.

{\subsection{Calculation of the probability ${{\Pr }^j}\left( {{X_{{{\rm{A}}_{\rm{c}}}}} =  - \left| {{{\rm{X}}_{{{\rm{A}}_{\rm{c}}}}}{\rm{,sb}}} \right.} \right)$}}
According to Eq.~(\ref{xbstate}) in Sec.~\ref{pma} of the main text, if we consider the basis matched case, the normalized state has the form:
\begin{equation}
\begin{array}{*{20}{l}}
\left| {{\Psi ^j}} \right\rangle  &= \left( {\frac{{{p_{\rm{Z}}}}}{{\sqrt {p_{\rm{Z}}^2 + p_{\rm{X}}^2} }}{{\left| {{0_{\rm{Z}}}} \right\rangle }_{{{\rm{A}}_{\rm{c}}}}}{{\left| {\Psi _{\rm{Z}}^j} \right\rangle }_{{\rm{A}},{\rm{E}}}}{{\left| {\Psi _{\rm{Z}}^j} \right\rangle }_{{\rm{B}},{\rm{E}}}} + \frac{{{p_{\rm{X}}}}}{{\sqrt {p_{\rm{Z}}^2 + p_{\rm{X}}^2} }}{{\left| {{1_{\rm{Z}}}} \right\rangle }_{{{\rm{A}}_{\rm{c}}}}}{{\left| {\Psi _{\rm{X}}^j} \right\rangle }_{{\rm{A}},{\rm{E}}}}{{\left| {\Psi _{\rm{X}}^j} \right\rangle }_{{\rm{B}},{\rm{E}}}}} \right){\left| {{0_{\rm{Z}}}} \right\rangle _{{{\rm{A}}_{{\rm{ba}}}}}}\\
&  = \left( {\frac{{{p_{\rm{Z}}}}}{{\sqrt {p_{\rm{Z}}^2 + p_{\rm{X}}^2} }}\frac{{{{\left|  +  \right\rangle }_{{{\rm{A}}_{\rm{c}}}}} + {{\left|  -  \right\rangle }_{{{\rm{A}}_{\rm{c}}}}}}}{{\sqrt 2 }}{{\left| {\Psi _{\rm{Z}}^j} \right\rangle }_{{\rm{A}},{\rm{E}}}}{{\left| {\Psi _{\rm{Z}}^j} \right\rangle }_{{\rm{B}},{\rm{E}}}} + \frac{{{p_{\rm{X}}}}}{{\sqrt {p_{\rm{Z}}^2 + p_{\rm{X}}^2} }}\frac{{{{\left|  +  \right\rangle }_{{{\rm{A}}_{\rm{c}}}}} - {{\left|  -  \right\rangle }_{{{\rm{A}}_{\rm{c}}}}}}}{{\sqrt 2 }}{{\left| {{1_{\rm{Z}}}} \right\rangle }_{{{\rm{A}}_{\rm{c}}}}}{{\left| {\Psi _{\rm{X}}^j} \right\rangle }_{{\rm{A}},{\rm{E}}}}{{\left| {\Psi _{\rm{X}}^j} \right\rangle }_{{\rm{B}},{\rm{E}}}}} \right){\left| {{0_{\rm{Z}}}} \right\rangle _{{{\rm{A}}_{{\rm{ba}}}}}}\\
&  = \left[ {\frac{{{p_{\rm{Z}}}}}{{\sqrt {2\left( {p_{\rm{Z}}^2 + p_{\rm{X}}^2} \right)} }}{{\left| {\Psi _{\rm{Z}}^j} \right\rangle }_{{\rm{A}},{\rm{E}}}}{{\left| {\Psi _{\rm{Z}}^j} \right\rangle }_{{\rm{B}},{\rm{E}}}} + \frac{{{p_{\rm{X}}}}}{{\sqrt {2\left( {p_{\rm{Z}}^2 + p_{\rm{X}}^2} \right)} }}{{\left| {\Psi _{\rm{X}}^j} \right\rangle }_{{\rm{A}},{\rm{E}}}}{{\left| {\Psi _{\rm{X}}^j} \right\rangle }_{{\rm{B}},{\rm{E}}}}} \right]{\left|  +  \right\rangle _{{{\rm{A}}_{\rm{c}}}}}{\left| {{0_{\rm{Z}}}} \right\rangle _{{{\rm{A}}_{{\rm{ba}}}}}}\\
&{\kern 11pt}  + \left[ {\frac{{{p_{\rm{Z}}}}}{{\sqrt {2\left( {p_{\rm{Z}}^2 + p_{\rm{X}}^2} \right)} }}{{\left| {\Psi _{\rm{Z}}^j} \right\rangle }_{{\rm{A}},{\rm{E}}}}{{\left| {\Psi _{\rm{Z}}^j} \right\rangle }_{{\rm{B}},{\rm{E}}}} - \frac{{{p_{\rm{X}}}}}{{\sqrt {2\left( {p_{\rm{Z}}^2 + p_{\rm{X}}^2} \right)} }}{{\left| {\Psi _{\rm{X}}^j} \right\rangle }_{{\rm{A}},{\rm{E}}}}{{\left| {\Psi _{\rm{X}}^j} \right\rangle }_{{\rm{B}},{\rm{E}}}}} \right]{\left|  -  \right\rangle _{{{\rm{A}}_{\rm{c}}}}}{\left| {{0_{\rm{Z}}}} \right\rangle _{{{\rm{A}}_{{\rm{ba}}}}}},
\end{array}
\end{equation}
where ${{{\left|  +  \right\rangle }_{{{\rm{A}}_{\rm{c}}}}}}$ and ${{{\left|  -  \right\rangle }_{{{\rm{A}}_{\rm{c}}}}}}$ are the two eigenstates of the quantum coin in the $\rm X_{A_c}$ basis with ${{{\left|  +  \right\rangle }_{{{\rm{A}}_{\rm{c}}}}}}=\frac{{{{\left| {{0_{\rm{Z}}}} \right\rangle }_{{{\rm{A}}_{\rm{c}}}}}}+{{{\left| {{1_{\rm{Z}}}} \right\rangle }_{{{\rm{A}}_{\rm{c}}}}}}}{\sqrt{2}}$ and ${{{\left|  -  \right\rangle }_{{{\rm{A}}_{\rm{c}}}}}}=\frac{{{{\left| {{0_{\rm{Z}}}} \right\rangle }_{{{\rm{A}}_{\rm{c}}}}}}-{{{\left| {{1_{\rm{Z}}}} \right\rangle }_{{{\rm{A}}_{\rm{c}}}}}}}{\sqrt{2}}$.

In each round, the conditional probability ${{{\Pr }^j}\left( {{X_{{{\rm{A}}_{\rm{c}}}}} =  - \left| {{{\rm{X}}_{{{\rm{A}}_{\rm{c}}}}}{\rm{,sb}}} \right.} \right)}$ is given by
\begin{equation}
\begin{array}{*{20}{l}}
{ {\Pr }^j \left( {{X_{{{\rm{A}}_{\rm{c}}}}} =  - \left| {{{\rm{X}}_{{{\rm{A}}_{\rm{c}}}}},{\rm{sb}}} \right.} \right)}\\
{ = [\frac{{{{\left( {{p_{\rm{Z}}}} \right)}^2}}}{{2\left( {p_{\rm{Z}}^2 + p_{\rm{X}}^2} \right)}}\langle \Psi _{\rm{Z}}^j|{{\left. {\Psi _{\rm{Z}}^j} \right\rangle }_{{\rm{A}},{\rm{E}}}}\langle \Psi _{\rm{Z}}^j|{{\left. {\Psi _{\rm{Z}}^j} \right\rangle }_{{\rm{B}},{\rm{E}}}} + \frac{{{{\left( {{p_{\rm{X}}}} \right)}^2}}}{{2\left( {p_{\rm{Z}}^2 + p_{\rm{X}}^2} \right)}}\langle \Psi _{\rm{X}}^j|{{\left. {\Psi _{\rm{X}}^j} \right\rangle }_{{\rm{A}},{\rm{E}}}}\langle \Psi _{\rm{X}}^j|{{\left. {\Psi _{\rm{X}}^j} \right\rangle }_{{\rm{B}},{\rm{E}}}}}\\
{\kern 12pt}{ - \frac{2{{p_{\rm{Z}}}{p_{\rm{X}}}}}{{\left( {p_{\rm{Z}}^2 + p_{\rm{X}}^2} \right)}}{\rm{Re}}\left( {\langle \Psi _{\rm{Z}}^j|{{\left. {\Psi _{\rm{X}}^j} \right\rangle }_{{\rm{A}},{\rm{E}}}}\langle \Psi _{\rm{Z}}^j|{{\left. {\Psi _{\rm{X}}^j} \right\rangle }_{{\rm{B}},{\rm{E}}}}} \right)]}\\
{ = \frac{1}{2}\left[ {1 - \frac{2{{p_{\rm{Z}}}{p_{\rm{X}}}}}{{p_{\rm{Z}}^2 + p_{\rm{X}}^2}}{\rm{Re}}\left( {\langle \Psi _{\rm{Z}}^j|{{\left. {\Psi _{\rm{X}}^j} \right\rangle }_{{\rm{A}},{\rm{E}}}}\langle \Psi _{\rm{Z}}^j|{{\left. {\Psi _{\rm{X}}^j} \right\rangle }_{{\rm{B}},{\rm{E}}}}} \right)} \right].}
\end{array}\label{XAc}
\end{equation}

Then we have that
\begin{equation}
\mathop {\max }\limits_{j \in \left\{ {1,2,...N} \right\}}  {\Pr }^j \left( {{X_{{{\rm{A}}_{\rm{c}}}}} =  - \left| {{{\rm{X}}_{{{\rm{A}}_{\rm{c}}}}},{\rm{sb}}} \right.} \right) = \frac{1}{2}\left\{ {1 - \frac{{2{p_{\rm{Z}}}{p_{\rm{X}}}}}{{p_{\rm{Z}}^2 + p_{\rm{X}}^2}}\mathop {\min }\limits_{j \in \left\{ {1,2,...N} \right\}} {\rm{Re}}\left( {\langle \Psi _{\rm{Z}}^j|{{\left. {\Psi _{\rm{X}}^j} \right\rangle }_{{\rm{A}},{\rm{E}}}}\langle \Psi _{\rm{Z}}^j|{{\left. {\Psi _{\rm{X}}^j} \right\rangle }_{{\rm{B}},{\rm{E}}}}} \right)} \right\}.
\end{equation}

\bibliography{MDIBib}

\end{document}